\documentclass[a4paper,12pt,oneside]{book}

\usepackage[a4paper,left=3cm,right=2cm,top=3cm,bottom=2cm]{geometry}

\usepackage{setspace}
\usepackage{amsmath,amsthm,amsfonts,amssymb}
\usepackage[mathcal]{eucal}
\usepackage{latexsym}
\usepackage{bm}
\usepackage{graphicx}
\usepackage[figurewithin=none,tablewithin=none]{caption}
\usepackage{float}
\usepackage{wasysym} 

\usepackage{amssymb}
\usepackage[all]{xy}
\usepackage{makeidx} 
\usepackage{epigraph}
\usepackage[Sonny]{fncychap}
\usepackage{tikz-cd}
\usepackage{hyperref}
\usepackage{calligra}
\usepackage[T1]{fontenc}
\usepackage{times}

\usepackage{verbatim} 

\makeindex 

\author{Osmar do Nascimento Souza\\ \texttt{}}

\title{Layers of planar hexagonal heterostructures
modeled by quantum graphs }
\theoremstyle{plain}
\newtheorem{theorem}{Theorem}[section]

\newtheorem{corollary}{Corollary}[section]
\newtheorem{lemma}{Lemma}[section]

\newtheorem{proposition}{Proposition}[section]

\theoremstyle{definition}
\newtheorem{definition}{Definition}[section]

\newtheorem{observation}{Observation}[section]

\newcommand{\R}{\mathbb{R}}
\newcommand{\N}{\mathbb{N}}
\newcommand{\Z}{\mathbb{Z}}
\newcommand{\C}{\mathbb{C}}

\makeatletter
\def\@evenhead{ 
	\underline{\hbox to \textwidth{{\rm \thepage} \sl \hfill \leftmark}}}
\def\@oddhead{
	\underline{\hbox to \textwidth{\hbox{}\sl \rightmark \hfill\rm\thepage}}}

\usepackage{xspace,setspace}
\usepackage[rm]{titlesec}
\usepackage{fancybox}

 \font\numberfont= pzcmi scaled 3000
\titleformat{\chapter}[display]
{\normalfont\Large}
{\filright
	\rule[32pt]{.7 \linewidth}{4pt}\hspace{-5pt}
	\shadowbox{
		\begin{minipage}{.15\linewidth}
			\begin{center}
				\textsl{\large\chaptertitlename}\\
				\vspace{1ex}
				{\numberfont \thechapter}\\
				\vspace{1ex}
			\end{center}
		\end{minipage}
	}
}
{-10pt}
{\filcenter \sl \Huge}
[\vskip-2.8ex\hfill\rule{.8\textwidth}{0.5pt}\\
\vskip-2.8ex\hfill\rule{.7\textwidth}{4pt}]
\titlespacing{\chapter}{0pt}{*4}{*1}

\newcommand{\dis}{\displaystyle}

\usepackage{tocloft, blindtext} 
\renewcommand{\cftfigpresnum}{Figura }      
\begin{document}

\thispagestyle{empty}
\begin{center}
{\bf\Large UNIVERSIDADE FEDERAL DE SÃO CARLOS \\  CENTRO DE CIÊNCIAS EXATAS E DE TECNOLOGIA  \\ PROGRAMA DE PÓS-GRADUAÇÃO EM MATEMÁTICA}

\end{center}\vspace{5cm}

\begin{center}

\textsc{\bf\Large OSMAR DO NASCIMENTO SOUZA }

\vspace{2cm}

{\bf\LARGE Layers of planar hexagonal heterostructures modeled by quantum graphs}
\end{center}

\vspace{2cm}

\vfill{\begin{center} {\large\bf S\~ao Carlos - SP\\ 
 \text{August of 2022}}
\end{center}

\newpage
\setcounter{page}{1}
\thispagestyle{empty}
\begin{center}
{\bf\Large UNIVERSIDADE FEDERAL DE SÃO CARLOS \\  CENTRO DE CIÊNCIAS EXATAS E DE TECNOLOGIA  \\ PROGRAMA DE PÓS-GRADUAÇÃO EM MATEMÁTICA}

\vspace{3cm}

\textsc{\bf\Large OSMAR DO NASCIMENTO SOUZA}

\vspace{2cm}

{\bf\LARGE Layers of planar hexagonal heterostructures modeled by quantum graphs}

\vspace{2cm}
\end{center}

\begin{flushright}
\begin{minipage}[b]{8,5cm}{\normalsize
Thesis submitted to PPGM/UFSCar as partial fulfillment of the requirements for the degree of
Doctor of Science.\\

Supervisor: Prof. Dr. César Rogerio de Oliveira}
\end{minipage}
\end{flushright}

\begin{center}
\vfill {\large\bf S\~ao Carlos - SP\\ 
 \text{August of 2022}}
\end{center}

\newpage 
\thispagestyle{empty}
\vspace*{0.75\textheight}

\begin{flushright}
 
To my sister Maria de F\'atima (\textit{in memorian}).

\end{flushright}


\chapter*{Acknowledgment}
\par I eternally thank God for all His Love, for the gift of life, for His infinite Kindness and Mercy and for not giving up on me, and to Mary, the Mother of God, for her ``Yes'', and for her mother's loving intercession.

\par I would like to say a huge thank you to my parents, Ant\^onio e Luiza, for their generosity in being open to life allowing the author, the 7$^{\mathrm o}$ child, to come in to this word; to my siblings Maria Ivani, Osvaldo, Maria Hilda, Maria de F\'atima, Maria Nazar\'e and Jos\'e, and to my dears nephews Pedro, Laura, Ver\^onica, Nayele, Maria Luiza and N\'adia. Anyway, i am very grateful to each and every one of my family for the encouragement and, especially, for the prayers.

I also thank the teachers of the Federal University of S\~ao Carlos with whom I had the opportunity to study each subject and, in particular, teacher C\'esar Rog\'erio de Oliveira, for the guindance, disposition, mathematical teachings and,above all, for never backing down from difficulties of work and always guide with patience, good humor and confidence. I extend my thanks to my fellow professors of the Mathematics course at the Federal University of Mato Grosso do Sul, Campus do Pantanal, for the encouragement of studies; to my big friend John Adriano Silva Azeredo for his advices and spiritual directions; to Win\'icius Jacinto Alves, Altair Santos de Oliveira Tosti, Giovana Alves, Tallyta Ananda, Marcelo Augusto, Maicom Douglas Varella Costa, Luan Barbosa, Thalyne Rocha, Ricardo Menotti, Gustavo Figueira and L\'ivia Massula, Nilton Mendes and to the mineiros C\'esar and Priscila for bearing patiently my faults and for the continuous friendship. I add to my thanks the Parish Street Ministry of Nossa Senhora de F\'atima for the learning in each pastoral meeting and Opus Dei in Ribeir\~ao Preto for the insights and direction; to Vin\'icius Louren\c{c}o da Rocha for his disposition and help in the studies and, finally, to friend Thales Fernando Vilamaior Paiva for the fellowship in the teaching work, academic studies and for the daily conversations about Faith and others virtues in the room 109 in the Department of Mathematics at UFSCar.
\thispagestyle{empty}
\\
\begin{titlepage}
\vspace*{15cm}
\begin{flushright}
\begin{minipage}{10cm}
	``There are people who want to know just to know, and that is curiosity; others, to achieve fame, and that is vanity; others, to get rich with their science, and that is a nasty business; others to be built, and that is prudence; others, to built others, and that is charity.''\, - Saint Augustine \vspace*{0.02cm}
\\
\end{minipage}
\end{flushright}
\end{titlepage}

\newpage
\chapter*{Resumo}
\setcounter{page}{5}
\thispagestyle{empty}

Iniciamos este trabalho revisando uma aplica\c{c}\~ao da teoria de grafos qu\^anticos peri\'odicos para modelar monocamada de materiais hexagonais com par\^ametros $\delta_a$ e $\delta_b$ associados aos tipos de \'atomos distintos situados em seus v\'ertices. Verificamos que materiais dessa natureza possuem lacunas em suas bandas espectrais e expressamos o tamanho dessa abertura em fun\c{c}\~ao desses par\^ametros. Nos Cap{\'{\i}}tulos \ref{capEstrHomAA} e \ref{capEstrHomAAnovo} estendemos essa modelagem para bicamadas iguais, empilhadas no tipo $AA$ e $AA'$, e no Cap{\'{\i}}tulo~\ref{capHetNBG} estudamos heteroestruturas com duas camadas mistas e os ``sandu{\'{\i}}che'' grafeno-nitreto de boro hexagonal: folha de grafeno entre duas folhas de hBN, e hBN entre grafenos. Em cada uma dessas configura\c{c}\~oes, usamos o operador de Schr\"odinger com suas respectivas condi\c{c}\~oes de contorno e introduzimos um par\^ametro de intera\c{c}\~ao fraca  $t_0$ entre as conex\~oes de diferentes camadas. Analisamos analiticamente a rela\c{c}\~ao de dispers\~ao obtida nesses modelos quanto \`a exist\^encia de toques c\^onicos ou parab\'olicos e confirmamos, em modelos rigorosos, resultados conhecidos na literatura f{\'{\i}}sica, a saber: bicamadas de hBN n\~ao possuem cones de Dirac, por\'em no empilhamento $AA$ identificamos a presen\c{c}a de toques parab\'olicos. No caso de bicamadas mistas, nosso estudo permite concluir que a inclus\~ao de uma camada de hBN sobre uma de grafeno pode induzir um gap na folha de grafeno e expressamos a largura desse gap em fun\c{c}\~ao dos par\^ametros $t_0$ e $\delta_a$. No estudo dos ``sandu{\'{\i}}ches'', hBN-grafeno-hBN e grafeno-hBN-grafeno, para certos valores particulares dos par\^ametros, \mbox{veri}ficamos que a inclus\~ao de uma folha de grafeno entre duas de hBN n\~ao elimina o gap do hBN, mas induz uma redu\c{c}\~ao na largura da lacuna espectral em uma ordem de grandeza; por outro lado, no caso grafeno-hBN-grafeno, o cone do grafeno na origem prevalece neste sandu{\'{\i}}che, por\'em tamb\'em provocou lacunas nos outros cones de Dirac do grafeno. Tais resultados podem ser justificados pelo fato de que, nessas heteroestruturas, os \'atomos de carbono terem interagido com outros \'atomos inequivalentes de hBN, nitrog\^enio e boro, provocando redu\c{c}\~ao ou aumento das lacunas. Por fim, no \'ultimo cap{\'{\i}}tulo, consideramos grafos qu\^anticos hexagonais e adaptamos nossa proposta para incluir um campo magn\'etico na folha de hBN. Demonstramos que se o fluxo magn\'etico for constante na rede hexagonal e for m\'ultiplo racional de $2\pi$, ent\~ao existir\~ao valores desse fluxo de forma que, para certas condi\c{c}\~oes de contorno nos v\'ertices (modelando o hBN), os toques c\^onicos na rela\c{c}\~ao de dispers\~ao do operador deixar\~ao de existir e garantimos a exist\^encia de lacunas.\\

\noindent \textbf{Palavras-chave:} Grafos qu\^anticos; Rede hexagonal; Nitreto de boro; Grafeno; Heteroestruturas; Operador de Schrodinger; Campo magn\'etico; Cones de Dirac.

\chapter*{Abstract}				
\thispagestyle{empty}

We started this work by reviewing an application of periodic quantum graph theory to model monolayer hexagonal materials with $\delta_a$ and $\delta_b$ parameters associated with the different types of atoms located in their vertices. We verified that materials of this nature have gaps in their spectral bands and express the size of this opening according to these parameters. In the Chapters \ref{capEstrHomAA} and \ref{capEstrHomAAnovo}, extend this modeling to equal bilayers, stacked in type $AA$ and $AA'$, and in the Chapters \ref{capHetNBG}, we studied heterostructures with two mixed layers and the ``sandwich'' hexagonal boron graphene-nitride: a single graphene sheet between two layers of hBN, and a single hBN sheet between two layers of graphene. In each of these configurations, we use the Schr\"odinger operator with its respective boundary conditions and we introduced a weak $t_0$ interaction parameter between the connections of different layers. We analyzed initially the dispersion relationship obtained in these models regarding the existence of conical or parabolic touches and confirmed, in rigorous models, known results in the physical literature, namely: hBN bilayers do not have Dirac cones, but, in $AA$ stacking we identified the presence of parabolic touches. In the case of mixed bilayers, our study allows us to conclude that the inclusion of an hBN layer over a graphene layer can induce a gap in the graphene sheet and we express the width of this gap according to the parameters $t_0$ and $\delta_a$. In the study of ``sandwiches'', hBN-graphene-hBN and graphene-hBN-graphene, for certain particular values of the parameters, we found that the inclusion of a single graphene sheet between two sheets of hBN does not eliminate the gap of the hBN, but induces a reduction in the width of the spectral gap in an order of magnitude; by on the other hand, in the case graphene-hBN-graphene, the graphene cone at the origin prevails in this sandwich, but it also caused gaps in the other Dirac cones of the graphene. Such results can be justified by the fact that, in these heterostructures, carbon atoms have interacted with other inequivalent hBN, nitrogen, and boron atoms, causing a reduction or increase of the gaps. Finally, in the last chapter, we consider hexagonal quantum graphs and we adapt our proposal to include a magnetic field in the hBN sheet. We demonstrate that if the magnetic flux is constant in the hexagonal network and is a rational multiple of $2\pi$, then there will be values of this flux such that, for certain boundary conditions at the vertices (modeling the hBN), the conical touches in the operator scattering relation will cease to exist and we guarantee the existence of gaps.


\noindent \textbf{Keywords:} Quantum graphs; Hexagonal lattice; Boron nitride; Graphene; Heterostructures; Schrodinger operator; Magnetic field; Dirac cones.

\newpage 

\renewcommand\listfigurename{Lista de Figuras \vspace{-0.5cm} \thispagestyle{empty}} 
\thispagestyle{empty}
\listoffigures{\thispagestyle{empty}}

\onehalfspacing

\newpage 
\tableofcontents{\thispagestyle{empty}}
\pagestyle{headings}

\ \


\chapter*{Introduction}\label{capIntrod}

\addcontentsline{toc}{chapter}{Introduction}
Among the many features and motivations for studying graphene, is the fact that it is a highly resistant material (more inclusive than diamond), excellent thermal and electrical conductor, as well as a strong candidate to replace silicon and revolutionize the technology industry, besides that it is basis for others materials, like graphite (pencil tip) which is made up of several layers of graphene (see Figure \ref{grafite_grafeno}; basics references, including historical aspects of this important material are \cite{castroEtAl,katsnelsonBook}. However, maybe the most important property of the graphene is the presence of Dirac cones located at a finite number of points in the Brillouin zone, called \textit{Dirac points}, which are of particular importance for the description of the eletronics properties of this material. They are located exactly where the valence band and  the conduction band of graphene touch, and this touch occurs in a conical shape.

It is worth mentioning that the presence of a Dirac cone in the dispersions relations indicate that, in the corresponding energy region, the behavior of an electron in the material would be effectively described by a one-dimensional Dirac operator, which would lead to relativistics behaviors of that electron, justifying the great conduction of graphene and others materials on what shuch cones are present. See, for example, \cite{castroEtAl,katsnelsonBook} and, for a mathematical treatment of this phenomenonin a continuous model  \cite{FWdirac}, which was discussed in details by Fefferman and Weunstein \cite{FW1} and, being technically much more complicated than the graph model, it becomes more difficult for generalizations the heterostructure and several layers (our main objectivein this Thesis). The first study to predict the Dirac cones in graphene is not recent, it was by Wallace~\cite{wallace}, using tight-binding approximations and published in 1947; however, the ``two-dimensional'' experimental management of this material only occurred in 2004, which resulkted in a Nobel Prize in Physics in 2010 for two authors, after they isolated graphene experimentally, fact that boosted the growth in interest in studying this material.

In summary, the single layer of graphene is a semimetal and, therefore, it does not have an intrinsic gap  and this limits the number of possible applications. Thus, such material can´t be used alone as an transistor, because the electricity will flow constantly, in other words, the electricty flow cannot be turned off like other materials like silicon. The abscence of gap 
between the conduction band and the valence band for the graphene, with the presence Dirac points, leads to difficulties in using them in electronics systems and transistors, because it  ``never turns off''. Therewith, some ways to induce a gap in the single layer of graphene have been explored.

\vspace{-0.4cm}
	\begin{figure}[H]
		\centering
		\includegraphics[width=0.4\linewidth]{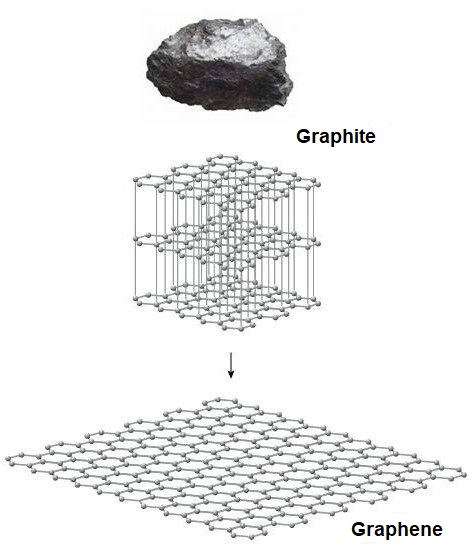}
		\caption{Graphene-graphite representation.}
		\label{grafite_grafeno}
	\end{figure}
On the other hand, it is known that single layers of hexagonal boron nitride -- which in the course of that work we will denote as hBN -- presents a great gap between the valence and the conduction bands, characterized as an insulator. In \cite{G.K.B.B.K}, the authors proposed to consider graphene-hexagonal boron nitride stacking (the interaction between layers is by van der Walls) as a way to induce gap in graphene. We mentioned here other very promising possibility, two ``twisted'' graphene layers (see ~\cite{TKV2019} and the references there cited there) which includes others aspects of interests, but it will not be covered  here.

From certain two-dimensional materials, differents heterostructures can be created, which essentially consist in using a $2D$ material (in fact a sheet with an atom thick) as a base and add another two-dimensional material on top of it. The resulting stack represents an artificial material assembled in a chosen sequence, as shown in Figure~\ref{FigWaals}. Covalent (strong) occur between atoms in the same layer, providing stability in the plane of $2D$ materials, while weak forces between atoms in overlapping layers (van der Waals bonds) hold the stack together. These staked materials are called \textit{van der Waals heterostructures} and the possibility of producing was demonstrated experimentally and recently \cite{FEIJO}. Throughout this work, when nothing is said, we will only use the term heterostructures to refer to  van der Waals heterostructure. 

	\begin{figure}[H]
		\centering
		\includegraphics[width=0.5\linewidth]{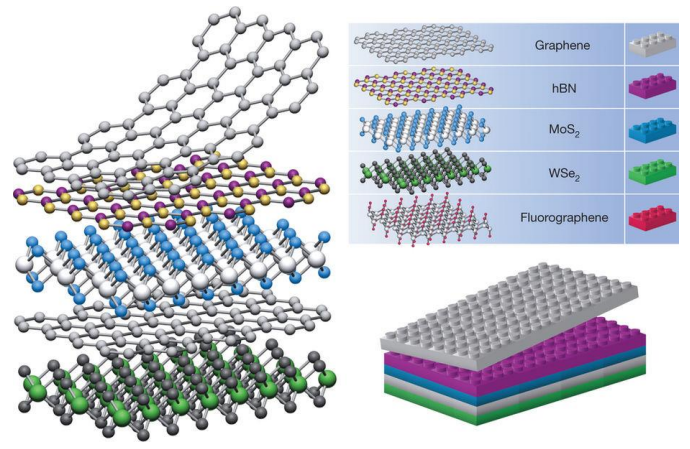}
		\caption{Van der Waals heterostructures. Comparison of Lego pieces with $2D$ materials, which can be stacked to form heterostructures linked by van der Waals forces (adapted from \cite{GEIM}).}
	\label{FigWaals}
	\end{figure}

 Theoretical and numerical works indicate that the eletronic properties of graphene can be modulated when considered these heterostructures formed by graphene-hBN for certain stacking configurations (which can lead to graphene-hBN devices with diferent mobilities \cite{MATOS}), so that it is possible predict the opening of gaps in the eletronic estructure of graphene since in the new structure graphene-hNB there are two types of atoms generating different potencials, making the carbon atoms inequivalent, that is, such atoms begin to interact with others atoms different from carbon and of different natures, such as nitrogen and boron.

There are several ways to stack $n$-layers of $2D$ hexagonal-shaped materials, like graphene or boron nitride, namely, $AB$ (also known as Bernal) and $AB'$ type stacking, however here we highlight the $AA$ and $AA'$ configurations, as shown in Figure~\ref{FigBetaDpoint}, where two atoms from thge sub-network $A$ and $B$ of the upper (layer labeled $A_2$ and $B_2$, respectively) are points located on the vertices of the hexagon, which we will say are vertices of type-$A$ and vertices of type-$B$, respectively.

\begin{figure}[H]
	\centering
  \includegraphics[width=0.6\linewidth]{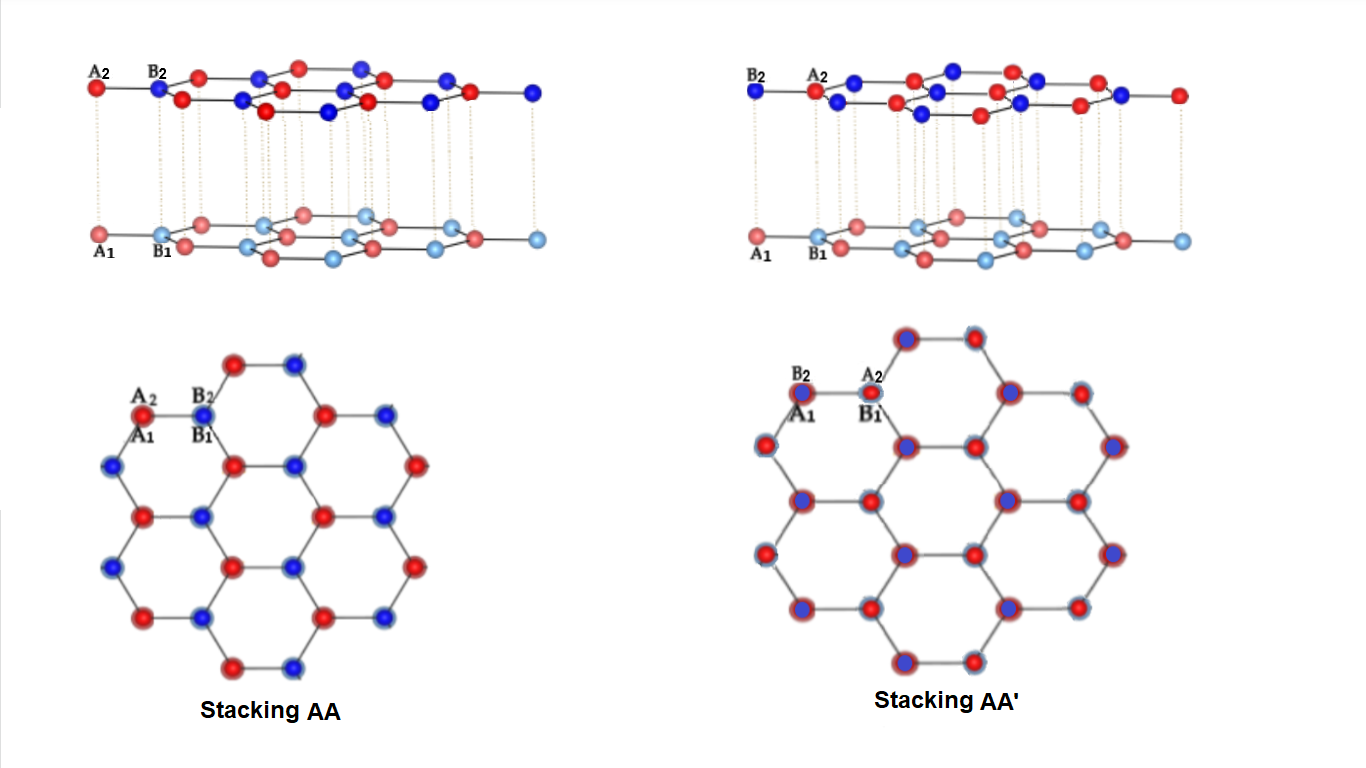}
\caption{Hexagonal bilayers - Side and top views, respectively, of the ~$AA$ and $AA'$ stacks.}
 \label{FigBetaDpoint}
 \end{figure}

\begin{itemize}

    \item Stacking of type~$AA$: all vertices are aligned so that a vertex of type-$A$ is located on a vertix of type-$A$; and on a vertex of type-$B$ there is one of type-$B$.

	\item Stacking of type~$AA'$: all vertices are ligned so that on a vertex of type-$A$ there is one of type-$B$, and on a vertex of type-$B$ there is one of type-$A$. \footnote{note that for graphene bilayers, the stacking types $AA$ and $AA'$ coincide.}
\end{itemize}

In \cite{ROCHA} the spectral characterization and excistence of Dirac cones were studied, using periodic quantum graph models, or even, through the Schr\"odinger with Neumann vertex condition, a graphen sheet was modeled, as weel as two and three layers of graphene stacked in the type $AB$ and also in $n$ graphene sheets in the  $AA$ stack. The study conclude that $n$-graphene layers have Dirac cones if aligned in the  $AA$ type and three sheets in the $AB$, but  they have only parabolic touches in bilayers~$AB$.

In relation to others mathematical results in literature, quantum graph models for graphene was also explored in ~\cite{shipman2020,FLS2021}. In the case of the $AA$ stack of graphene, in any numbers of layers, the presence of Dirac cones was mathematically demonstrated in ~\cite{shipman2020,FLS2021}, in  studies that focused in the (ir)reducibility of the dispersion relations; adapting the elegants arguments of symmetry of ~\cite{BC2018} another proof can be obtained, originally to the one-layer continuous model of Fefferman and Weinstein  mentioned above. We complete reinforcing that even in the case $AA$ for graphene (with Neumann boundary conditions), and including graphite, the cones were demonstrated in ~\cite{ROCHA, OLIRO}, but 
assuming that the interaction parameter between the layers, $t_0$, was small enough.  In the case of the $AB$ stack, the cones were mathematically discussed~\cite{deORochaAB} only in cases of three layers (with the existence cones, regardless of the value of the interaction parameter) and two layers (absence of cones and presence of parabolic touches).

 With this in mind, our study proposal here is to consider quantum graph models~\cite{KBlivro} (which simplify technicalities and result in several rigorous analytical results) of heterostructures with bi- and tri-layers identical or mixed,  such as layers of boron nitride together with graphene. In particular, our work, through a mathematical approach, in addition to contenplanting (within the necessary simplifications) systems found in laboratories and in nature, can go a step further and analyze the possible presence of Dirac cones in heterostructures with two and three mixed layers not yet discussed in the literature (as far as we know): graphene-hBN sandwiches formed by graphene between two sheets of hexagonal boron nitride, and hBN between two sheets of graphene.

 Motived by structures of this nature we propose to analytically analyze the existence or not of touches in the dispersions relations of operators associated with various configurations. These touches can be characterized as Dirac cones or parabolic touches. Intuitively, an Dirac cone\footnote{In chapter 1, we present the precise definition of Dirac cone, see\eqref{Cones de Dirac}} is a point at which two spectral bands touch each other linearly -- at least in a first order approximation -- while a parabolic touch is a point at which two spectral bands gently touch each other, in a parabolic way. 

 In \cite{KP1}, a layer of graphene was modeled using a periodic quantum Schr\"odinger graph with Neumann vertex condition, that is, in terms of Robin condition \eqref{Neumann} with parameter  $\delta_v = 0$ for all vertices, while in \cite{ROCHA,deORochaAB,OLIRO} such modeling is done considering $n$ sheets of graphene. Just like in \cite{ROCHA}, here we will also use a more general boundery condition, that is, a constant that is not necessarily zero $\delta_v$ at the vertices, to model the sime type of atom (different types of atoms will be modeled by differents values of~$\delta_v$).

  Figure \ref{fig:nit} illustrates the two-dimensional hexagonal boron nitride with hexagonal structure composed of nitrogen and boron atoms at its vertices. Let's say that the nitrogen atoms are located on the vertices of type-$A$, while the boron atoms are at the vertices of type-$B$. Thus, we use different values of the $\delta_v$ parameter in the Robin \eqref{Neumann} condition as way of distinguishing these atoms in the model: if $\delta_v$ is a vertice of type-$A$ (or type-$B$) we choose $\delta_v = \delta_a \quad  \big(\mbox{or}\,\,\delta_v = \delta_b\big)$, with $\delta_a \neq \delta_b$.

\begin{figure}[H]
		\centering
		\includegraphics[width=0.4\linewidth]{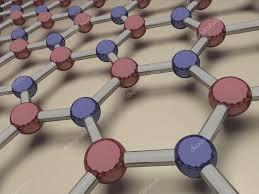}
		\caption{Representation of a hexagonal boron nitride sheet.}
	\label{fig:nit}
	\end{figure}

	{On the other hand, the hBN seen as a solid has a layered structure similar to graphite\footnote{{hexagonal boron nitride, due to its structure in layers, simillar to graphite, is also called "white graphite", however it has some properties entirely different from graphite; fact that arouses scientific interest in studying it, including our own interest.}} and is described by sequencial layers stacked in the form $AA'$, with the layers located one above the others, Figure \ref{hBN3D}.}

		\begin{figure}[H]
		\centering
		\includegraphics[width=0.5\linewidth]{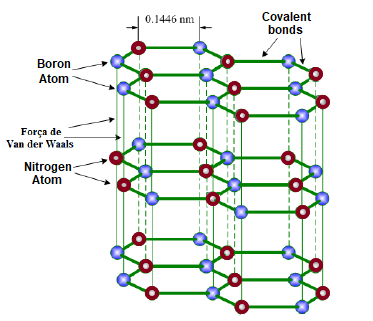}
		\caption{hBN structure seen as a 3D material: in each layer, boron and nitrogen atoms are linked by strong bonds, and these layers are also linked by van der Waals forces (Adapted from \cite{KKopeli}).}
	\label{hBN3D}
	\end{figure}

However, for the purposes of our work proposal, we will focus on the analysis of configurations relevant to the study motivation: models of the form $AA$ and $AA'$ for two sheets of hexagonal-shaped materials, such as hBN and graphene, two and three distinct layers. With this in mind, in Chapter\ref{capPrelimina}, we present in detail basic and necessary concepts that can be applied to model an hBN sheet (and consequently graphene); in the Chapters~\ref{capEstrHomAA} and~\ref{capEstrHomAAnovo}, we cover configurations of the type~$AA$ and $AA'$ for two identical $2D$ layers with two different types of atoms in the hexagonal net and, then, in the Chapter\ref{capHetNBG} we study these cases for two distinct sheets, the heterostructures; in particular, in its last two sections, we attack the central problem of this work: van der Walls heterostructures with three distinct layers that, in particular, can be applied to the graphene-hBN sandwich, which contributes to the current discussion regarding the production of materials composed of graphene and hexagonal boron nitride, which tt is still a point of debate and unresolved, with conflicting results in the literature.

Before proceeding, we emphasize that this is a mathematical work with an applied bias. With this in mind, we sometimes mix the nomenclature\footnote{For example, when the term \textbf{hexagonal boron nitride} is mentioned in the text, we are not necessarily referring to hBN, which is a real material, but rather a simple modeling proposal to portray it from different areas, which we consider natural.}


\chapter{Preliminary}\label{capPrelimina}

We begin this chapter with a quick abstract of the hexagonal structures which discribe a single sheet of two-dimensional materials like graphene and boron nitride, as already discussed in \cite{KP1} (which was based on~\cite{ALM2004}) and \cite{ROCHA}, respectively, is which concepts and preliminary results for the development of this work are also presented.
   
 \section[Monocamadas]{Boron nitride and graphene monolayers}  
 Quantum graphs are metric graphs with a self-adjoint operator on each edge; they are one-dimensional singular varieties together with  differential self-adjoint operator defined on them. We then present, in this section, the contruction of quantum graphos that can be proposed to represent a single hexagonal layer of graphene and hBN materials. In particular, similar to the work developed in \cite{ROCHA,deORocha2022}, we will focus, for the moment, on configurations composed of a hexagonal boron nitride sheet. We highlight that in ~\cite{ROCHA,deORocha2022} the Dirac operator was used in the vertices, while here we will use Schr\"odinger one.
 
 Let $E_1=\left(\frac{3}{2},\frac{\sqrt{3}}{2}\right)$ and $E_2=\left(0,\sqrt{3}\right)$ be  vectors in $\mathbb{R}^2$ and the points $A=(0,0)$, $B=(1,0)$. Consider the triangiular network
  $r=\Z E_1 \oplus \Z E_2$ and the subnets $r_{_A} = r + A$ and $r_{_B}= r + B$. So the hexagonal network $G$ that discribe one boron nitride sheet is $$
  G = r_{_A} \cup r_{_B} = \Big\{x \in \R^2: x=p_1E_1+p_2E_2+z, z \in W\Big\},
  $$ being $p=(p_1,p_2)\,\in \Z^2$ and $W=\{f, g, h, v_1,v_2\} $ the \textbf{fundamental network domain} of $G$, which is the set where the two verticies $v_1,\,v_2$ and the three edges $f,g$ e $h$ are conveniently directed, acordding to the Figuire~\ref{regiaow}. In general, a fundamental domain is the smaller domain (not unique) where the entire network is recovered from translations through the base vectores $E_1$ e $E_2$ (generators).

\begin{figure}[H]
	\centering
	\includegraphics[width=0.4\linewidth]{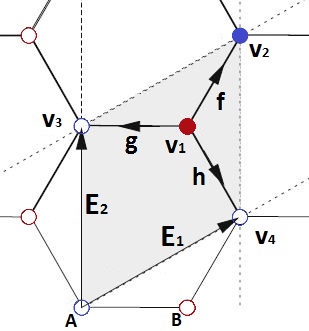}
	\caption{Hexagonal network $G$ of hBN (the same as graphene): the fundamental domain $W$ and the generating vectors of the network $E_1$ and $E_2$.}
\label{regiaow}
\end{figure}

We emphasize that in all our modeling, we are always assuming that all adges, both horizontal and vertical, has length~$1$. In particular, for graphene it is considered that the carbon atoms are located at the verticies of $G$ and the covalents bonds are represented by adges of length $1$; while for boron nitride, the verticies are interspersed with nitrogen atoms ($N$) anjd boron atoms ($B$) -- more precisely, a type of atom in the subnet~$r_B$ --, and with edges also assumed to be of length~1.

  We will denote, respectively, by $\mathcal{E}(G)$ and $\mathcal{V}(G)$ the set of edges and the set of verticies de $G$. We identify each edge with the interval $[0,1]$, so that we induce from $[0,1]$ a metric on $G$ and then we can consider $G$ embedded in $\R ^2$. So,

 \begin{itemize}
 	\item we have a natural measure in $G$, denoted by $dx$;
 	\item we can defien functions in $G$, denoted by $u=\{u_e\}_{e\, \in \, \mathcal{E}(G)}$;
 	\item we can differentiate and integrate functions in $G$. 
 \end{itemize}
 The Hilbert space is also defined:
 \begin{equation*}
 	L^2(G,{\C})=\bigoplus_{e\,\in \,\mathcal{E}(G)} L^2(e,{\C}), \quad  \mbox{with}\quad  L^2(e,{\C}) \approx L^2\big([0,1],{\C}\big).
 \end{equation*}
  Note that $$u \in L^2(G,{\C}) \Leftrightarrow \|u\|^2_{_{L^2(G,{\C})}} {:=}\dis\sum_{e \,\in\, E(G)}\!\!\|u_e\|^2_{{L^2(e,{\C})}}<\infty\, .$$
 
therefore, in this work a quantum graph is a metric graph $G$ equipped with an diferential operator~$H$ acting on $L^2(G,{\C})$, and with conditions of contour at the vertices so that~$H$ is self-adjoint. In each $e \in \mathcal{E}(G)$, the Schr\"odinger differential operator is considered,

\begin{equation}\label{Schro} 
 Hu_e = -\Delta u_e + q_0u_e\, ,
 \end{equation}
 with $q_0:[0,1]\rightarrow \R$ pair function (in relation to the center $x=\frac{1}{2}$) and  continuous, that is, $q_0(x) = q_0(1-x) \,\, \mbox{in} \,\, [0,1]$; 
 \begin{enumerate}
 	\item ${\mathrm {dom}}(H) \subset \mathcal{H}^2(G,{\C})$ \quad \quad  \Big( Sobolev space in $G$\Big);
 	\item $\dis\sum_{e \, \in \, \mathcal{E}(G)}\|u_e\|^2_{_{\mathcal{H}^2(e,{\C})}}<\infty$;
 	\item \textit{Robin condition at the vertex}
 	\begin{equation}\label{Neumann}\\
 		\left\{
 		\begin{array}{llll}
 			(i) \quad \quad \quad  u_{e_1}(v) \,\, = \,\, u_{e_2}(v) , \hspace{0.3cm} \forall \,\, e_1,e_2 \in \mathcal{E}_v(G) \,\, (\mbox{Continuity at the vertex})\\	
 			(ii) \dis\sum_{e \,\in \, \mathcal{E}_v(G)} \!\! u'_e(v)  = \delta_v u(v)\,, \hspace{1cm} (\mbox{Total flow})
 		\end{array}
 		\right.
 	\end{equation}
 \end{enumerate}
with $\delta_v$ a real number, $\mathcal{E}_v(G)$ the set of edges such that $v$ is commom vertex, and $u'_e(v)$ a derivative of $u_e$ along $e$ at the vertex $v$, which is is positive if $v$ is the initial vertex of $e$, and negative otherwise. When the context is clear, we will omit the subscript ``$e$'' to simplify the note. 


\section{Floquet Theory}

For each quasimoment $\theta= (\theta_1, \theta_2)$ in the Brillouin zone $B=[-\pi,\pi]^2$, there are Floquet operators, $H(\theta) = -\Delta + q_0 $, acting in the fundamental domain $W$, satisfying~\eqref{Neumann} and the cyclic condition Floquet for periodic operators:
 \vspace{0.3cm}
 \begin{equation}\label{Cond_Floquet}
 u(x+p_1E_1+p_2E_2)=e^{ip\cdot\theta}u(x)=e^{i(p_1\theta_1+p_2\theta_2)}u(x), 
 \end{equation}
 for all $p=(p_1,p_2)\in\mathbb{Z}^2$ and $x\in G$. It is known that  $H^\theta$ has a purely discrete spectrum~\cite{KP2},that is, $\sigma(H(\theta))=\big\{\lambda_k(\theta)\big\}_{k\geq 1},$ 
 com $\lambda_1(\theta) \leq \lambda_2(\theta) \leq \cdots  \leq \lambda_k(\theta) \leq  \cdots \rightarrow \infty $, when $k \to \infty $.
 
 We call  the union of the images of the functions $\theta \mapsto \{\lambda_k(\theta)\}_{k\geq 1}$. From Floquet Theory~\cite{EAST,KP3,REED}, we know that the spectrum of the Schr\"odinger operator $H$ defined in \ref{Schro}, $\sigma(H)$, \'is given by
 \begin{eqnarray}\label{EspGraf}
 	\sigma(H)=\bigcup_{\theta \,\in \, B}\sigma(H(\theta))\,.
 \end{eqnarray}
  Therefore, by \eqref{EspGraf}, to obtain $\sigma(H)$ just determine $\sigma(H(\theta))$, that is,  the spectrum problem of $H$ in the periodic graph is tranferred to the spectrum of $H(\theta)$ in a fundamental domain, so we have to solve the eigenvalue problem
 \begin{eqnarray}\label{AutoV}
 	H(\theta)u=\lambda u,
 \end{eqnarray}
 with $\lambda \in \R, u \in {\mathrm {dom}}(H(\theta)), \theta \in B$, being $u \in L^2(W)$ a function non-trivial that fulfills the \eqref{Neumann} boundary condition. 
 
As each edge $e  \in  \mathcal{E}(G)$ is identified with the interval $[0, 1]$, we have the relations $v_1 \sim \!0$ and $v_i \sim \!1$, being $v_1,v_i$ vertices, for $i = 2, 3, 4$, as shown in Figure~\ref{regiaow}. Therefore, we write the continuity conditions at the vertex $(i)$ in \eqref{Neumann} and the cyclic  condition \eqref{Cond_Floquet} in the  domain fundamental $W$, that is, 
 	\begin{equation}\label{Floquet}\\
 	\left\{
 	\begin{array}{llll}
   u_{a_1}(1)&=& u(v_2)=u(v_3+E_1)=e^{i\theta_1}u(v_3)=e^{i\theta_1}u_{a_2}(1)\\	
 	u_{a_1}(1)&=& u(v_2)= u(v_4+E_2)=e^{i\theta_2}u(v_4)=e^{i\theta_2}u_{a_3}(1)
 	\end{array}.
 	\right.
 	\end{equation}
 
 To exemplify what has been exposed so far, we will briefly review the modeling (proposed in our research group) of a boron nitride sheet, which consequently also applies to a graphene sheet for particular parameter choices, so More details can be found at ~\cite{ROCHA,KP1}. Using \eqref{Floquet}, we can rewrite the  \eqref{Neumann} and cyclic Floquet conditions 
 \eqref{Cond_Floquet} in $W$ as
  \begin{equation}\label{Sist.1}\\
  \left\{
  \begin{array}{llll}
  u_{a_1}(0)=u_{a_1}(0)=u_{a_1}(0)=:A_1\\
  u'_{a_1}(0) + u'_{a_2}(0) + u'_{a_3}(0)=\delta_aA_1\\	
  u_{a_1}(1)=e^{i\theta_1}u_{a_2}(1)=e^{i\theta_2}u_{a_3}(1)=:B_1\\
  -u'_{a_1}(1)-e^{i\theta_1}u'_{a_2}(1)-e^{i\theta_2}u'_{a_3}(1)=\delta_bB_1
  \end{array}.
  \right.
  \end{equation}
  Consider the auxiliary operator Schr\"odinger--Dirichlet $H^{D}$ in $[0,1]$, \begin{equation}\label{H-DiriSchro}
	H^{D}\!\!u=\frac{d^2u(x)}{dx^2}+q_0(x)u(x)\,,
	\end{equation} with Dirichlet boundary conditions $u(0) = u(1) = 0$. Denoting by $\sigma(H^{D})$ its spectrum, it is shown that, for each,  $\lambda \notin  \sigma(H^{D})$, there are functions \footnote{We will omit the letter $\lambda$ in the notations $\varphi_{\lambda,0},\varphi_{\lambda,1}$ when there is no need to emphasize its dependence on~$\lambda$.} linearly independent, $\varphi_{\lambda,0}:=\varphi_{_0}, \varphi_{\lambda,1}:=\varphi_{_1}$ so that they solve the eigenvalue problem
   \begin{equation}\label{Prob.AutoVal}
   H(\theta)\varphi=\lambda \varphi,\, \qquad \lambda \, \in \R, \varphi \neq 0,
   \end{equation}
    \begin{equation}
     \varphi_{_0}(0)=1=\varphi_{_1}(1)\,,  \varphi_{_1}(0)=0=\varphi_{_0}(1)\,\,\mbox{e}\,\, \varphi'_{_1}(x)=-\varphi'_{_0}(1-x)\,,\, \forall \,\, x \in [0,1].
    \end{equation}
   
   Furthermore, on each edge $a_i,\,i=1,2,3$ we obtain the representation
  \begin{equation}
  \left\{
  \begin{array}{lll}\label{Sist.2}
  u_{a_{1}}(x)&=&A_1\varphi_{_0}(x)+B_1\varphi_{_1}(x)\\
  u_{a_{2}}(x)&=&A_1\varphi_{_0}(x)+e^{-i\theta_1}B_1\varphi_{_1}(x)\\
  u_{a_{3}}(x)&=&A_1\varphi_{_0}(x)+e^{-i\theta_2}B_1\varphi_{_1}(x)
  \end{array}, 
  \right.
  \end{equation}
   so that this linear combination fulfills the continuity conditions in  \eqref{Sist.1} and the eigenvalue problem  \eqref{Prob.AutoVal}. Combining  \eqref{Sist.1} and \eqref{Sist.2}, we obtain

\begin{equation}
\left\{
\begin{array}{ll}\label{Sist.3}
(-3\eta-\alpha_a)A_1+\bar{F}B_1=0 \\
FA_1+(-3\eta-\alpha_b)B_1=0
\end{array}, 
\right.
\end{equation}
with $F=F(\theta):=1+e^{i\theta_1}+e^{i\theta_2}$, $\bar{F}$ its conjugate complex, $\eta:=\dfrac{\varphi'_{_1}(1)}{\varphi'_{_1}(0)}$ and $\alpha_i:=\dfrac{\delta_i}{\varphi'_{_1}(0)}\,,i=a,\,b$ (note that $\varphi'_{_1}(0)\ne0$).
Or, we can write \eqref{Sist.3} as $M(\eta(\lambda), \theta)\! \cdot X=\textbf{0}$, with
$$ 
M(\eta(\lambda),\theta)=\left(\begin{array}{cccc}
-3\eta-\alpha_a &	\bar{F} 	\\
F &  -3\eta-\alpha_b   
\end{array}\right),\,\quad X=(A_1,B_1)^{T}\,,\,\textbf{0}=(0,0)^T \,.$$
Furthermore, 
\[
p(\eta):=\det(M(\eta(\lambda),\theta)))=9\eta^2+3(\alpha_b+\alpha_a)\eta+\alpha_b\alpha_a-|F|^2
\] \'is a polynomial in $\eta(\lambda)$ whose roots are

\begin{equation}\label{R0AA'}
r^{\pm}(\theta)=\frac{-(\alpha_b+\alpha_a)\pm\sqrt{(\alpha_b-\alpha_a)^2+4|F|^2}}{6}\,.
\end{equation}

In other words, we see that there is  $\theta \in B$ so that \footnote{For simplicity, when there is no confusion, we will just use the notation $\eta(\theta)$ or $\eta(F)$ \big(since  $F=F(\theta)$\big) instead of  $\eta(\lambda)$. In fact, to $\lambda \notin \sigma(H^D)$, $\lambda \in \sigma(H) \Leftrightarrow $ exist $\theta \in B$ such that $\eta(\lambda)$ is a function of $\theta$, that is, $\eta(\lambda)=r(\theta)$, with $r(\theta)$ being a root of $P(\eta)=0$. This fact gives the dispersion relation of $H$.} $\det M(\eta(\lambda),\theta) = 0$, being $F=F(\theta)$ and, therefore, the representation \eqref{Sist.2}  solves the problem of eigenvalues \eqref{Prob.AutoVal} and, by \eqref{EspGraf}, it follows that  $\lambda \in \sigma(H)$, according to the following proposition, whose demonstration, can be found in \cite{ROCHA}, \cite{KP2}.

\begin{proposition}\label{prop_chave}
Let $\lambda \notin \, \sigma(H^{D})$.  Then $\lambda \in \, \sigma(H)$ if, and only if, there exists $\theta \, \in \,B$ such that $$\det(M(\eta(\lambda),\theta))=0\,.$$
\end{proposition}

\begin{observation}
For  $\lambda \in \sigma(H^{D}) $,  it is possible to construct~\cite{KP1} (infinite) eigenfunctions of~$H$, each one supported on a single hexagon, satisfying the continuity boundary conditions (together with Dirichlet); consequently,  $\lambda$ \ is an eigenvalue of~$H$. 
\end{observation}

We want to interpret these roots in terms of the original potential $q_0(x)$ em $[0,1]$. To this end, we consider two auxiliary operators: the Schr\"odinger-Dirichlet operator  $H^{D}$ given by  \eqref{H-DiriSchro} and the Hill operator, $H^{\mathrm{per}}$ in $\R$, which periodically extends $q_0(x)$,  given as in \eqref{Schro}, for the entire line $\R$,  with the resulting periodic potential $q_p$:
	\begin{equation}\label{Hill}
	H^{\mathrm{per}}u(x)=\frac{d^2u(x)}{dx^2}+q_p(x)u(x)\,, \quad u \in L^2(\R).
	\end{equation}
We are interested in the spectral problem
\begin{equation}\label{Hper}
H^{\mathrm{per}}u=\lambda u \,.
\end{equation}
Let us then consider the \textit{monodromy matrix} $M(\lambda)$ of $H^{\mathrm{per}}$ given by (see \cite{Floquettheory4BES})
\begin{equation}\label{Monodromia}
\begin{pmatrix}
\varphi(1) \\
\varphi'(1)
\end{pmatrix}=M(\lambda)\begin{pmatrix}
\varphi(0) \\
\varphi'(0)
\end{pmatrix}\,,
\end{equation}
with $\varphi$ being a solution to the problem \eqref{Hper}. The matrix $M(\lambda)$ shifts the solution  $\big(\varphi(0)\,\,\varphi'(0)\big)^T$ by the period of $q_{_p}$, equal to 1 in this case. It is
\[
d(\lambda)={\mathrm{tr}}(M(\lambda))
\] the discriminant of the Hill operator  $H^{\mathrm{per}}$. Of the main results and properties on the spectrum of $H^{\mathrm{per}}$ and $d(\lambda)$, we highlight the relationship between the spectra of $H^{\mathrm{per}}$ and its discriminant $d(\lambda)$:
\begin{equation}\label{EspHper}
\sigma(H^{\mathrm{per}})=\{\lambda \in \R: |d(\lambda)|\leq 2\}\,.
\end{equation}

Furthermore, it is also known that  $\sigma(H^{\mathrm{per}})$ is purely absolutely continuous and that $\sigma(H^{\mathrm{per}})$ is the union of intervals closed $B_k$,called  $\sigma(H^{\mathrm{per}})$ bands, so that, for $\lambda \in B_k, d'(\lambda) \neq 0$ and $d:B_k \rightarrow [-2,2] $ is a homeomorphism, for each $k$ (see \cite{KP1}, Proposition 3.4, and also \cite{Floquettheory4BES}, \cite{EAST}, \cite{KP3}, \cite{MAGNUS}, \cite{REED}).

In particular, the roots $r^{\pm}$ in \eqref{R0AA'} are related to the discriminant $d(\lambda)$ doof the periodic Hill operator $H^{\mathrm{per}}$, as shown in the following results.

\begin{lemma}\label{Lema2.2.2} (Lemma 2.2.2 in \cite{ROCHA})
	 Let $\lambda \notin \sigma(H^{^D})$ and $d(\lambda)$ be the discriminant of $H^{\mathrm{per}}$. So
	\begin{equation}
	\eta(\lambda) = \frac{d(\lambda)}{2}\,,
	\end{equation}
	where $\eta(\lambda)$ is a root in \eqref{R0AA'}.
\end{lemma}

\begin{lemma}\label{Lema2.2.3} (Lemma 2.2.3 in \cite{ROCHA})
	Let $\alpha_b=\alpha_a$. We have
	\begin{itemize}
		\item[(a)] If $\alpha_a \in [0,3]$, then $|r^+(\theta)|\leq 1\,,\,\, \forall \,\, \theta \, \in B_d$, 
		\item[(b)] If $\alpha_a \in [-3,0]$, then $|r^-(\theta)|\leq 1\,,\,\, \forall \,\, \theta \, \in B_d$,
		\item[(c)] If $\alpha_a \in [-3,3]$, then $r^+(\theta_D)=r^-(\theta_D)\,$, with $\theta_D=(\frac{2\pi}{3},-\frac{2\pi}{3})$, occur in $(-1,1)$. In particular, the only case where $|r^+(\theta)|\leq 1$ and $|r^-(\theta)|\leq 1$ occurs when $\alpha_b=\alpha_a=0$, in other words, the standard Neumann vertex condition and the original modeling~\cite{KP1} \textbf to represent the graphene material.
	\end{itemize} 
\end{lemma}

\begin{lemma}\label{Lema2.2.4} (Lemma 2.2.4 in \cite{ROCHA})\,
	Let $\alpha_b \neq \alpha_a$. Em \eqref{R0AA'}, we have
	\begin{itemize}
		\item[(a)] If $\alpha_b \in [-3,\infty]$ \,and $\frac{-3\alpha_b}{3+\alpha_b}\leq \alpha_a \leq 3$, then $|r^+(\theta)|\leq 1\,,\,\, \forall \,\, \theta \, \in B_d$.
		\item[(b)]  If $\alpha_b \in [-\infty, 3]$ \,and $-3\leq \alpha_a \leq \frac{-3\alpha_b}{3-\alpha_b}$, then $|r^-(\theta)|\leq 1\,,\,\, \forall \,\, \theta \, \in B_d$.
	\end{itemize} 
\end{lemma}

\begin{observation}\label{Crucial}
The Lemma \ref{Lema2.2.2} tells us that given  $\lambda \notin \sigma(H^D)$, $\lambda \in \sigma(H)$ if, and only if, $r(\theta)=\frac{1}{2}d(\lambda)$ is a solution of $\det(M(\eta(\lambda)))=0$. However, this information  alone  is not enough to describe the $\sigma(H)$ spectrum. We saw that  \eqref{EspHper} presents a complete characterization of $H^{\mathrm{per}}$ in terms of  $d(\lambda)$, this is, 
$\sigma(H^{\mathrm{per}})=\{\lambda \in \R: |d(\lambda)|\leq 2\}$ and, therefore, it is necessary to demand that $|r(\theta)|\leq 1$  in order to relate the spectra $\sigma(H)$ and $\sigma(H^{\mathrm{per}})$. That said, we see that the conditions in Lemmas \ref{Lema2.2.3} and \ref{Lema2.2.4} imply $|r(\theta)|\leq 1$ and, finally, the Spectral characterization of  $H$ can be guaranteed.
\end{observation}

These lemmas, together with the next theorems, further tell us that, under certain conditions of $\alpha_b$ and $\alpha_a$, the dispersion relation of $H$ has cone by Dirac; this is, a point 
$(\lambda, \theta_D)$ at which two spectral bands touch linearly, at least in a lower-order approximation:
$$\lambda(\theta)-\lambda(\theta_D) \approx \pm\gamma\!\cdot\! |\theta-\theta_D|\,,$$
for some $\gamma>0$, as we present precisely below.
\begin{definition}( Dirac Cones )\label{Cones de Dirac}\,\\
	A point $(\lambda, \theta_D)$  in the dispersion relation, for $\lambda \in \R$ and $\theta_D \in B$, is said \textbf{ Dirac cone } if it exists a constant $\gamma \neq 0$ such that 
	\begin{equation}\label{ConesDirac}
	\lambda(\theta) - \lambda(\theta_D) + \mathcal{O}((\lambda(\theta)-\lambda(\theta_D))^2) = \pm \!\gamma\! \cdot |\theta - \theta_{D}|+ \mathcal{O}(|\theta - \theta_{D}|^2).
	\end{equation}
	In this case, $\theta_D$ is said to be a D-point and, in  \eqref{ConesDirac}, the ``--'' \,\,and ``+''\, are signals for the bands of valence and conduction, respectively.
\end{definition}

The description of the spectral structure of the $H$ operator in $G$ is given by the following theorem.

\begin{theorem}\label{Teo2.2.1} (Theorem 3.6 in \cite{KP1})\,
	\begin{itemize}
		\item[(a)] The continuous singular spectrum of $H$, $\sigma_{\mathrm{sc}}(H)$, is empty.
		\item[(b)] Except, possibly for a countable quantity of $\lambda$, the dispersion relation of $H$ is given by
		\begin{equation}\label{R0AA'1}
		d(\lambda)=\frac{-(\alpha_b+\alpha_a)\pm\sqrt{(\alpha_b-\alpha_a)^2+4|F|^2}}{3}\,,
		\end{equation}
		with $\alpha_b$ and $\alpha_a$ taken so that $|r^{\pm}(\theta)|\leq 1$, for all $ \theta \in B$ and $r^{\pm}(\theta)$ given in~\eqref{R0AA'}.
		\item[(c)] The absolutely continuous spectrum of $H$, $\sigma_{\mathrm{ac}}(H)$, coincides, as a set, with $\sigma(H^{\mathrm{per}})$, that is, it has a "band-gap"\, structure \,and \begin{equation}\label{EspAbsCont}
		\sigma_{\mathrm{ac}}(H)=\{\lambda \in \R: |d(\lambda)|\leq 2\}\,,
		\end{equation} being $d(\lambda)$ is the discriminant of  $H^{per}$.
		\item[(d)] The pure point spectrum of $H$, $\sigma_{\mathrm{pp}}(H)$, coincides with $\sigma({H^D})$.
	\end{itemize}
\end{theorem}

\begin{observation}
	The countable set considered in item (b) of Theorem \ref{Teo2.2.1} is, in fact, formed by the eigenvalues of the operator $H$ with Dirichlet boundary conditions.
\end{observation}

\begin{theorem}\label{Teo2.3.1} (Theorem 2.3.1 em \cite{ROCHA})\,
	\begin{itemize}
		\item[(a)] If $\alpha_a=\alpha_b$, the dispersion relation of $H$ has Dirac cones for $\theta_0=\pm(\frac{2\pi}{3},- \frac{2\pi}{3 })$ (Figure \ref{Uma-folhab02n02}).
		\item[(b)] If $\alpha_a \neq \alpha_b$, the dispersion relation of $H$ does not have Dirac cones (Figure \ref{Uma-folhab02n05}).
	\end{itemize}
\end{theorem}

The idea of proving Theorem \ref{Teo2.3.1} is interesting and will serve as a basis for demonstrations throughout this work, in the cases of layers with two and three sheets of boron nitride. Thus, we will present the idea of your demonstration, omitting some technical accounts.

\begin{proof}
Essentially, the argument consists of taking $\theta \in B_d = \{(\theta_1,\theta_2) \in B: \theta_2=-\theta_1\}$ e $\lambda \notin \sigma(H^D)$\, \big(then $d'(\lambda)\neq 0$\big) and expanding the functions $r^{\pm}$ around the point $\theta_D=\pm(\frac{2\pi}{3},-\frac{2\pi}{3})$ and $d$ around $\lambda(\theta_0)$, with $\theta_0:=\pm\frac{2\pi}{3}$, thus obtaining
\begin{equation}
 r^{\pm}(\theta) -r^{\pm}(\theta_D)=\pm\frac{\sqrt{3}}{3}\big|\theta_1-\theta_0\big|+\mathcal{O}\big(|\theta_1-\theta_0|^2\big),
\end{equation}
and 
\begin{equation}\label{DerConeDirac}
	d(\lambda(\theta)) - d(\lambda(\theta_D))=d'(\lambda(\theta_D))\big(\lambda(\theta)-\lambda(\theta_D)\big)+ \mathcal{O}\big((\lambda(\theta)-\lambda(\theta_D))^2\big)\,.
\end{equation}

Combining these two expansions with the definition of a Dirac cone, we obtain
\begin{equation}
\lambda(\theta) - \lambda(\theta_D)+\mathcal{O}\big((\lambda(\theta)-\lambda(\theta_D))^2\big)=\pm \frac{2\sqrt{3}}{3d'(\lambda(\theta_D))}\big|\theta-\theta_D\big|+ \mathcal{O}\big(|\theta-\theta_D|^2\big)\,,
\end{equation}
so that in \eqref{ConesDirac} we obtain the non-zero constant,
\begin{equation}
\gamma=\frac{2\sqrt{3}}{3d'(\lambda(\theta_D))}\,.
\end{equation}

In the case where $\alpha_b \neq \alpha_a$, to demonstrate the non-existence of Dirac cones, it is enough to note that, by the Lemmas \ref{Lema2.2.2} and \ref{Lema2.2.4}, the functions $r^+(\theta)$ and $r^-(\theta)$ never touch each other for $\alpha_b \neq \alpha_a$. Furthermore, a simple calculation shows us that the derivative of $r^{\pm}$ vanishes at $\theta_D$ and, consequently, $r^{\pm}$ does not exhibit linear behavior around $\theta_D$.
\end{proof}

\begin{observation}\label{ConstGama} In general, the following relation is valid between the constant $\gamma$, obtained from the definition of Dirac Cones, and the constant $\gamma_{_D}$, obtained from the Taylor expansion:
	 $$\gamma=\dfrac{2\gamma_{_D}}{d'(\lambda(\theta_D))}\,,$$
	desde que $d'(\lambda(\theta_{D})) \neq 0$.
\end{observation}

A consequence of this theorem can be obtained if we consider that, in a hexagonal lattice, at the vertices of type-$A$ there is a nitrogen atom (N) and at the vertices of type-$B$ there is a boron atom (B), or that is, if we try to model boron nitride\footnote{We remember that here such modeling for hBN is a very simplified proposal for this material.}. Then, putting $\delta_N$ and $\delta_B$ the parameters (in the boundary conditions) associated with N and B, respectively, we have the following corollary.

\begin{corollary}\label{Rel-Disp-hBN} The dispersion relation of the operator that models boron nitride does not have Dirac cones if $\delta_N \neq \delta_B$ under vertex conditions. For $\delta_N = \delta_B$ (that would be graphene), the relation has Dirac cones.
\end{corollary}	

 \begin{figure}[h]
	\centering
	\includegraphics[width=0.3\linewidth]{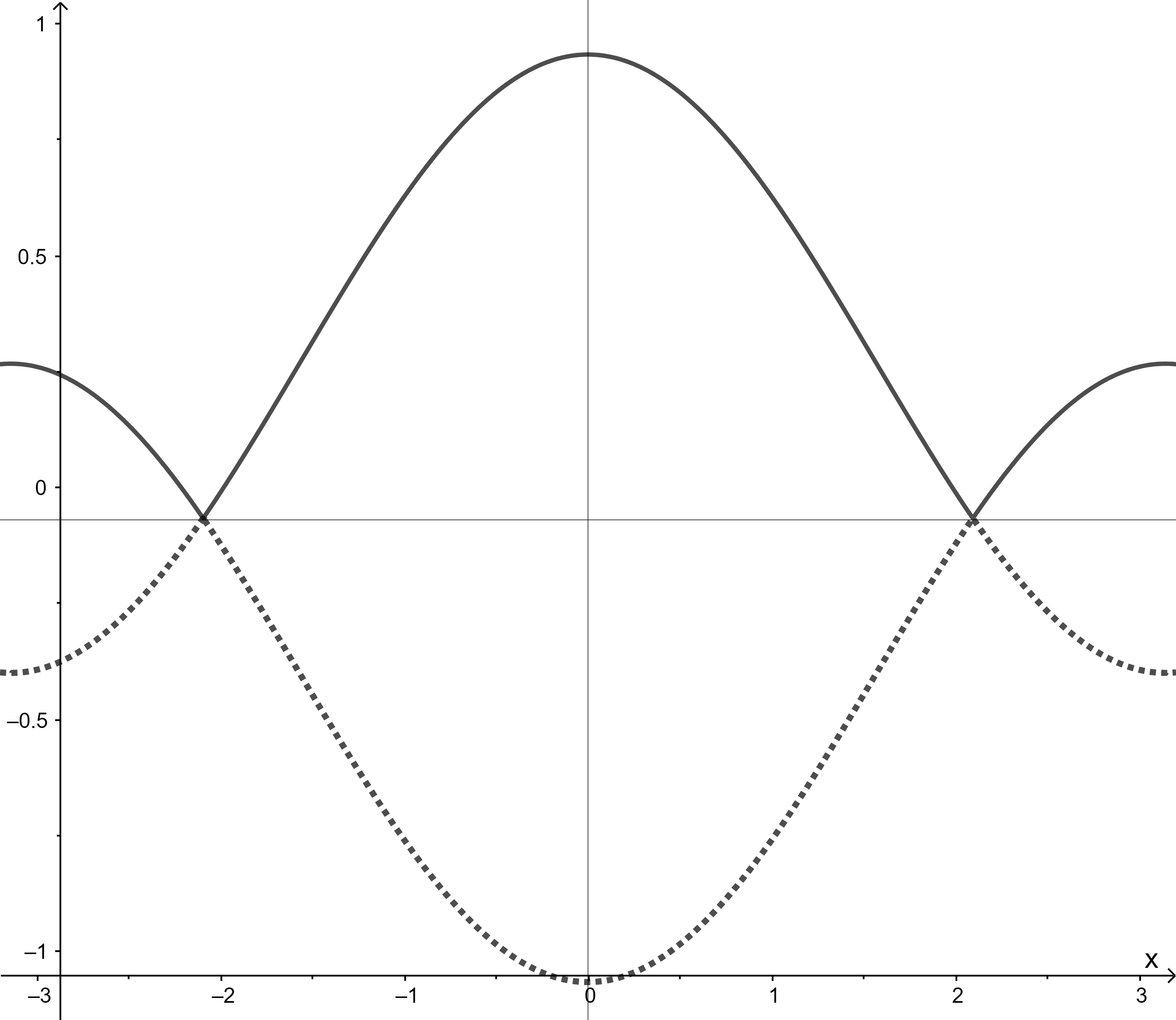}
	\caption{$\alpha_N =\alpha_B=0.2$; Dirac cone at $F=0$ is formed when the parameters $\alpha_N$ and $\alpha_B$ coincide.}
	\label{Uma-folhab02n02}
\end{figure}

\begin{figure}[h]
	\centering
	\includegraphics[width=0.3\linewidth]{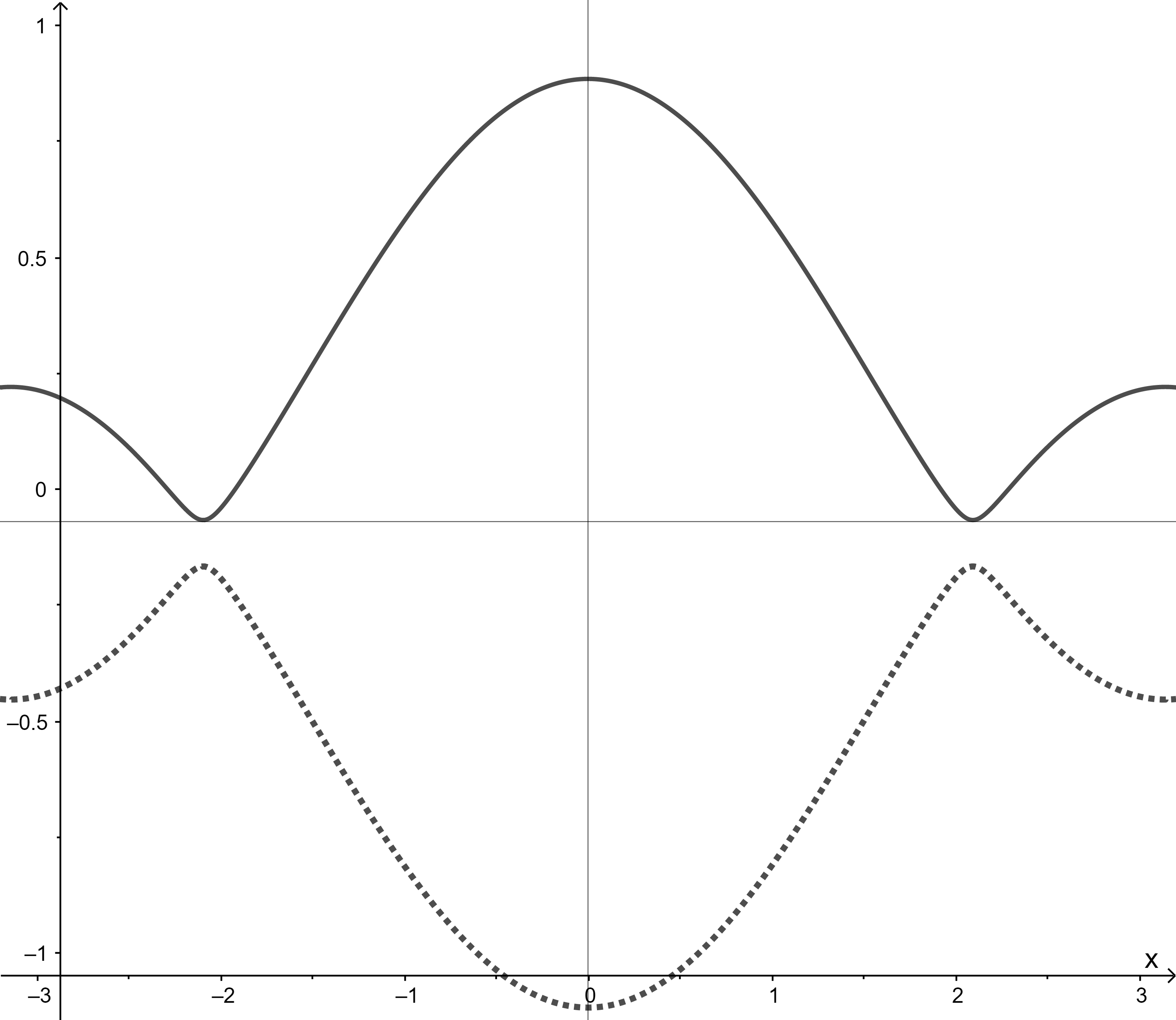}
	\caption{$\alpha_N \neq \alpha_B$: $\alpha_N=0.5,\alpha_B=0.2$; when the parameters $\alpha_N$ and $\alpha_B$ are distinct, gaps appear where there were Dirac cones}
	\label{Uma-folhab02n05}
\end{figure}

Note that we can determine the width of the gap\footnote{We do not have clear indications of what would be the values of the parameters $\delta_N$, $\delta_B$ or $\delta_C$ that \,``best describe'' the hBN and graphene, which makes it difficult to make a more precise numerical comparison of the results obtained in our model with those in the experimental physics literature.} when the atoms are distinct. In fact, for $\alpha_b=\alpha_a$ the conical touch occurs when $\theta=\dfrac{2\pi}{3}$, that is, $F=0$. Thus, taking $F=0$ in \eqref{R0AA'}, we obtain that the gap is given by
\begin{equation}\label{Gap0}
	g(\alpha_a,\alpha_b)=\dfrac{|\alpha_a-\alpha_b|}{3}\,.
\end{equation}
In particular, for $\alpha_b=-\alpha_a=1$, we have a gap of width ${2}/{3}\approx 0.667$. 

Throughout this work, we will make extensive use of this relationship to justify the presence of Dirac cones in dispersion relations of two-and three-layer stacks, including heterostructures. In general, to demonstrate that a point $(\lambda,\theta)$ in the dispersion relation is a Dirac cone, we will show that there is a constant $\gamma_{_D}\neq 0$ determined by the Taylor expansion, as we saw in this example, without directly mentioning the \eqref{ConstGama} relation above.

In the next chapter we extend the monolayer study, described here, to address stacks composed of two identical $2D$ hexagonal layers, generalizing some cases of graphene stacking described in \cite{deORochaAB,OLIRO}.


\chapter{Structures of type $AA$}\label{capEstrHomAA}

 Let us remember that the graphene bilayer consists of two vertically stacked graphene sheets (with a certain type of stacking) that weakly interact with each other, held together by van der Waals interactions. It is mainly found in the so-called $AB$ \cite{ROCHA} stacking, however there are other alternative stacking configurations, in which one layer is translated or rotated by an angle in relation to the other.

As described in \cite{SOUZA}, in these systems, in which layers of a material are stacked in van der Waals structures, it is possible to exist regions that are rich in $AB$ stacking and regions that mainly exhibit $AA$ stacking. In particular, graphene bilayers aligned in the $AA$ type have attracted theoretical interest from researchers because of experimental investigations that show the presence of this type of stacking in certain samples and suggest its experimental feasibility.

In this chapter we study $AA$ stackings with two equal layers in a hexagonal shape in which we take two different types of atoms interspersed at the vertices, of each hexagon, in each layer. We begin, in the first section, by considering configurations of two identical sheets stacked so that each vertex of type-$A$ on the lower sheet is overlapped and connected to a vertex of type-$A$ on the upper sheet, and each vertex of type-$B$ of the lower layer is superimposed and connected to a vertex of type-$B$ in the overlying sheet, as shown in Figure \ref{AA1}. We will say that two atoms are equivalent if they are identical (for example, both of type-$A$), otherwise we will say that they are inequivalent.

\vspace{-0.5cm}
\begin{figure}[H]
	\centering
	\includegraphics[width=0.7\linewidth]{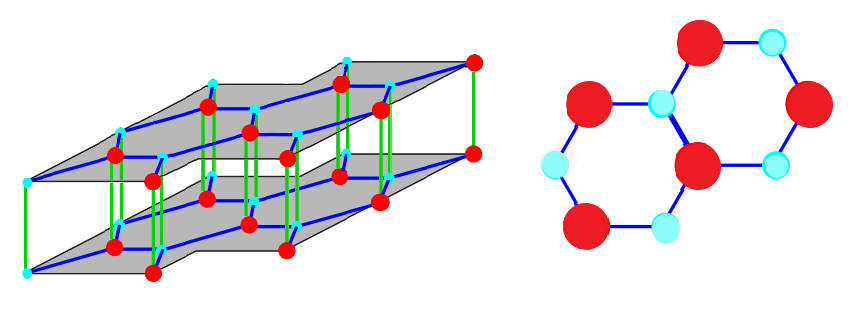}
	\caption{Side and top views, respectively, of material composed of two identical hexagonal layers aligned in type $AA$.}
	\label{AA1}
	\end{figure}

Notice in Figure \ref{AA1} that each atom in a sheet interacts weakly with another equivalent atom in the other layer. We emphasize that we do not aim here to judge whether such a stacking model makes physical sense or describes any real material found in nature. As it is a mathematical model, we are, \textit{a priori}, only interested in investigating the spectral properties and the possible presence of Dirac cones for such a hypothetical material in the form of a van der Waals heterostructure.

\section[Structure with two identical leaves]{Structure with two identical leaves}

 In this model we consider two identical layers stacked in the form AA such that an atom in the lower (upper) layer, in addition to interacting strongly with three other distinct atoms (first neighbors) in the same layer, also interacts weakly with another equivalent atom (i.e. similar atom) of the upper (lower) sheet, as shown in Figure \ref{AA2}.

\begin{figure}[H]
	\centering
	\includegraphics[width=0.6\linewidth]{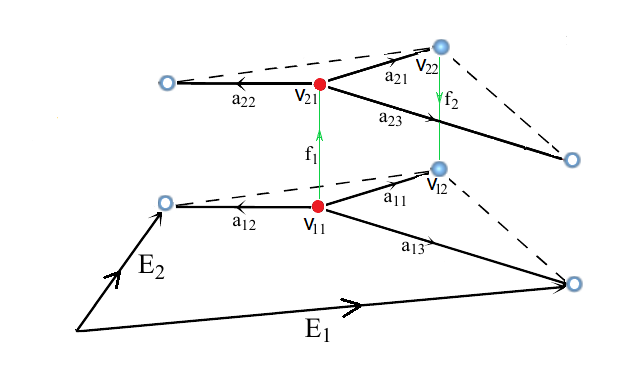}
	\caption{Material composed of two hexagonal layers stacked in the type $AA$.}
	\label{AA2}
\end{figure}

We will denote by $\bar{\mathcal{E}}(G^{AA})$ the set of edges in each leaf and by $\hat{\mathcal{E}}(G^{AA})$ the set of edges between two consecutive sheets; then we have the disjoint union
\[
 \mathcal{E}(G^{AA}) =  \bar{\mathcal{E}}(G^{AA}) \cup \hat{\mathcal{E}}(G^{AA}).
\] 
For each vertex $v$ of the fundamental domain of $G^{AA}$, the Robin conditions presented in Chapter \ref{capPrelimina} can be adapted and applied to two layers, and become what we will call \textit{ modified Neumann vertex condition}:
\begin{equation}\label{Robin}\\
	\left\{
	\begin{array}{ccll}
		(i)\quad  u_{a_1}(v) \quad  = \quad  u_{a_2}(v) \quad  =\quad  \dfrac{u_{f_1}(v)}{t_0}&=&\dfrac{u_{f_2}(v)}{t_0}, \hspace{0.5cm} \forall \quad  a_1,a_2,f_1,f_2 \in \mathcal{E}_v(G);\\	
		(ii) \dis\sum_{a \in \mathcal{E}_v(G)} \!\! u'_a(v) \pm \!\! \dis\sum_{f \in \hat{\mathcal{E}}_v(G)} t_0 u'_f(v)  &=& \delta_v u(v)
	\end{array}\,,
	\right. 
\end{equation}
where 
\begin{itemize}
	\item $\delta_v$ a real number;
	\item  $u'_a(v)$ the derivative of $u_a$ along the edge $a$ at the vertex $v$;
	\item  $0 < t_0 \leq 1$ a (weak) interaction parameter between consecutive leaves; sign $``+$\,'' in $\pm u'_a(v)$ if $v$ is a starting point of the edge, sign $``-$\,'' if it is the end point;
    \item ``\,$a$\,'' represents the edges in the leaves and $``\,f$\,'' the edges between two consecutive leaves.
\end{itemize}

In particular, in $(ii)$ of \eqref{Robin}, when $\delta_v=0$ for any $v \in \mathcal{V}(G)$, we have the Neumann vertex condition.

\begin{observation}\label{AA-AA'} We use a quantum graph model that, taken in its particular form, models graphene, as well as hexagonal boron nitride, of two or three sheets with $AA'$ and $AA$ stacking. We consider the Schr\"odinger operator described in \eqref{Schro} with the modified Neumann vertex condition, which requires weighted continuity of functions and for this we consider the parameter $t_0$, in the interval $(0,1]$, which also models the weak interaction between consecutive leaves. The proposed idea is that the smaller $t_0$ is, the weaker the interaction between such overlapping layers would be; this comes from the following reasoning: from \eqref{Robin}, the smaller $t_0$, the smaller the value of $|u_{f}(v)|$ that connects consecutive layers and the smaller the contribution of its derivative to the total flow in each vertex. We emphasize that we do not consider the value $t_0 = 0$, as in this case we have a singular limit and it is understood, intuitively, that the layers stop interacting. 

We also note that we could have considered two parameters $t_a$ and $t_b$, instead of a single parameter $t_0$, to model the two weak interactions between two consecutive hexagonal layers $\hexagon_1$ and $\hexagon_2$: $t_a$ between atoms $A$ of the network $\hexagon_1$ with atoms $A$ of the network $\hexagon_2$, and $t_b$ between atoms of type $B$. In this more general case, as discussed in the Appendix \ref{Apendice}, basically the results are repeated, but with different values of the parameters.

On the other hand, although in $AA$ stacking it is pertinent to consider two connection parameters $t_a$ and $t_b$ in the model, in $AA'$ stacking this becomes unnecessary, as we will see in more detail in the following chapter, because in $AA'$ (according to our modeling proposal) a single weak connection occurs: type $A$ atoms from the $\hexagon_1$ network with type $B$ atoms from the $\hexagon_2$ network and vice versa (we then use a single parameter $t_0:=t^{b}_{a}$).
\end{observation}

In the $AA$ stacking, the fundamental domain for two layers is formed by six horizontal edges, two vertical edges and the four initial and final vertices, that is, 
$$ 
W_2^{AA}=\{a_{11}, a_{12}, a_{13}, a_{21}, a_{22}, a_{23}, f_1, f_2, v_{11}, v_{12}, v_{21}, v_{22}\}\,.
$$ 

Let $H_2^{AA}$ be the Schr\"odinger operator associated with stacking of the type $AA$ with two identical layers. As already mentioned in the previous chapter and with analogous notation, we know that $H_2^{AA}(\theta)$has a purely discrete spectrum \cite{KP1}, denoted by $\sigma(H_2^{AA}(\theta))=\{\lambda_k(\theta)\}_{k\geq 1}$. The union of the images of the functions $\theta \mapsto \{\lambda_k(\theta)\}$ is the dispersion relation of $H_2^{AA}$, which determines its spectrum
\begin{equation}
	\sigma(H_2^{AA})=\bigcup_{\theta \in \mathcal{B}_d}\sigma(H_2^{AA}(\theta))\,.
\end{equation}
As we have already described in Chapter \ref{capPrelimina}, from the Floquet cyclic condition, we have
\begin{equation}
	u_{a_{k2}}(1)=e^{i\theta_1}u_{a_{k1}}(1) \quad \mbox{e}\quad u_{a_{k3}}(1)=e^{i\theta_2}u_{a_{k1}}(1), \quad k=1,2\,,
\end{equation}
and with the conditions of continuity at the vertices,
\begin{equation}
	\left\{
	\begin{array}{llll}
		u_{a_{11}}(0)&=&u_{a_{12}}(0)=u_{a_{13}}(0)=\dfrac{u_{f_{1}}(0)}{t_0}=:A_1\\
		u_{a_{11}}(1)&=&e^{i\theta_1}u_{a_{12}}(1)=e^{i\theta_2}u_{a_{13}}(1)=\dfrac{u_{f_{2}}(0)}{t_0}=:B_1\\
		u_{a_{21}}(0)&=&u_{a_{22}}(0)=u_{a_{23}}(0)=\dfrac{u_{f_{1}}(0)}{t_0}=:A_2\\
		u_{a_{21}}(1)&=&e^{i\theta_1}u_{a_{22}}(1)=e^{i\theta_2}u_{a_{23}}(1)=\dfrac{u_{f_{2}}(0)}{t_0}=: B_2
	\end{array},\right.
\end{equation}
By Robin condition \eqref{Robin}, we have
\begin{equation}\label{1}
	\left\{
	\begin{array}{llll}
		u'_{a_{11}}(0) + u'_{a_{12}}(0) + u'_{a_{13}}(0) + t_0u'_{f_{1}}(0)&=&\delta_aA_1\\
		-u'_{a_{11}}(1) - e^{i\theta_1}u'_{a_{12}}(1) - e^{i\theta_2}u'_{a_{13}}(1) - t_0u'_{f_{2}}(1)&=&\delta_bB_1\\
		u'_{a_{21}}(0) + u'_{a_{22}}(0) + u'_{a_{23}}(0) - t_0u'_{f_{1}}(1)&=&\delta_aA_2\\
	-u'_{a_{21}}(1) - e^{i\theta_1}u'_{a_{22}}(1) - e^{i\theta_2}u'_{a_{23}}(1) + t_0u'_{f_{2}}(0)&=&\delta_bB_2
	\end{array}.
	\right.
\end{equation}
Let $\lambda \notin \sigma (H^D) $. Then there are two linearly independent solutions $\varphi_0=\varphi_{\lambda,0}$ and $\varphi_1=\varphi_{\lambda,1}$ of the problem
\begin{equation}\label{Prob.AutVal2}
	-\dfrac{d^2\varphi(x)}{dx^2}+q(x)\varphi(x)=\lambda\varphi(x)\,,
\end{equation}
such that  
\begin{equation}
	\varphi_0(0)=1=\varphi_1(1)\,,\quad \varphi_0(1)=0=\varphi_1(0)\quad \mbox{e}\quad \varphi'_1(x)=-\varphi'_0(1-x)\,,\quad x \in [0,1]\,.
\end{equation}
Since each edge of $W_2^{AA}$ is identified with the range $[0,1]$, we define $ \varphi_{\lambda,k}$ on each edge (with the same notation $ \varphi_{\lambda, k}$). Therefore, for each $\lambda \notin \sigma(H^D)$, we represent
\begin{equation}
	\left\{
	\begin{array}{llll}\label{3}
		u_{a_{k1}}&=&A_k\varphi_{\lambda,0}+B_k\varphi_{\lambda,1}\\
		u_{a_{k2}}&=&A_k\varphi_{\lambda,0}+e^{-i\theta_1}B_k\varphi_{\lambda,1}\\
		u_{a_{k3}}&=&A_k\varphi_{\lambda,0}+e^{-i\theta_2}B_k\varphi_{\lambda,1}\\
		u_{f_{1}}&=&t_0(A_1\varphi_{\lambda,0}+A_2\varphi_{\lambda,1})\\
		u_{f_{2}}&=&t_0(B_2\varphi_{\lambda,0}+B_1\varphi_{\lambda,1})
	\end{array}, \quad  k = 1,2\
	\right..
\end{equation}
With this, we can verify that the continuity conditions and \eqref{Prob.AutVal2} are satisfied. Now we have to check the ``Total-Flow'' condition in \eqref{1}. Using the representation of each function $u_{e}$ in \eqref{3}, we have
\begin{equation}
	\left\{
	\begin{array}{llll}\label{4}
		(3+t_0^2)\varphi'_{\lambda,0}(0)A_1+\bar{F}(\theta)\varphi'_{\lambda,1}(0)B_1+t_0^2\varphi'_{\lambda,1}(0)A_2&=&\delta_aA_1\\
		F(\theta)\varphi'_{\lambda,0}(1)A_1+(3+t_0^2)\varphi'_{\lambda,1}(1)B_1+t_0^2\varphi'_{\lambda,0}(1)B_2&=&-\delta_bB_1\\
		-t_0^2\varphi'_{\lambda,0}(1)A_1+\big(3\varphi'_{\lambda,0}(0)-t_0^2\varphi'_{\lambda,1}(1)\big)A_2+\bar{F}(\theta)\varphi'_{\lambda,1}(0)B_2&=&\delta_aA_2\\
		-t_0^2\varphi'_{\lambda,1}(0)B_1+\bar{F}(\theta)\varphi'_{\lambda,0}(1)A_2+\big(3\varphi'_{\lambda,1}(1)-t_0^2\varphi'_{\lambda,0}(0)\big)B_2&=&-\delta_bB_2
	\end{array}, 
	\right.
\end{equation}
with $F(\theta):=1+e^{i\theta_1}+e^{i\theta_2}$ and $\bar{F}(\theta)$ its complex conjugate. Like $\varphi'_{\lambda,0}(1)=-\varphi'_{\lambda,1}(0)$, $\varphi'_{\lambda,0}(0)=-\varphi'_{\lambda,1}(1)$ and, being $\varphi'_{\lambda,1}(0)\neq 0$, we put 
\[
\eta(\lambda):=\dfrac{\varphi'_{\lambda,1}(1)}{\varphi'_{\lambda,1}(0)},
\] multiplying the second and fourth lines by -1 in \eqref{4}, we obtain the system
$$
M_2^{AA}(\eta(\lambda),\theta)X=0, \quad  \mbox{com}\quad X=[A_1\,\, B_1\,\, A_2\,\,B_2]^T
$$ 
and 
\begin{equation*}
	M_2^{AA}(\eta(\lambda),\theta)=\left(\begin{array}{cccc}
		-T\eta-\alpha_a &	\bar F  & t_0^2  &  0	 \\
		F &  -T\eta-\alpha_b  & 0 & t_0^2  \\
		t_0^2 &  0  & -T\eta-\alpha_a & \bar F  \\
		0 &  t_0^2  & F & -T\eta-\alpha_b 
	\end{array}\right) ,  
\end{equation*}
where $T:=3+t_0^2$, $F=F(\theta)$, $\eta = \eta(\lambda)$ and $\alpha_k:=\dfrac{\delta_k}{\varphi'_{\lambda,1}(0)}$, $k \in \{a,b\}$.
Putting $P(\eta):=\det(M^{AA}_2(\eta(\lambda),\theta)$, we have 
$$
\begin{array}{lll}\label{Pol.AA}
P(\eta)\!\!&\!\!=\!\!&\!\!T^4\eta^4\!+\!2(\alpha_b\!+\!\alpha_a)T^3\eta^3\!+\!\big[(\alpha_a\!+\!\alpha_b)^2\!-\!2(|F|^2\!-\!\alpha_a\alpha_b\!+\!t_0^4)\big]T^2\eta^2\\
	&-& \!\!2(\alpha_a+\alpha_b)\big[|F|^2\!-\!\alpha_b\alpha_a\!+\!t_0^4\big]T\eta+\big[|F|^2\!-\!\alpha_b\alpha_a\big]^2\!+\!t_0^4\big[t_0^4\!-\!\alpha^2_n-\alpha^2_b-2|F|^2\big]\\
	&=& \big(T^2\eta^2+(\alpha_a+\alpha_b-2t_0^2)T\eta-|F|^2+t_0^4-(\alpha_b+\alpha_a)t_0^2+\alpha_b\alpha_a\big)\cdot\\
	&\,&\big(T^2\eta^2+(\alpha_a+\alpha_b+2t_0^2)T\eta-|F|^2+t_0^4+(\alpha_a+\alpha_b)t_0^2+\alpha_b\alpha_a   \big)
\end{array}
$$ 
and it was possible to find its roots \footnote{Here, the superscripts $\pm$ refer to the first symbol, while the subscripts $\pm$ refer to the second.}, which are 
\begin{equation}\label{raizes1}
	r^{\pm}_{\pm}(\theta)=-\dfrac{\alpha_b+\alpha_a\pm 2t_0^2 \pm\sqrt{(\alpha_a-\alpha_b)^2+4|F|^2}}{2T}\,.
\end{equation}
That is, there is $\theta \in \mathcal{B}$ such that $\det\big(M^{AA}_2(\eta(\lambda),\theta)\big)=0$; therefore, the representation \eqref{3} solves the problem of eigenvalues \eqref{Prob.AutVal2} and by Proposition \ref{prop_chave}, we conclude that $\lambda \in \sigma(H_2^{AA})$.

\section[Analysis of the dispersion relation]{Analysis of the dispersion relation}

Let analyze the behavior of the \eqref{raizes1} functions found in the previous section, two-layer stacking. From now on, we study the Dirac cones for two consecutive layers, making the values of $\gamma$ in each cone explicit. In this case, in fact, for identical bi- or tri-layers it is enough to work with the right side of \eqref{ConesDirac}, as the discussion for the left side repeats the monolayer case made above in \eqref{DerConeDirac}. We will check the presence of Dirac Cones in this model, as well as parabolic touches. Through a direct calculation, we see that the function
$$|F(\theta)|^2= 1 + 8\cos\left(\frac{\theta_1-\theta_2}{2}\right)\cos\left(\frac{\theta_1}{2}\right)\cos\left(\frac{\theta_2}{2}\right)$$ has as its image the interval $[0,9]$, reaches a maximum at $(0,0)$ and minimum at $\pm\left(\frac{2\pi}{3},-\frac{2\pi }{3}\right)$; Furthermore, the functions $\pm|F|$ touch each other at $0$ only when $\theta=\pm\left(\frac{2\pi}{3},-\frac{2\pi}{3} \right)$. Therefore, to analyze the image of the functions $r^{\pm}_{\pm}$, simply take $\theta$ on the diagonal $B_d$ of the Brillouin zone,
$$B_d=\{(\theta_1,\theta_2) \in [-\pi,\pi]^2: \theta_2=-\theta_1\}\,$$
and with that, we have $F(\theta)= 1+2\cos(\theta_1)$.


Initially, for $\alpha_a=\alpha_b$ in \eqref{raizes1}, we have
\begin{equation}\label{Dirac1}
	r^{\pm}_{\pm}(\theta)=\dfrac{-\alpha_b\pm t_0^2\pm|F|}{3+t_0^2} \,.
\end{equation}	

\begin{lemma}\label{Disp}
Under \eqref{Dirac1} conditions, we have for the images of the respective functions
	\begin{equation*}
		{\mathrm{Img}}\big(r^+_+(\theta)\big)= \left[\frac{-\alpha_b+t_0^2}{3+t_0^2},\frac{-\alpha_b+t_0^2+3}{3+t_0^2}\right], \hspace{1cm} 
		{\mathrm{Img}}\big(r^+_-(\theta)\big)= \left[\frac{-\alpha_b+t_0^2-3}{3+t_0^2},\frac{-\alpha_b+t_0^2}{3+t_0^2}\right],
	\end{equation*}
	\begin{equation*}
		{\mathrm{Img}}\big(r^-_+(\theta)\big)= \left[\frac{-\alpha_b-t_0^2}{3+t_0^2},\frac{-\alpha_b-t_0^2+3}{3+t_0^2}\right],
		\hspace{1cm}
		{\mathrm{Img}}\big(r^-_-(\theta)\big)= \left[\frac{-\alpha_b-t_0^2-3}{3+t_0^2},\frac{-\alpha_b-t_0^2}{3+t_0^2}\right]\,.
	\end{equation*}
Consequently:
	\begin{enumerate}
		\item 
		$\min\left(r^+_+(\theta)\right)\!\!=\!\!\dfrac{-\alpha_b+t_0^2}{3+t_0^2}$ and $\max\left(r^+_+(\theta)\right)\!\!=\!\!\dfrac{-\alpha_b+t_0^2+3}{3+t_0^2}$
		reached at $\theta=\pm\big(\!-\frac{2\pi}{3},\frac{2\pi}{3}\big)$ and $\theta\!=\!(0 ,0)$, respectively. \\
		\item 
		$\min\left(r^+_-(\theta)\right)=\dfrac{-\alpha_b+t_0^2-3}{3+t_0^2}$ and  $\max\left(r^+_-(\theta)\right)=\dfrac{-\alpha_b+t_0^2}{3+t_0^2}$
		reached at $\theta=(0,0)$ e $\theta=\pm\big(\!-\frac{2\pi}{3},\frac{2\pi}{3}\big)$, respectively; \\
		\item 
		$\min\left(r^-_+(\theta)\right)\!\!=\!\!\dfrac{-\alpha_b-t_0^2}{3+t_0^2}$ e $\max\left(r^-_+(\theta)\right)\!\!=\!\!\dfrac{-\alpha_b-t_0^2+3}{3+t_0^2}$
		reached at $\theta=\pm\big(\!-\frac{2\pi}{3},\frac{2\pi}{3}\big)$ and $\theta=(0,0)$, respectively. \\
		\item 
		$\min\left(r^-_-(\theta)\right)=\dfrac{-\alpha_b-t_0^2-3}{3+t_0^2}$ and  $\max\left(r^-_-(\theta)\right)=\dfrac{-\alpha_b-t_0^2}{3+t_0^2}$
		reached at $\theta=(0,0)$ and $\theta=\pm\big(\!-\frac{2\pi}{3},\frac{2\pi}{3}\big)$, respectively.  
		
	\end{enumerate}
	
\end{lemma}
Finally, we arrive at the following theorem, which establishes results regarding the dispersion relation of $H^{AA}_2$.
\begin{theorem}\label{Disp4}(Conclusion on the dispersion relationship of $H^{AA}_2$) Fix $t_0 \in (0,1)$ and $\alpha_b \in \R$ such that $|r^{\pm}_{\pm}(\theta)|\leq 1$ in \eqref{raizes1} and let $\alpha_a=(1-u)\cdot\big(\!\!-2t_0^2+\alpha_b\big)+u\cdot\big(2t_0^2+ \alpha_b\big)$. The dispersion relation of the $H^{AA}_2$ operator is given by
	\begin{equation*}\label{Disp3}
		r^{\pm}_{\pm}(\theta)=\dfrac{\pm\sqrt{|F|^2 +t_0^4(1-2u)^2}\pm t_0^2+t_0^2(1-2u)-\alpha_b}{T}\,,
	\end{equation*} and we conclude that
	\begin{enumerate}
		\item[(a)] For $\alpha_a=\alpha_b$, there are Dirac cones (Figure \ref{CasoAA-BN-BNa}).
		\item[(b)] For $\alpha_a=-2t_0^2+\alpha_b$ or $\alpha_a=2t_0^2+\alpha_b$, there are parabolic touches (Figure \ref{CasoAA-BN-BNb}).
		\item[(c)] For $\alpha_a \notin [-2t_0^2+\alpha_b,2t_0^2+\alpha_b]$, the dispersion relation has gaps (Figure \ref{CasoAA-BN-BNc}).
		\item[(d)] For $0<u<1$ and $\alpha_a\neq \alpha_b$, the curves only intersect, but there are no parabolic or conic touches (Figure \ref{CasoAA-BN-BNd}).
		\end{enumerate}
\end{theorem}
We will only give an idea of the demonstration here.

\begin{proof}
Let take $\theta \! \in\! B_d\!=\!\big\{(\theta_1,-\theta_1): \theta_1\! \in \![-\pi,\pi]\big\}$. For $\alpha_a=\alpha_b$, we will apply the Taylor expansion to the function $r^{\pm}_{+}$ in \ref{Dirac1} at the point $\theta_D=\pm(\frac{2\pi }{3},-\frac{2\pi}{3})$. Note that, putting $\theta_0:=\pm\frac{2\pi}{3}$, we have
$$F(\theta)= 1+2\cos(\theta_1)=\pm\sqrt{3}(\theta_1-\theta_0)+\mathcal{O}\big((\theta_1-\theta_0)^2\big)\,.
$$
Since $T:=3+t_0^2$, we have
$$
r^{\pm}_+(\theta):=\dfrac{-\alpha_b\pm t_0^2+|F(\theta)|}{T}= \dfrac{-\alpha_b\pm t_0^2+\sqrt{3}|\theta_1-\theta_0|}{T}+\mathcal{O}\big(|\theta_1-\theta_0|^2\big)
$$
and 
$$
r^{\pm}_+(\theta_D):=\dfrac{-\alpha_b\pm t_0^2}{T}\,.
$$
Therefore,
\begin{equation*}\label{Disp1}
	r^{\pm}_+(\theta)-r^{\pm}_+(\theta_D)=\pm\dfrac{\sqrt{3}}{T}|\theta_1-\theta_0|+\mathcal{O}\big(|\theta_1-\theta_0|^2\big)\,.
\end{equation*} 
Similarly, for $r^{\pm}_{-}$, we obtain
\begin{equation*}\label{Disp2}
	r^{\pm}_-(\theta)-r^{\pm}_-(\theta_D)=\mp\dfrac{\sqrt{3}}{T}|\theta_1-\theta_0|+\mathcal{O}\big(|\theta_1-\theta_0|^2\big)\,.
\end{equation*}  

For $\alpha_a=2t_0^2+\alpha_b$ (or $\alpha_a=-2t_0^2+\alpha_b)$, we see that the functions $r^{+}_{-}$ and $r^{-} _{+}$ touch when $\theta_p=\pm(\frac{2\pi}{3},-\frac{2\pi}{3})$, or even, $r^{+}_ {-}(\theta_p)=r^{-}_{+}(\theta_p)=0$ so that there is the parabolic touch $(\theta_p,0)$. In fact, in \ref{raizes1}, taking $\theta \in B_d$, $F=F(\theta)=1+\cos(\theta)$ and putting $\alpha_a=\pm2t_0^2+\alpha_b$, we get
\begin{equation*}
	r^{\pm}_{\mp}(\theta)=-\dfrac{-\alpha_b-t_0^2\pm t_0^2\mp\sqrt{t_0^4+|F|^2}}{3+t_0^2}\,.
\end{equation*}  
If $\theta \neq \theta_p$, a direct calculation shows us that $\dfrac{dr^{\pm}_{\mp}}{d\theta}$ cancels out in $\pm\frac{2\pi }{3}$. Furthermore, to obtain the behavior of the functions $r^{\pm}_{\mp}$ around $\pm\frac{2\pi}{3}$, we apply the relation $(1+x)^ {\frac{1}{2}} \approx 1 +\frac{x}{2}-\frac{x^2}{8}$, for small $x$, in the expression $\sqrt{t_0^4 +|F|^2}$ and we conclude that $r^{\pm}_{\mp}$ has parabolic behavior around $\pm\frac{2\pi}{3}$, that is, next of $F=0$.

For the other cases, in particular, when $\alpha_a \notin [-2t_0^2+\alpha_b,2t_0^2+\alpha_b]$, a simple calculation shows us that the functions $r^{\pm}_{\pm}$ do not touch each other regardless of the values of $\theta$, for any $\theta \in B$.
\end{proof}
Below are graphic images of some cases obtained from the theorem above.	
		\begin{figure}[H]
		\centering
		\includegraphics[width=0.5\linewidth]{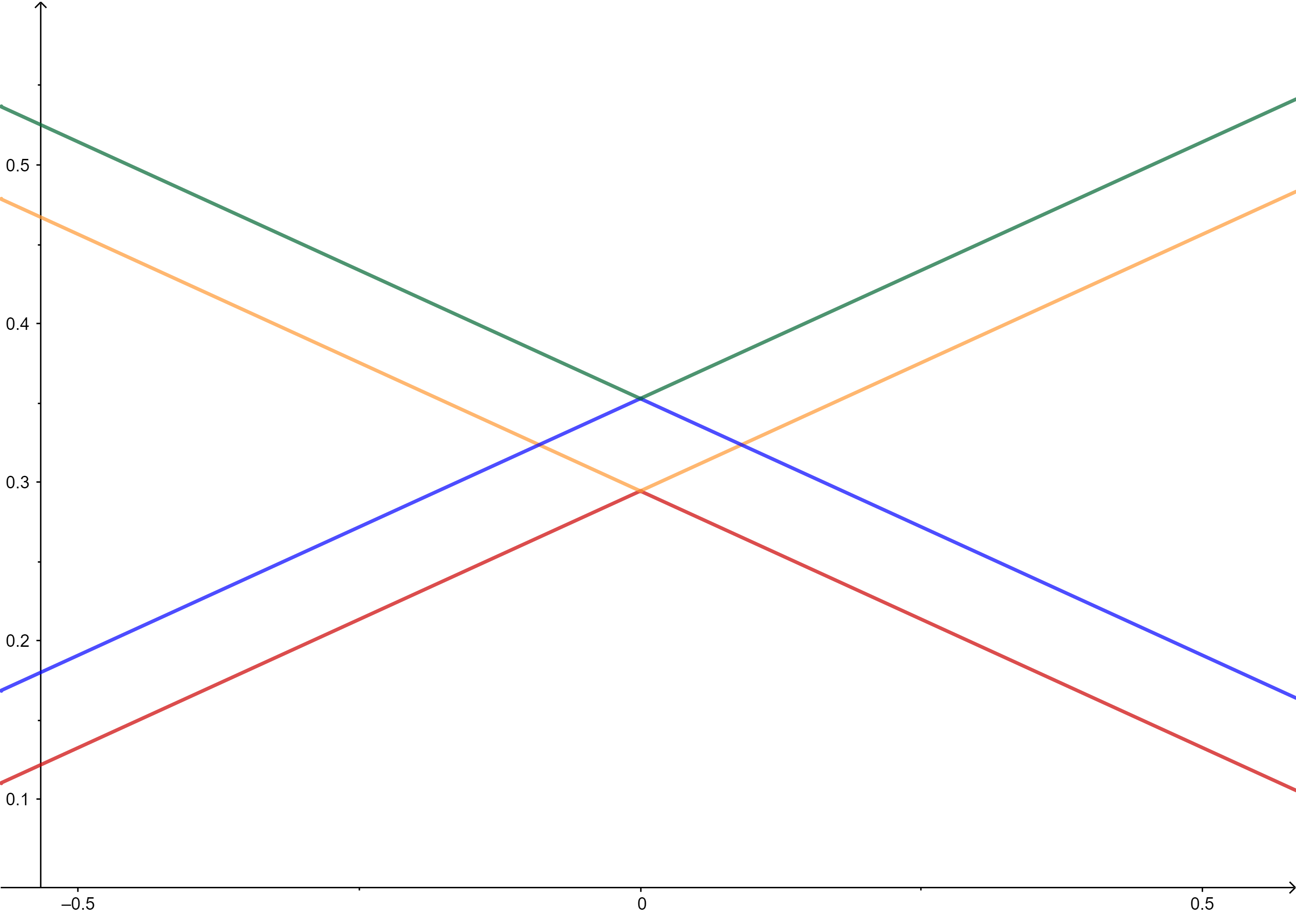}
		\caption{Dirac cone if the atoms are equal; here $\alpha_a=\alpha_b=-1\,,t_0=0.3\,.$}
	\label{CasoAA-BN-BNa}
	\end{figure}
	
	\begin{figure}[H]
		\centering
		\includegraphics[width=0.5\linewidth]{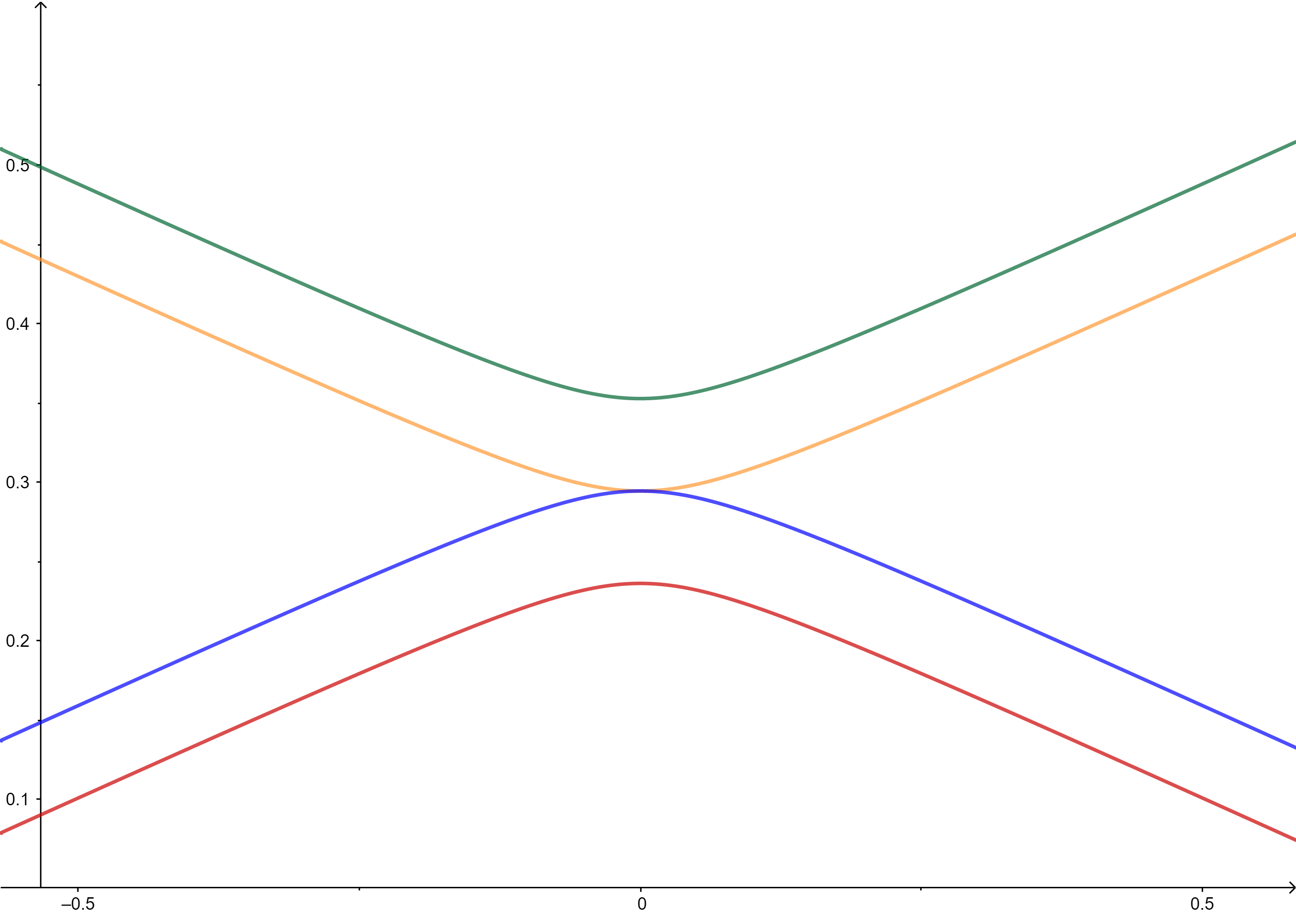}
		\caption{Parabolic touch; here $\alpha_a = -1\,,\alpha_b=2t_0^2+\alpha_a\,,t_0=0.3\, .$}
	\label{CasoAA-BN-BNb}
	\end{figure}
	
	\begin{figure}[H]
		\centering
		\includegraphics[width=0.5\linewidth]{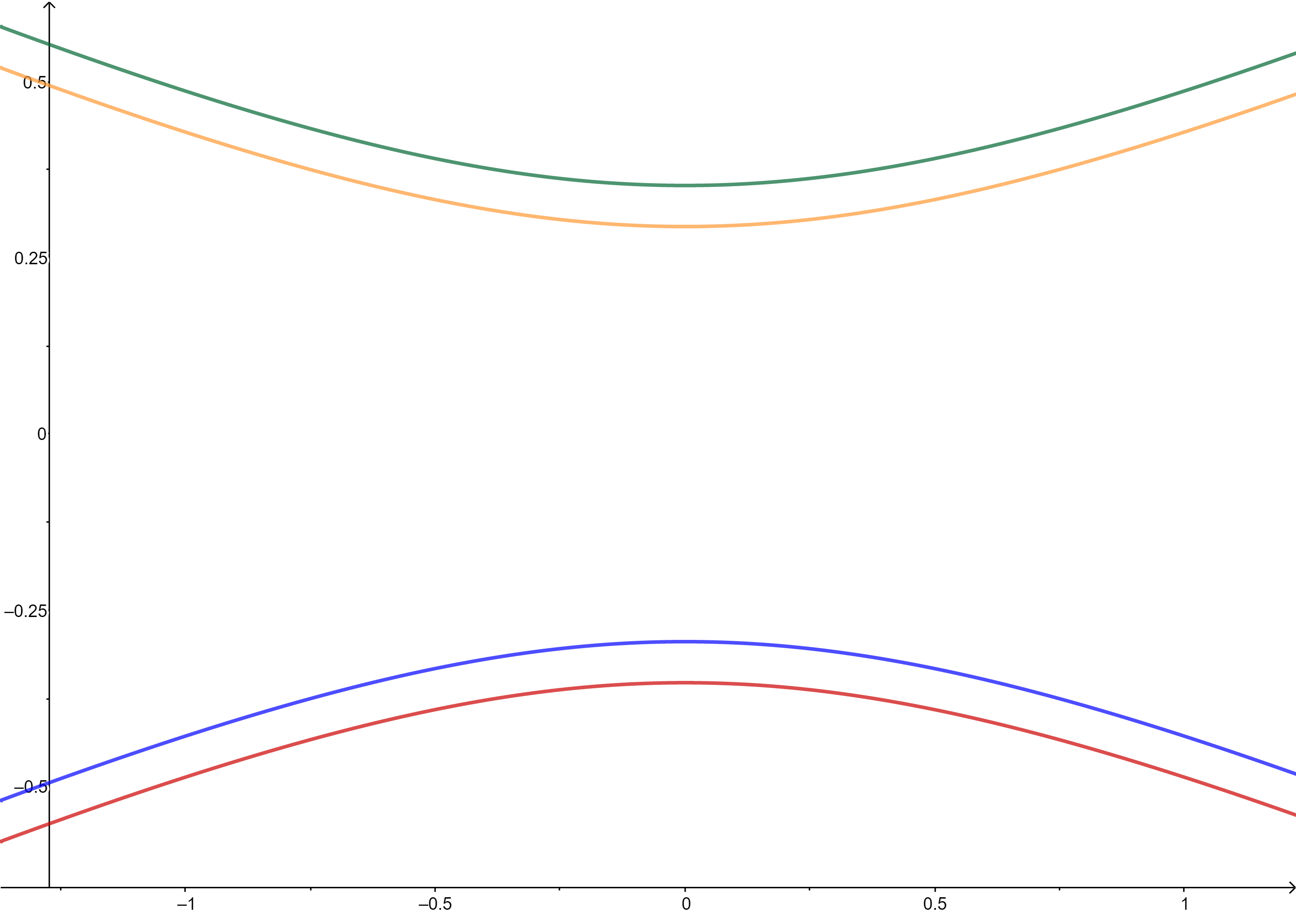}
		\caption{Gap of 0.589 in the dispersion relation at the origin for $\alpha_a=-1,\,\alpha_b=1\,,t_0=0.3\,.$}
	\label{CasoAA-BN-BNc}
	\end{figure}
	
	\begin{figure}[H]
		\centering
		\includegraphics[width=0.5\linewidth]{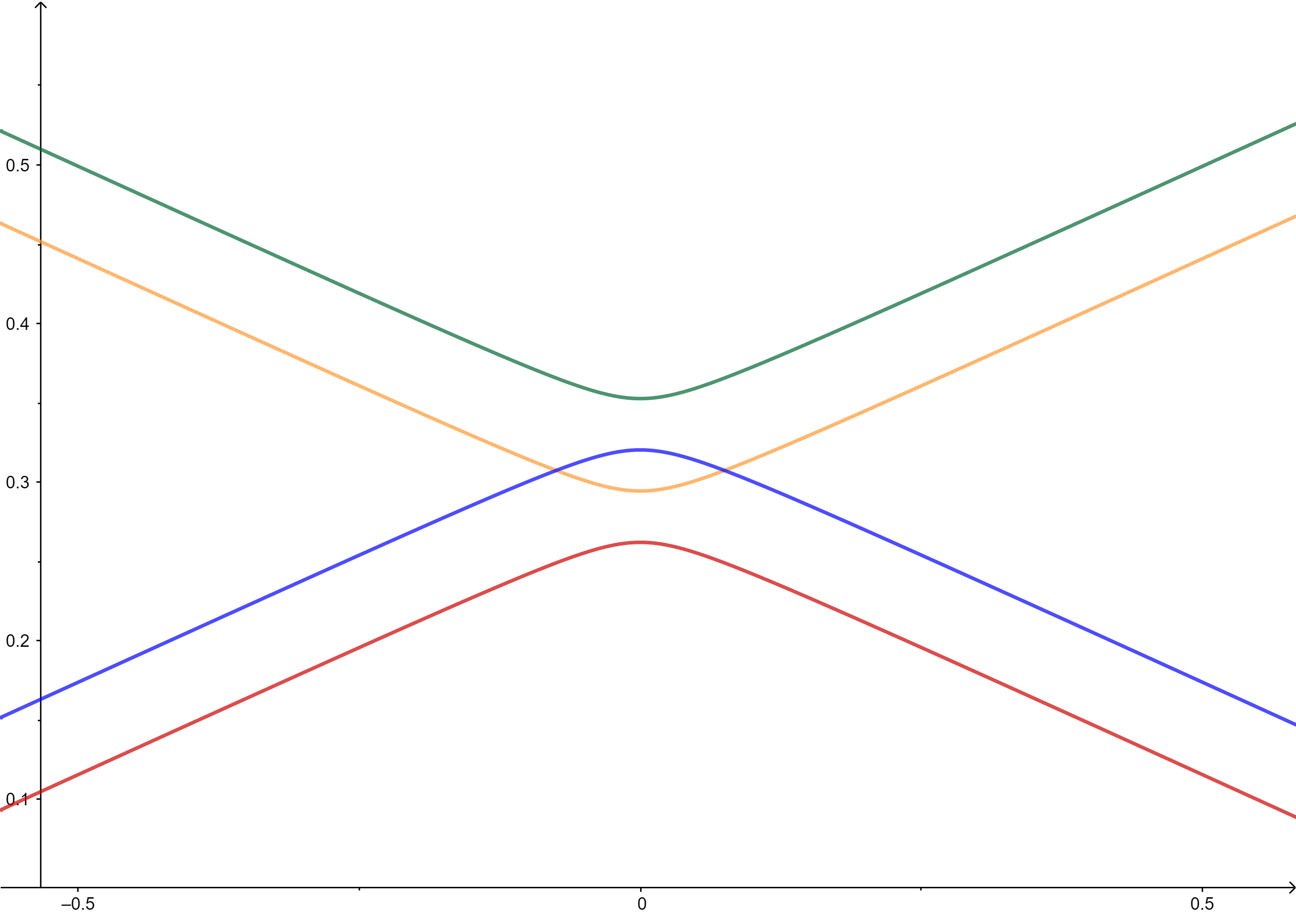}
		\caption{The curves intersect, but there are no parabolic or conic touches: $\alpha_a\!=\!-1,\alpha_b\!=\!-0.9,t_0\!=\!0.3\,.$}
	\label{CasoAA-BN-BNd}
	\end{figure}
	
	We summarize these results in the following section.
	
	\section{Analysis and conclusions}
	
	We saw in this chapter that in a van der Waals heterostructure, stacked in the ~$AA$ type with two equal hexagonal layers, in which two types of atoms are considered interspersed at the vertices of each hexagon, the following situations basically occur: (a) if the atoms are identical ($\alpha_a=\alpha_b$), the model has a Dirac cone; (b) has parabolic touches if $\alpha_b=\pm\,2t_0^2\,+\,\alpha_a$; (c) has gaps otherwise; 
(d) has no conical or parabolic touches for $\alpha_b \neq \alpha_a$ in the interval $(-2t_0^2\,+\,\alpha_a, 2t_0^2\,+\,\alpha_a)$. Note also that for (only formally) $t_0 = 0$ (in which case there is no interaction between the layers and the bilayers are reduced to the monolayer), we recover the results for the monolayer seen in the previous chapter. In particular, according to our model, heterostructures of two sheets of boron nitride stacked in the $AA$ type do not have Dirac cones, but can have parabolic touches.
	
\section{Appendix: the parameter $t_0$}\label{Apendice}

Let us remember that, in the previous section, we used different values of the $\delta_v$ parameter in the Robin \eqref{Neumann} condition as a way of distinguishing the atoms in the model: if $\delta_v$ is associated with a vertex of type-$A$ (or type-$B$), we choose $\delta_v = \delta_a$ $ \big(\mbox{ou}\,\,\delta_v = \delta_b\big)$, with $\delta_a \neq \delta_b$ if A$\ne$B. Furthermore, in stacking $AA$ we admit the same parameter $t_0$ simulating the weak connections between vertices of consecutive layers, that is, both to vertices of type-$A$ and to those of type-$B$, we associate a single parameter $t_0$. Let's now generalize this modeling: for bonds between type A atoms we will associate an interaction parameter $t_{a}$, and for bonds between type~B atoms, we will associate $t_{b}$.

In this case, the Floquet cyclic condition and the continuity conditions at the vertices will be given by 
\begin{equation}
	\left\{
	\begin{array}{llll}
		u_{a_{11}}(0)&=&u_{a_{12}}(0)=u_{a_{13}}(0)=\dfrac{u_{f_{1}}(0)}{t_{a}}=:A_1\\
		u_{a_{11}}(1)&=&e^{i\theta_1}u_{a_{12}}(1)=e^{i\theta_2}u_{a_{13}}(1)=\dfrac{u_{f_{2}}(1)}{t_{b}}=:B_1\\
		u_{a_{21}}(0)&=&u_{a_{22}}(0)=u_{a_{23}}(0)=\dfrac{u_{f_{1}}(1)}{t_{a}}=:A_2\\
		u_{a_{21}}(1)&=&e^{i\theta_1}u_{a_{22}}(1)=e^{i\theta_2}u_{a_{23}}(1)=\dfrac{u_{f_{2}}(0)}{t_{b}}=:B_2
	\end{array}.\right.
\end{equation}
Again, introducing the parameters $\delta_{a}, \delta_{b}$ to the respective vertices type-$A$ and type-$B$, using the Robin condition \eqref{Robin}, we obtain the Total Flow equations
\begin{equation}
	\left\{
	\begin{array}{llll}
		u'_{a_{11}}(0) + u'_{a_{12}}(0) + u'_{a_{13}}(0) + t_{a}u'_{f_{1}}(0)&=&\delta_aA_1\\
		-u'_{a_{11}}(1) - e^{i\theta_1}u'_{a_{12}}(1) - e^{i\theta_2}u'_{a_{13}}(1) - t_{b}u'_{f_{2}}(1)&=&\delta_bB_1\\
		u'_{a_{21}}(0) + u'_{a_{22}}(0) + u'_{a_{23}}(0) - t_{a}u'_{f_{1}}(1)&=&\delta_aA_2\\
	-u'_{a_{21}}(1) - e^{i\theta_1}u'_{a_{22}}(1) - e^{i\theta_2}u'_{a_{23}}(1) + t_{b}u'_{f_{2}}(0)&=&\delta_bB_2
	\end{array}.
	\right.
\end{equation}
Similarly, for each $\lambda \notin \sigma(H^D)$, we represent
\begin{equation}
	\left\{
	\begin{array}{llll}
		u_{a_{k1}}&=&A_k\varphi_{0}+B_k\varphi_{1}\\
		u_{a_{k2}}&=&A_k\varphi_{0}+e^{-i\theta_1}B_k\varphi_{1}\\
		u_{a_{k3}}&=&A_k\varphi_{0}+e^{-i\theta_2}B_k\varphi_{1}\\
		u_{f_{1}}&=&t_{a}(A_1\varphi_{0}+A_2\varphi_{1})\\
		u_{f_{2}}&=&t_{b}(B_2\varphi_{0}+B_1\varphi_{1})
	\end{array}, \quad  k = 1,2\
	\right.\,.
\end{equation}
After some calculations, we obtain the system
$$
M_{ab}^{AA}(\eta(\lambda),\theta)X=0, \quad  \mbox{com}\quad  X=[A_1\,\, B_1\,\, A_2\,\,B_2]^T
$$ 
and 
\begin{equation*}
	M_{ab}^{AA}(\eta(\lambda),\theta)=\left(\begin{array}{cccc}
		-T_a\eta-\alpha_a &	\bar F  & t_a^2  &  0	 \\
		F &  -T_b\eta-\alpha_b  & 0 & t_b^2  \\
		t_a^2 &  0  & -T_a\eta-\alpha_a & \bar F  \\
		0 &  t_b^2  & F & -T_b\eta-\alpha_b 
	\end{array}\right) ,  
\end{equation*}
where $T_a:=3+t_a^2,T_b:=3+t_b^2, F=F(\theta)$, $\eta = \eta(\lambda)$ e $\alpha_k:=\dfrac{\delta_k}{\varphi'_{1}(0)}$, $k \in \{a,b\}$.
Putting $P(\eta)\! :=\det(M_{ab}^{AA}(\eta(\lambda),\theta))$, we have 
\begin{eqnarray*}
	P(\eta)\!\!&=\!\!&\big[T_aT_b\eta^2+\big((\alpha_a-t_a^2)T_b+(\alpha_b-t_b^2)T_a\big)\eta+t_a^2t_b^2-\alpha_at_b^2-\alpha_bt_a^2+\alpha_a\alpha_b-|F|^2\big]\\
	\hspace{0.5cm}&\cdot\!&\! \big[T_aT_b\eta^2+\big((\alpha_a+t_a^2)T_b+(\alpha_b+t_b^2)T_a\big)\eta+t_a^2t_b^2+\alpha_at_b^2+\alpha_bt_a^2+\alpha_a\alpha_b-|F|^2\big].
\end{eqnarray*}	

In particular, note that for $t_a=t_b$, we recover the polynomial \eqref{Pol.AA} and the relation \eqref{raizes1}, and the results above coincide with those obtained previously, as predicted: there will be a cone if $\alpha_a=\alpha_b$, and there will be a gap if $\alpha_a\neq\alpha_b$.

Suppose then $t_a\neq t_b$. Let us analyze two particular cases, initially.

\textbf{Case $\alpha_a=t_a^2$ and $\alpha_b=t_b^2$}. We state that a Dirac cone occurs. In fact, in this case, we obtain the polynomial
\begin{eqnarray*}
	P(\eta)\!=\!\big(T_aT_\eta^2-|F|^2\big)\big(T_aT_b\eta^2+2(T_at_b^2+T_bt_a^2)\eta+4t_a^2t_b^2-|F|^2\big)
\end{eqnarray*}	
whose roots are
\begin{eqnarray*}
r^{\pm}(F)\!=\! \pm\dfrac{|F|}{T_aT_b}
\quad \mbox{and}\quad 
r_{\mp}(F)\!=\!\mp \dfrac{\sqrt{(T_bt_a^2-T_at_b^2)^2+4|F|^2}}{T_aT_b}-\dfrac{T_at_b^2+T_bt_a^2}{T_aT_b}
\end{eqnarray*}	
Clearly, we see that $r^{\pm}$ has conical behavior, whose touch occurs for $F=0$, that is, at $r^{+}(0)=0=r^{-}(0) $, while in $r_{\mp}$ there are no touches, since $t_a\neq t_b$. An analogous result is obtained for $\alpha_a=-t_a^2$ and $\alpha_b=-t_b^2$.

\textbf{Case $\alpha_a=t_a^2$ and $\alpha_b=-t_b^2$}. We claim that there is a parabolic touch. In effect, here, we obtain the polynomial
\begin{eqnarray*}
P(\eta)\!=\!\big(T_aT_b\eta^2+2T_bt_a^2\eta-|F|^2\big)\big(T_aT_b\eta^2-2T_at_b^2\eta-|F|^2)
\big)
\end{eqnarray*}
and its roots are
\begin{eqnarray*}
r^{\pm}(F)\!=\! \pm \dfrac{\sqrt{T_a^2t_b^4+T_aT_b|F|^2}}{T_aT_b}-\dfrac{T_at_b^2}{T_aT_b}
\quad \mbox{and}\quad 
r_{\mp}(F)\!=\!\mp \dfrac{\sqrt{T_a^2t_b^4+T_aT_b|F|^2}}{T_aT_b}+\dfrac{T_at_b^2}{T_aT_b}\,.
\end{eqnarray*}	
With this, we see that $r^{\pm}$ and $r_{\mp}$ have parabolic behavior, whose touch occurs at $r^{+}(0)=r_{-}(0)$. Similarly, we obtain a similar result if $\alpha_a=-t_a^2$ and $\alpha_b=t_b^2$. 

In general, the roots of $P$ are given by
\begin{eqnarray*}
r^{\pm}(F)\!\!&=\!\!&\pm \dfrac{\sqrt{(\alpha_a-t_a^2)^2T_b^2+(\alpha_b-t_b^2)^2T_a^2    -2T_aT_b\big[(\alpha_a-t_a^2)(\alpha_b-t_b^2)-2|F|^2\big]}}{2T_aT_b}\\ 
&-&\dfrac{(\alpha_a-t_a^2)T_b+(\alpha_b-t_b^2)T_a}{2T_aT_b}
\end{eqnarray*}	
and 
\begin{eqnarray*}
r_{\mp}(F)\!\!&=\!\!&\mp \dfrac{\sqrt{(\alpha_a+t_a^2)^2T_b^2+(\alpha_b+t_b^2)^2T_a^2    -2T_aT_b\big[(\alpha_a+t_a^2)(\alpha_b+t_b^2)-2|F|^2\big]}}{2T_aT_b}\\
&-&\dfrac{(\alpha_a+t_a^2)T_b+(\alpha_b+t_b^2)T_a}{2T_aT_b} 
\end{eqnarray*}	
 
Through a direct calculation and graphical representation of the dispersion relation, with distinct $t_a,t_b \in (0,1]$, analyzing the functions $r^{\pm}$ and $r^{\mp}$ above, we get the result below. 

\begin{theorem}\label{Teorem\~ao} Let $t_a,t_b \in (0,1]$ be distinct, weak interaction parameters between type~A and type~B atoms in different layers, respectively, of two hexagonal lattices. Consider the dispersion relation associated with the type structure $AA$. Then
\begin{itemize}
    \item[(a)] If $\alpha_a=t_a^2$ \big(or $\alpha_a=-t_a^2$\big) \,and $\alpha_b=t_b^2$ \big(or $\alpha_b=-t_b^2$\big), conical touch will occur in $F=0$ (Figure \ref{Cone_ab}).
    \item[(b)] If $\alpha_a\!=\!t_a^2$ \big(or $\alpha_a\!=\!-t_a^2$\big) \,and $\alpha_b\!=\!-t_b^2$ \big(or $\alpha_b=t_b^2$\big), there will be a parabolic touch $F\!=\!0$ (Figure \ref{Parabolico_ab}).
    \item[(c)] If $|\alpha_a|\neq t_a^2$ or $|\alpha_b|\neq t_b^2$, conical and parabolic touches no longer exist and the dispersion relation will have gaps (Figure \ref{Gap.-1-1_1-1}).
\end{itemize}
\end{theorem}
 
\begin{figure}[H]
		\centering
		\includegraphics[width=0.5\linewidth]{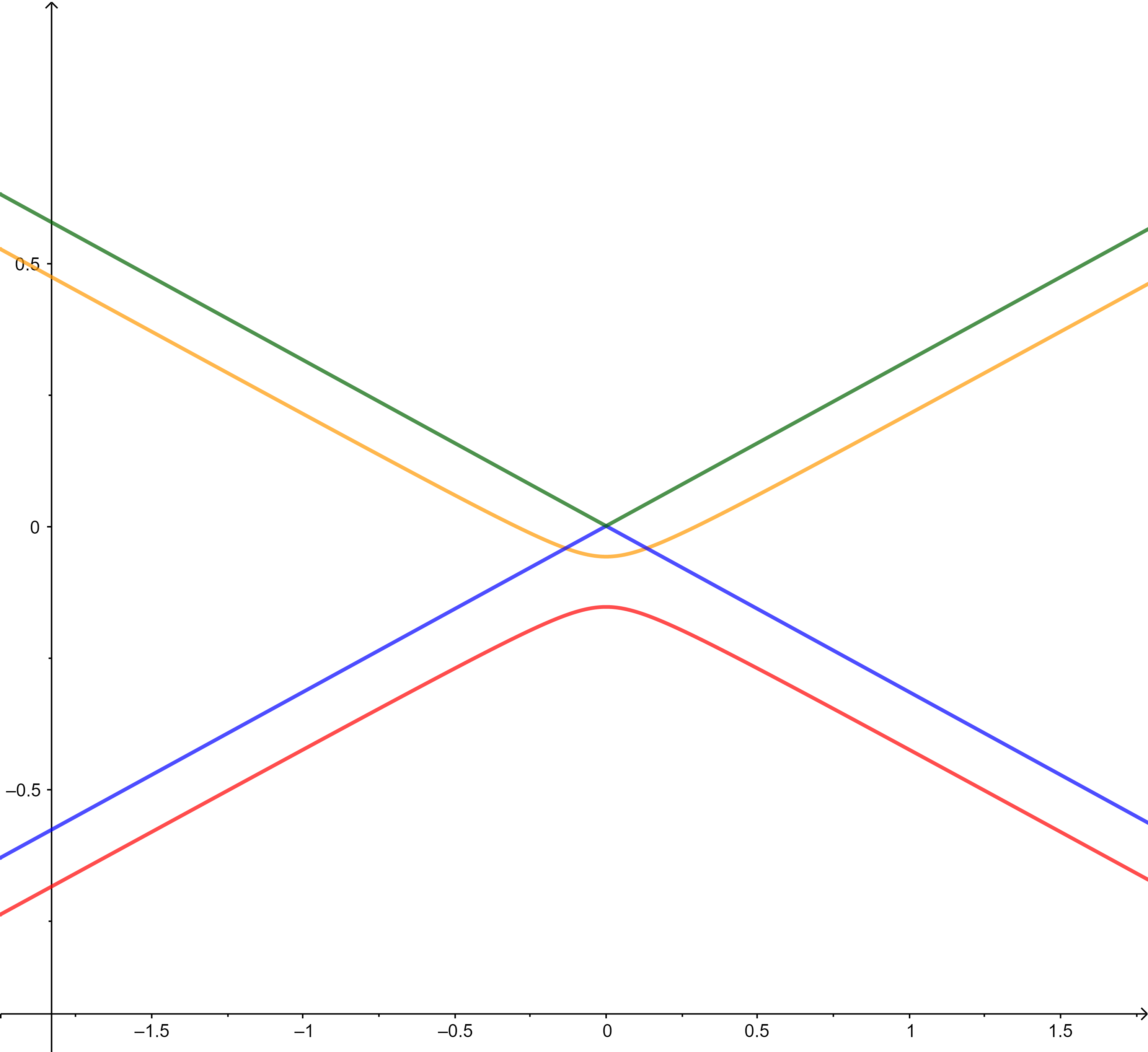}
		\caption{Dirac cone, $\alpha_a=0.25$ and $\alpha_b=0.09$\,.}
	\label{Cone_ab}
	\end{figure}

\begin{figure}[H]
		\centering
		\includegraphics[width=0.5\linewidth]{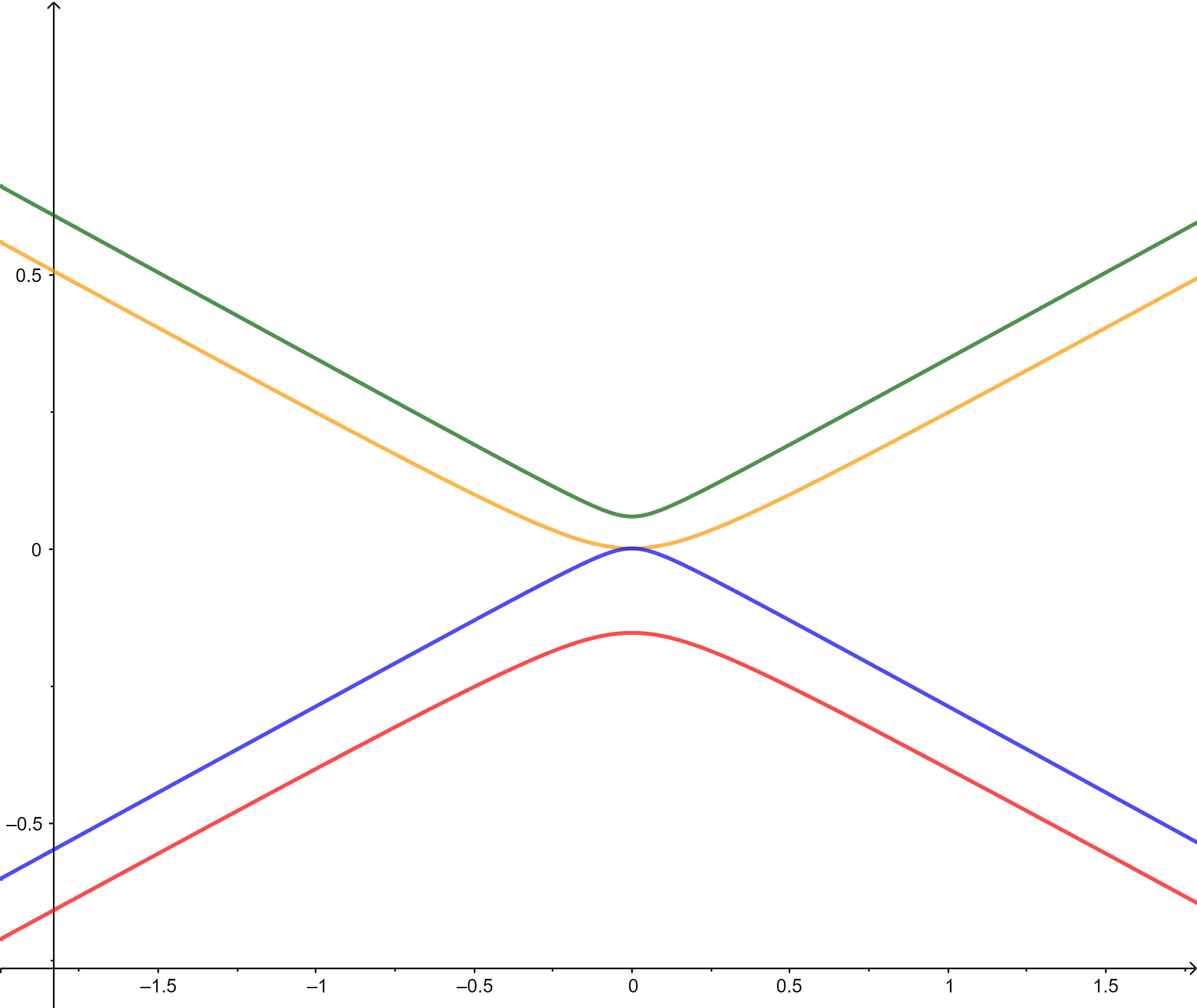}
		\caption{Parabolic Touch, $\alpha_a=0.25$ and $\alpha_b=-0.09$\,.}
	\label{Parabolico_ab}
	\end{figure}
	
	\begin{figure}[H]
		\centering
		\includegraphics[width=0.9\linewidth]{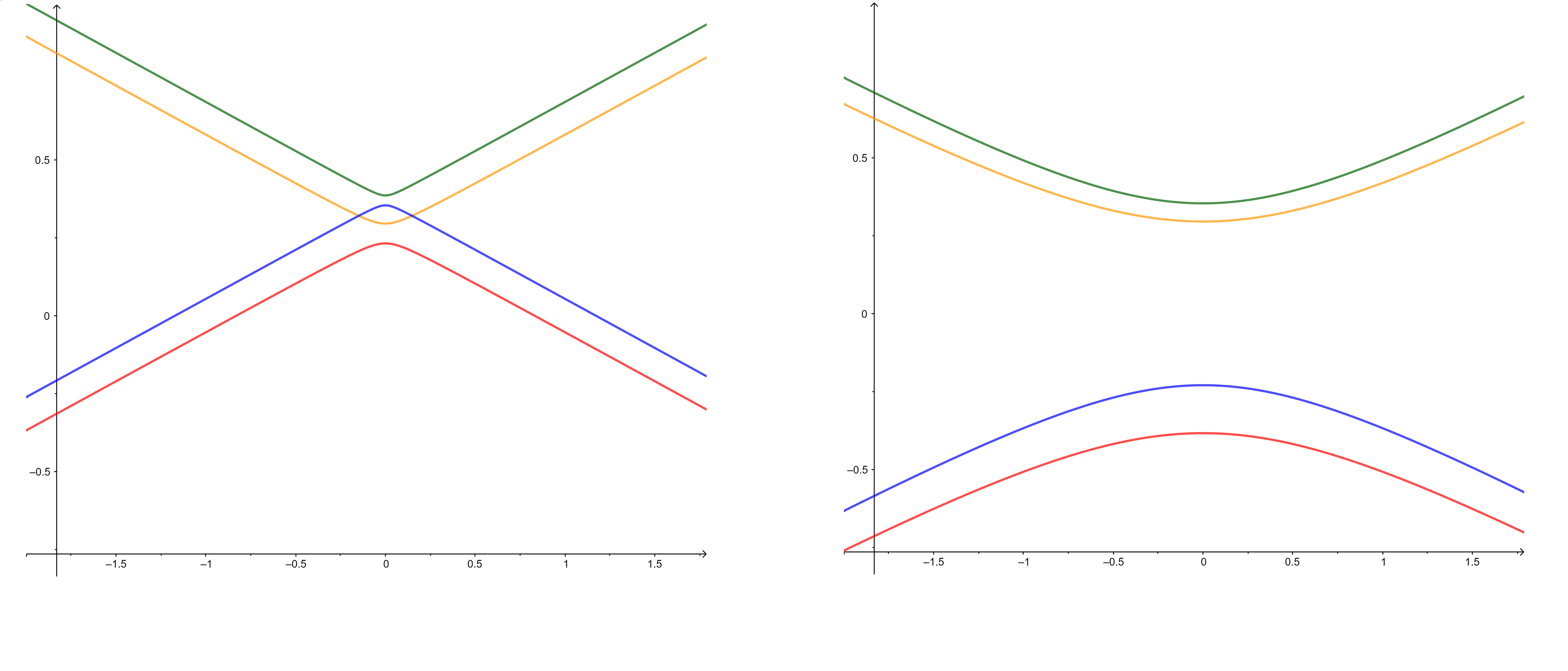}
		\caption{Gaps: left with $\alpha_a=\alpha_b=-1$; on the right with $\alpha_a=1$ and $\alpha_b=-1$.}
	\label{Gap.-1-1_1-1}
	\end{figure}

\begin{observation}\label{observacao-crucial}


In stacking $AA$ it is pertinent to consider two connection parameters $t_a$ and $t_b$ in the model, as explained above, since in this case two weak interactions occur between the hexagonal layers $\hexagon_1$ and $\hexagon_2$. However, in the $AA'$ stacking, as we will see in more detail in the following chapter, a single weak bond occurs: atoms A from the network $\hexagon_1$ with atoms B from the network $\hexagon_2$ and vice versa (we then use a single parameter $t^{b}_{a}=:t_0$).

Note that if we took the parameters $\alpha_a$ and $\alpha_b$ equal, such modeling with two different parameters would be reduced to the model with a single $t_0$.


\end{observation}	

 A comparison of these results with the $AA'$ stacking case appears in Observation \ref{AA-AA'}.

\begin{observation}\label{Obs-Cesar}
$AA$ stacking for hBN layers appears to be rare in nature; thus, the wealth of behaviors described here for this case were not found in the physical literature, remaining as a proposal for future experiments.
\end{observation}
 
 
 \chapter{Type structures $AA'$}\label{capEstrHomAAnovo}
 
 
 Let us remember that hexagonal boron nitride is composed of alternating boron $(B)$ and nitrogenene $(N)$ atoms arranged in a structure similar to graphene, however, with strong connections
 in a plane and a weak van der Waals force between the layers.
 
 In \cite{FANG} these layers are arranged in the $AA'$ stacking with $B$ atoms placed directly above the $N$ atoms. Therefore, similarly to the previous chapter, boron nitride bilayers can be considered, in which the nitrogen atoms interact with the boron atoms (and vice versa) both in the interaction the strong one (represented by the horizontal edges) as well as the van der Waals one (represented by the vertical edges), as shown in Figure \ref{BN-BN}.
 
 \begin{figure}[H]
 	\centering
 	\includegraphics[width=0.6\linewidth]{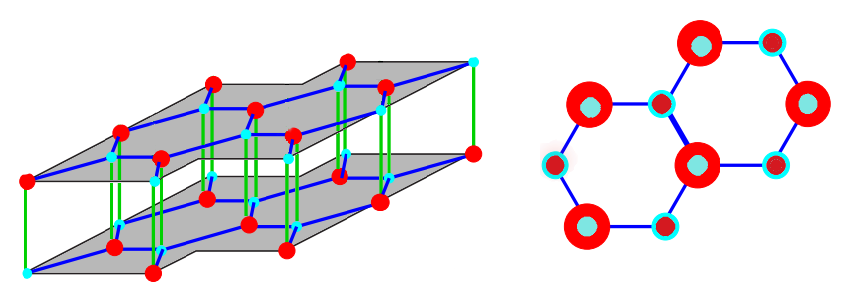}
 	\caption{Stacking $AA'$ of two hexagonal layers with 2 distinct atoms at the vertices, as in hBN. Side and top views, respectively.}
 	\label{BN-BN}
 \end{figure}
 However, in this chapter we will study stacks formed by two layers of the same material with a hexagonal shape, modeled by a quantum graph, in this case $AA'$. Clearly, in the case of bilayers, when one of them only has the same type of atom at the vertices, for example in the case of graphene, the configurations $AA$ and $ AA'$ coincide.
 
 \section[Structure with two identical leaves]{Structure with two identical leaves}
 
 To apply our model – in which quantum graph theory, Floquet theory, modified Neumann conditions are used – to this stacking, we will consider that the layers are arranged so that a type-A vertex (red color) of a leaf is connected to a type-B vertex (blue color) on the overlying leaf and vice versa, illustrated in Figure \ref{Aa0BN-BN}. To vertices of type-$A$ (type-$B$) we associate a contour parameter $\delta_a$ (\,$\delta_{b}$\,) in the condition flux; again, this is the proposal to differentiate the types of atoms.
 \vspace{-0.1cm}
 \begin{figure}[h]
 	\centering
 	\includegraphics[width=0.6\linewidth]{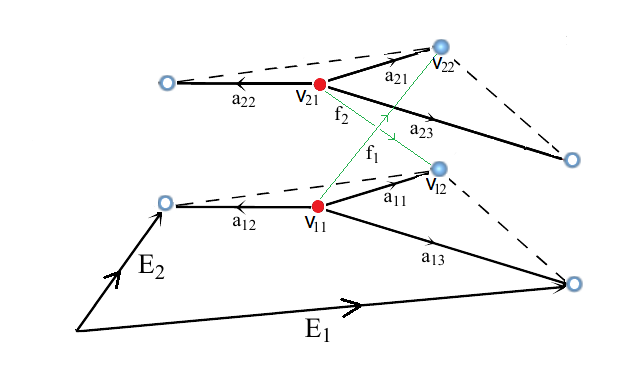}
 	\caption{Hexagonal bilayers with 2 distinct atoms at the vertices. Type-$A$ vertices are represented by red, and type-$B$ by blue.}
 	\label{Aa0BN-BN}
 \end{figure}
 
 As in the previous chapter, here we consider the fundamental domain given by
 $$ 
 W_2^{AA'}=\{a_{11}, a_{12}, a_{13}, a_{21}, a_{22}, a_{23}, f_1, f_2, v_{11}, v_{12}, v_{21}, v_{22}\}\,.
 $$ 
 
 Let us denote by $H_2^{AA'}$ the Schrödinger operator, associated with $AA'$ type stacking.
 with two layers. The spectrum of each operator $H_2^{AA'}(\theta)$ is purely discrete \cite{KP1}, which we denote $\sigma(H_2^{AA'}(\theta))=\{\lambda_k(\theta)\}_{k\geq 1}$; furthermore, the union of the images of the functions $\theta \mapsto \{\lambda_k(\theta)\}$ (he Dispersion Relation of $H_2^{AA'}$) determines its spectrum 
 \begin{equation}
 	\sigma(H_2^{AA'})=\bigcup_{\theta \in \mathcal{B}_d}\sigma(H_2^{AA'}\!(\theta)).
 \end{equation}
 
 With this, our objective now is to determine the spectrum of $H_2^{\!AA'}\!(\theta)$. To do so, let us remember that the cyclic Floquet conditions, together with the continuity conditions at the vertices \eqref{Robin} applied to the fundamental domain $W_2^{AA'}$, we get \\
 \begin{equation}
 	\left\{
 	\begin{array}{lll}
 		u_{a_{11}}(0)&=&u_{a_{12}}(0)=u_{a_{13}}(0)=\dfrac{u_{f_{1}}(0)}{t_0}=:A_1\\
 		u_{a_{11}}(1)&=&e^{i\theta_1}u_{a_{12}}(1)=e^{i\theta_2}u_{a_{13}}(1)=\dfrac{u_{f_{2}}(1)}{t_0}=:B_1\\
 		u_{a_{21}}(0)&=&u_{a_{22}}(0)=u_{a_{23}}(0)=\dfrac{u_{f_{2}}(0)}{t_0}=:A_2\\
 		u_{a_{21}}(1)&=&e^{i\theta_1}u_{a_{22}}(1)=e^{i\theta_2}u_{a_{23}}(1) = \dfrac{u_{f_{1}}(1)}{t_0}=:B_2
 	\end{array}\,.\right.
 \end{equation}
 
 Applying the modified Neumann vertex condition \eqref{Robin}, we obtain the equations of
 \begin{equation}\label{Sistema21}
 	\left\{
 	\begin{array}{llllll}
 		u'_{11}(0)+ u'_{12}(0)+u'_{13}(0)+t_0 u'_{f_{1}}(0) &=& \delta_{a}A_1\\
 		-u'_{11}(1)-e^{i\theta_1}u'_{12}(1)-^{i\theta_2}u'_{13}(1) -t_0u'_{f_{2}}(1) &=& \delta_{b}B_1\\
 		
 		u'_{21}(0)+ u'_{22}(0)+u'_{23}(0) +t_0u'_{f_{1}}(0) &=& \delta_{a}A_2\\
 		- u'_{21}(1) - e^{i\theta_1}u'_{22}(1) - e^{i\theta_2}u'_{23}(1) - t_0u'_{f_{2}}(1) &=& \delta_{b}B_2
 	\end{array}
 	\right.\,.
 \end{equation}
 
 Let $\lambda \notin \sigma (H^D) $. Then there are two linearly independent solutions $\varphi_0=\varphi_{\lambda,0}$ and $\varphi_1=\varphi_{\lambda,1}$ of the problem
 \begin{equation}\label{AutoV2}
 	-\dfrac{d^2\varphi(x)}{dx^2}+q(x)\varphi(x)=\lambda\varphi(x)\,,
 \end{equation}
 such that  
 \begin{equation}\label{AutoV2'}
 	\left\{
 	\begin{array}{ll}
 		\varphi_0(0)=1=\varphi_1(1)\\
 		\varphi_0(1)=0= \varphi_1(0)
 	\end{array}
 	\right.
 	\,\, \mbox{and} \quad \varphi'_1(x)=-\varphi'_0(1-x)\,\,, x \in [0,1]\,.
 \end{equation}
 
 As every edge of $W_2^{AA'}$ is identified with the range $[0, 1]$, we define $\varphi_{\lambda,k}$ on each edge (with the same notation $ \varphi_{\lambda,k}$). Then, for each $\lambda \notin \sigma(H^D)$, we write
 \begin{equation}
 	\left\{
 	\begin{array}{llllll}\label{CombLinearLII}
 		u_{a_{k1}}&=&A_k\varphi_{0}+B_k\varphi_{1}\\
 		u_{a_{k2}}&=&A_k\varphi_{0}+B_k\varphi_{1}e^{-i\theta_1}\\
 		u_{a_{k3}}&=&A_k\varphi_{0}+B_k\varphi_{1}e^{-i\theta_2}\\
 		u_{f_{1}}&=&t_0(A_1\varphi_{0}+B_2\varphi_{1})\\
 		u_{f_{2}}&=&t_0(A_2\varphi_{0}+B_1\varphi_{1})\\
 	\end{array}, \quad  k = 1,2
 	\right.\,\,,
 \end{equation}
 we can verify that the continuity conditions in \eqref{Robin} are satisfied. Now we are left
 check Robin's condition \eqref{AutoV2}. Using the representation of each function $u_{e}$ in \eqref{CombLinearLII} with the total flow equations in \eqref{Sistema21}, we have 
 \begin{equation}
 	\left\{
 	\begin{array}{llllll}\label{Sist22}
 		(3+t_0^2)\varphi'_{0}(0)A_1+\bar{F}\varphi'_{1}(0)B_1+t_0^2\varphi'_{1}(0)A_2&=&\delta_aA_1\\
 		
 		F\varphi'_{0}(1)A_1+(3+t_0^2)\varphi'_{1}(1)B_1+t_0^2\varphi'_{0}(1)B_2&=&-\delta_{b}B_1\\
 		
 		t_0^2\varphi'_{1}(0)B_1+(3\varphi'_{0}(0)-t_0^2\varphi'_{0}(0))A_2+F\varphi'_{1}(0)B_2&=&\delta_aA_2\\
 		
 		t_0^2\varphi'_{0}(1)A_1+\bar{F}\varphi'_{0}(1)A_2+(3\varphi'_{1}(1)+t_0^2\varphi'_{1}(1))B_2&=&-\delta_{b}B_2
 		
 	\end{array}, 
 	\right.
 \end{equation}
 with $F=F(\theta):=1+e^{i\theta_1}+e^{i\theta_2}$ and $\bar{F}(\theta)$ the conjugate complex of $F(\theta)$. 
 As $\varphi'_{\lambda,0}(1)=-\varphi'_{\lambda,1}(0)$, $\varphi'_{\lambda,0}(0)=-\varphi'_{\lambda,1}(1)$ and, where $\varphi'_{\lambda,1}(0)\neq 0$, putting 
 \[
 \eta=\eta(\lambda):=\dfrac{\varphi'_{\lambda,1}(1)}{\varphi'_{\lambda,1}(0)},
 \] $T:=3+t_0^2$, $ \alpha_k:=\dfrac{\delta_k}{\varphi'_{\lambda,1}(0)}$, with $k \in \{a,b\}$, and multiplying the even rows by -1 in \eqref{Sist22}, we obtain the following equivalent system
 $$
 M_2^{AA'}\!(\eta(\lambda),\theta)\cdot X=0, \quad  \mbox{with}\quad  X:=[A_1\,\, B_1\,\, A_2\,\,B_2]^T
 $$ 
 and 
 $$ 
 M^{AA'}_2(\eta(\lambda),\theta):=\left(\begin{array}{cccccc}
 	-T\eta-\alpha_a &	\bar F  & 0 &  t_0^2 	\\
 	F &  -T\eta-\alpha_{b}  & t_0^2  & 0  \\
 	0 &  t_0^2   & -T\eta-\alpha_a &  \bar F  \\
 	t_0^2  &  0 &  F & -T\eta-\alpha_{b} 
 \end{array}\right).
 $$
 The determinant of the matrix $M^{AA'}_2(\eta(\lambda),\theta)$ is given by
 
 $$
 \begin{array}{lll}
 	\det(M^{\!AA'}_2(\eta(\lambda),\theta) &\!\!=\!\!& \Big(T^2\eta^2+(\alpha_a+\alpha_{b})T\eta+\alpha_a\alpha_{b}-F^2-2t_0^2\mathcal{Re}(F)-t_0^4\Big)\\
 	&\cdot\!\!\! &\!\!\!\Big(T^2\eta^2+(\alpha_a+\alpha_{b})T\eta+\alpha_a\alpha_{b}-F^2+2t_0^2\mathcal{Re}(F)-t_0^4\Big)
 \end{array}\,,
 $$
 where $\Re(F)=1+\cos(\theta_1)+\cos(\theta_2)$ the real part of the complex number $F=F(\theta_1,\theta_2)$. The roots of $\det(M^{\!AA'}_2)$ are 
 \begin{equation}\label{RAA'1}
 	r^{\pm}(F)=\frac{-\alpha_a - \alpha_{b}\pm \sqrt{4F^2-8t_0^2\Re(F)+4t_0^4+(\alpha_a-\alpha_{_b})^2}}{2T} 
 \end{equation}
 and  
 \begin{equation}\label{RAA'2}
 	r_{\mp}(F)=\frac{-\alpha_a - \alpha_{b} \mp \sqrt{4F^2+8t_0^2\Re(F)+4t_0^4+(\alpha_a-\alpha_{b})^2} }{2T}\,.
 \end{equation}
 
 With this, we see that there is $\theta \in \mathcal{B}$ such that $\det(M_2^{\!AA'}(\eta(\lambda),\theta)=0$; that is, the representation \eqref{CombLinearLII} solves the eigenvalue problem \eqref{AutoV2} and, therefore, by the Proposition \ref{prop_chave} and \cite{KP2}, we conclude $\lambda \in \sigma(H_2^{\!AA'})$. 
 
 In the following section, we will analyze these functions for the existence of conical or parabolic touches in the operator dispersion relation $H^{\!AA'}_2$. Note that we have abandoned the notation $r(\theta)$ in which the roots explicitly depend on $\theta$, and we started writing the roots with the new notation $r(F)$ in which dependence on $\theta$ is implicit, since $F=F(\theta)$. This change facilitates the analysis of the dispersion relation, as well as the construction of its graphs, as we will see later.
 
 \section[Dispersion relationship analysis]{Dispersion relationship analysis}
 
 In this section we analyze the roots found in piles of the type $AA'$ of two consecutive sheets. 
 In a similar way to Chapter \ref{capPrelimina}, from now on, when nothing is said, we will always 
 consider $\theta \in B_d\!=\!\{(\theta_1,\theta_2)\! \in \![-\pi,\pi]^2\!:\! \theta_2\!=\!-\theta_1\}$, that is, we restrict the quasimoment $\theta$ to the diagonal of the Brillouin zone $B$. With this, $F$ becomes real, that is,
 and the roots in \eqref{RAA'1} and 
 \eqref{RAA'2} can be written in the form
 \begin{equation}\label{RAA'3}
 	r^{\pm}_{\mp}(F)=\frac{-\alpha_a - \alpha_{b}\pm \sqrt{4(F \mp t_0^2)^2+(\alpha_a-\alpha_{b})^2}}{2(3+t_0^2)} \,\,.
 \end{equation}
 
 Let us remember that the case $\alpha_{b}=\alpha_a=0$ reduces to what is discussed in \cite{ROCHA}, with two sheets of graphene. So, let us initially assume that $\alpha_{b}=\alpha_a\neq 0$; then in \eqref{RAA'3}, we get 
 \begin{equation}\label{Dirac0}
 	r^{\pm}_{\mp}(F)=  \dfrac{-\alpha_{b}\pm \left|F \mp t_0^2\right|}{3+t_0^2} \,, 
 \end{equation}	
 with $t_0 \in (0,1)$ fixed.
 
 With this, we see that the analysis remains analogous to the study carried out in Chapter \ref{capEstrHomAA}, in case of stacking $AA$, just as the results and demonstrations are easily adapted from the Lemma \ref{Disp} and of the Theorem \ref{Disp4}. We then reach the following conclusion about the $AA'$ type stacking with two atoms alternating at the vertices.
 
 \begin{theorem}(Operator $H^{AA'}_2$ dispersion relation)\, Consider $t_0 \in (0,1)$ fixed and $r^{\pm}_{\pm}$ as in \eqref{RAA'3}. Then
 	\begin{enumerate}
 		\item[(a)] If $\alpha_a=\alpha_{b} \neq 0$, the operator dispersion relation $H^{AA'}_2$ always has Dirac cones in \footnote{That is, when $\theta_D=\pm \left(\arccos\big(\frac{t_0^2-1}{2}\big),-\arccos\big(\frac{t_0^2-1}{2}\big)\right)$, there is $F_D:=F(\theta_D)=\pm \left(t_0^2,-t_0^2\right)$.} $F_D=\pm \left(t_0^2,-t_0^2\right)$ (Figure \ref{AaBN-BNa}).
 		
 		\item[(b)] For $\alpha_a\neq\alpha_{b}$, the dispersion relation of $H^{AA'}_2$ it has no touch, that is, it always has gaps (Figure \ref{AaBN-BNb}).
 		
 	\end{enumerate}	  
 	
 \end{theorem}
 
 \begin{figure}[H]
 	\centering
 	\includegraphics[width=0.4\linewidth]{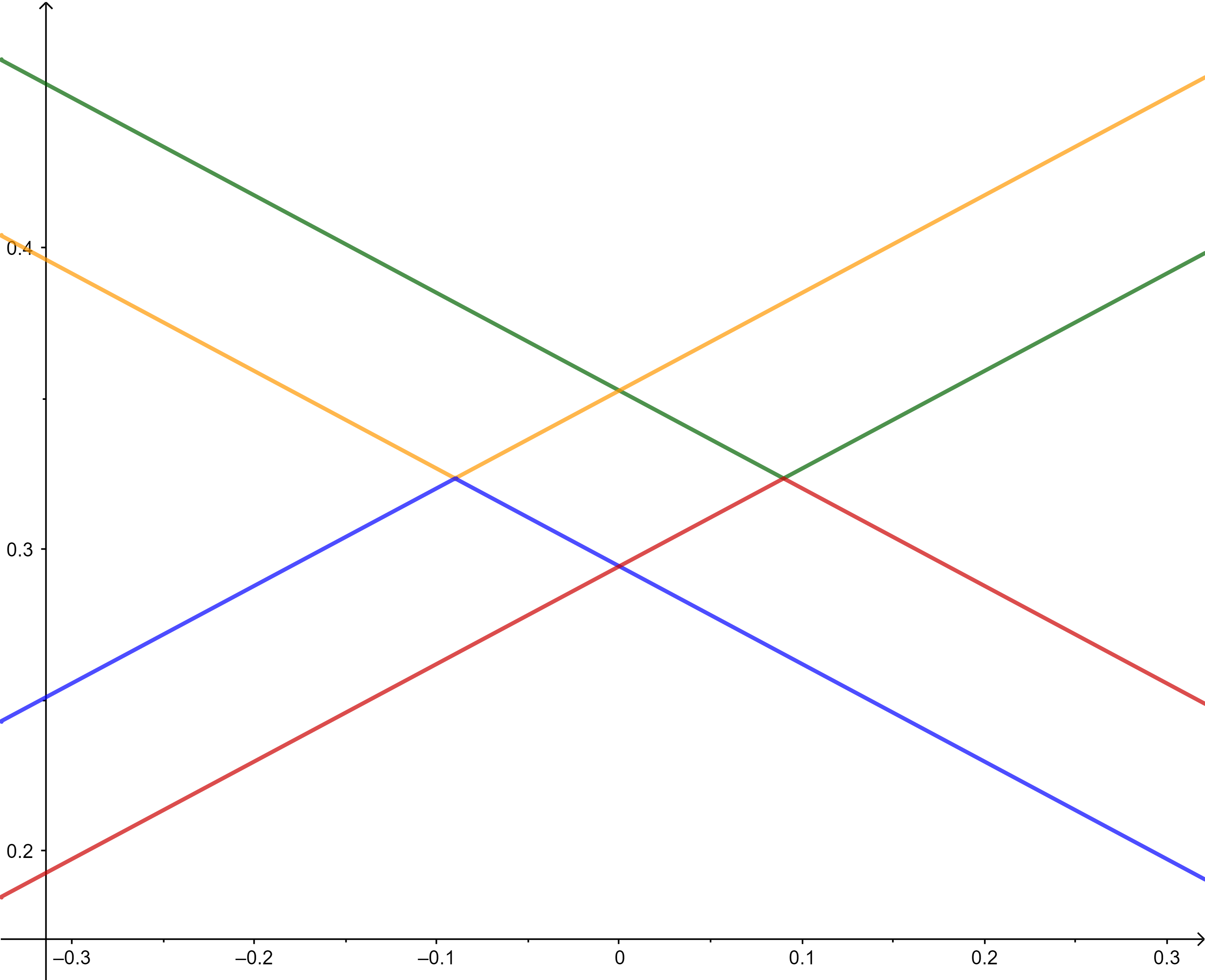}
 	\caption{Two Dirac cones when the atoms are equal; $\,\alpha_a =\alpha_{b}=-1\,,t_0=0.3$.}
 	\label{AaBN-BNa}
 \end{figure}
 
 \begin{figure}[H]
 	\centering
 	\includegraphics[width=0.4 \linewidth]{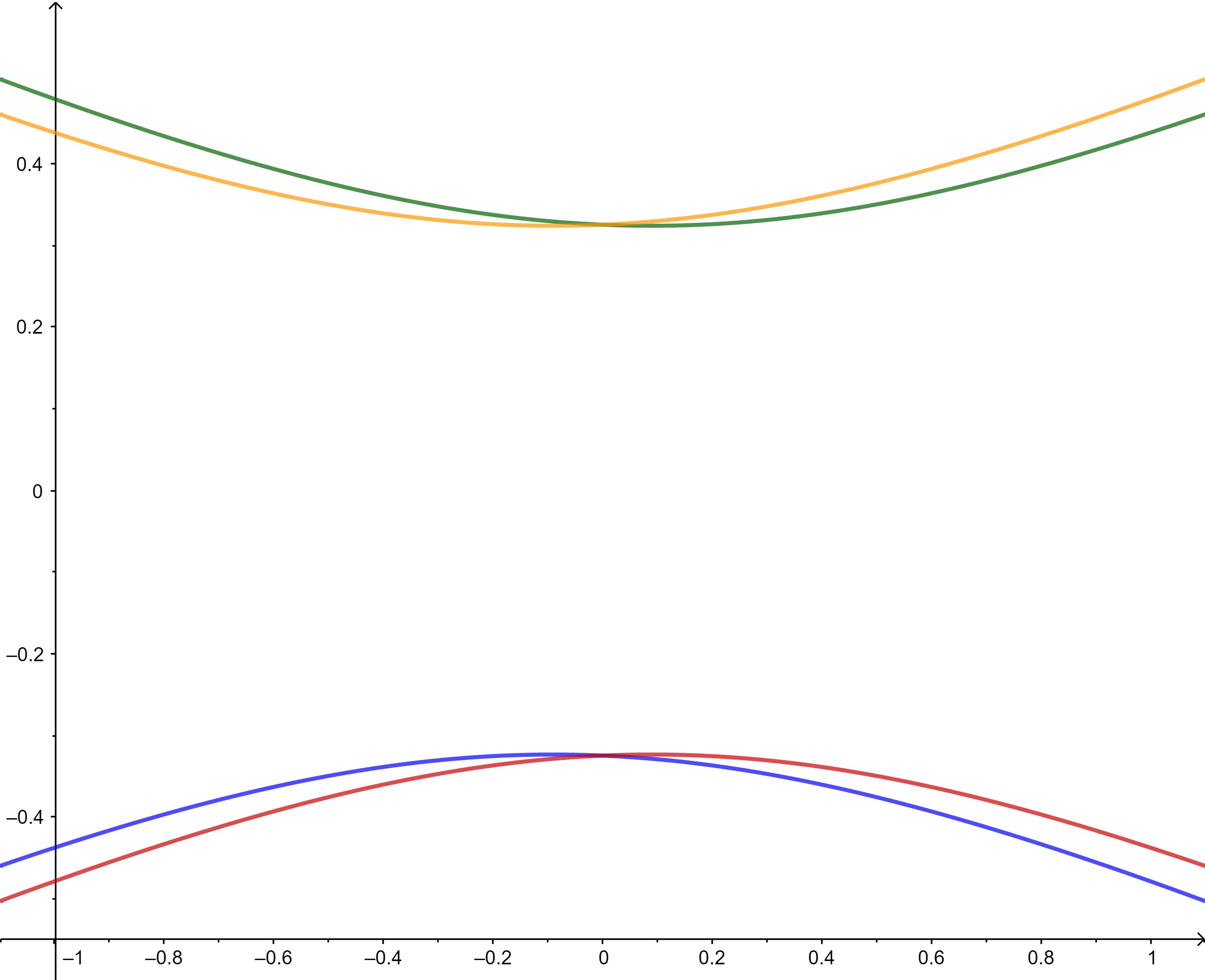}
 	\caption{Gaps if atoms are distinct; $\alpha_{b} =1,\,\alpha_a=-1\,,t_0=0.3$.}
 	\label{AaBN-BNb}
 \end{figure}
 
 In the following section, we summarize these results.
 
 \section{Analysis and conclusions}
 
 In the calculations and analyzes carried out so far, any material formed by hexagons with two atoms interspersed at the vertices was considered and modeled with a quantum graph. However, item (b) of the theorem above tells us that if, in particular, we apply these results to boron nitride, since the boron and nitrogen atoms are distinct, we obtain the following conclusion:
 
 \textit{If $\alpha_a$ and $\alpha_{b}$ represent the boundary conditions associated with nitrogen and boron, respectively, we conclude that the dispersion relation of the $hBN$ operator, in $AA'$ stacking, has no touch, or even that hexagonal boron nitride (bilayer) always has gaps.}
 
 This is in agreement with the results found in the experimental physics literature \cite{FANG}; A material that has gaps between its valence bands is said to be an insulator. In particular, boron nitride is an insulator, which is a well-known fact, corroborating the above results for two layers in this configuration.
 
 \begin{observation}\label{AA-AA'}
 	There is a small and interesting distinction between the results presented by $AA$ and $AA'$ models when considering $hBN$ bilayers.
 	
 	\begin{enumerate}
 		\item The $AA'$ model eliminates the parabolic touches that occur in stacking $AA$.
 		\item If $\alpha_{b}=\alpha_a$,
 		\,the $AA'$ model horizontally displaces the Dirac cones of $AA$ by $t_0^2$.
 		
 		\item While in $AA$ stacking we found a gap of 0.589, in $AA'$ the gap (where the parabolic touch occurred) is 0.64987 and (where the conical touch occurred) is 0.64725, as shown in Figures \ref{CasoAA-BN-BNc} and \ref{AaBN-BNb}. 
 		{Finally, we note that we did not find experimental studies in the physical literature considering two layers of hBN stacked in the type $AA$, a fact that we makes it impossible to make more precise analyzes and comparisons of our study with others.}
 	\end{enumerate}  
 	
 	Note also that, although $AA$ stacking makes sense because it is a mathematical model, such a structure does not describe the hexagonal boron nitride material seen as a solid composed of n-layers, as an essential characteristic of this material is that the nitrogen interacts with the boron atom and vice versa, both in the strong interaction in the same layer and in the van der Walls interaction, between the sheets \cite{FANG}, \cite{G.K.B.B.K}. However, modeling with $AA'$ stacking presents a better proposal for describing hBN, as each boron atom interacts with a nitrogen atom and vice versa, whether in a weak or strong interaction, so that the property of the dispersion relationship of a monolayer in having holes is preserved in type $AA'$ stacking for bilayers, a fact corroborated by \cite{FANG}.
 	
 	
 \end{observation}
 
 \begin{observation}\label{ObsBNBNBN}
 	An analogous calculation (not presented here) considering three sheets, that is, adding another layer of the same material to the $AA'$ stack of two sheets, shows us that the dispersion relation obtained always has gaps for $\alpha_{b}=-\alpha_a$.
 \end{observation}
 
 In the next chapter, we will address stacks with two and three distinct two-dimensional layers, that is, cases of heterostructures.
 
 
 \chapter[Heterostructures]{Heterostructures: boron nitride and graphene}\label{capHetNBG}
 
 Now we will use quantum graphs to model systems formed by some combined layers of graphene and boron nitride, particularly the one studied in \cite{G.K.B.B.K}, that is, some heterostructures with two and three layers. In the aforementioned article, the authors considered a heterostructure with a graphene layer on (a block of) boron nitride, so that the entire analysis and conclusion are carried out considering only the interaction between the graphene layer and the boron nitride layer; see Figure \ref{BN-Graf}.
 \vspace{-0.55cm}
 \begin{figure}[h]
 	\centering
 	\includegraphics[width=0.7\linewidth]{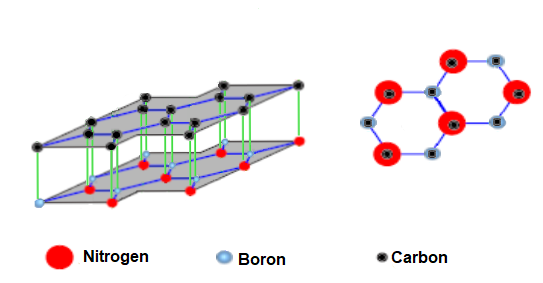}
 	\caption{A two-layer van der Waals heterostructure: side and top views, respectively.}
 	\label{BN-Graf}	
 \end{figure}

 \section{Distinct bilayers}
 
 Initially, we will stick to the two-layer heterostructure: on the bottom sheet, we fix a layer formed by hexagons with two atoms (mathematically represented by the parameters $\delta_a$ and $\delta_{b}$ in the Robin~\eqref{Robin} boundary conditions ) interspersed at the vertices of each hexagon; in the upper one, we take a hexagonal sheet with an atom ($\delta_{_C}$) at each vertex. In particular, when convenient, we will also use the constants $\delta_{_B}, \delta_{_N}$ and $\delta_{_C}$ to indicate the boundary condition attributed to the vertices, modeling with boron, nitrogen and carbon atoms respectively.  
 
 \begin{figure}[h]
 	\centering
 	\includegraphics[width=0.6\linewidth]{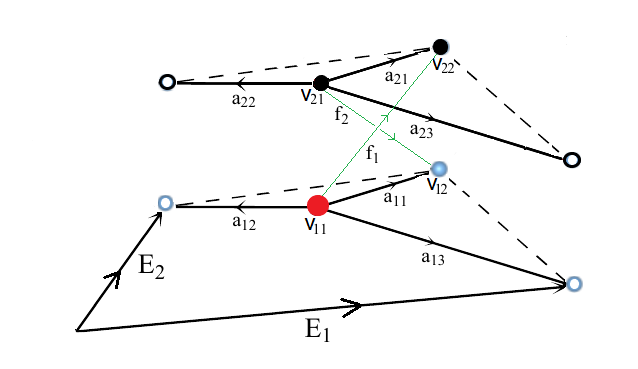}
 	\caption{Fundamental domain $W$ of a heterostructure bilayer formed by 8 edges and 4 vertices.}
 	\label{AaBN-G-MAT}
 \end{figure}
 
 As in the previous chapter, we considered the fundamental domain given by
 $$ 
 W=\{a_{11}, a_{12}, a_{13}, a_{21}, a_{22}, a_{23}, f_1, f_2, v_{11}, v_{12}, v_{21}, v_{22}\}\,.
 $$ 
 
 In a similar way to the previous chapters, we apply Floquet cyclic conditions and continuity conditions to the fundamental domain $W$ and obtain\\
 \begin{equation}
 	\left\{
 	\begin{array}{lll}
 		u_{a_{11}}(0)&=&u_{a_{12}}(0)=u_{a_{13}}(0)=\dfrac{u_{f_{1}}(0)}{t_0}=:A_1\\
 		u_{a_{11}}(1)&=&e^{i\theta_1}u_{a_{12}}(1)=e^{i\theta_2}u_{a_{13}}(1)=\dfrac{u_{f_{2}}(1)}{t_0}=:B_1\\
 		u_{a_{21}}(0)&=&u_{a_{22}}(0)=u_{a_{23}}(0)=\dfrac{u_{f_{2}}(0)}{t_0}=:A_2\\
 		u_{a_{21}}(1)&=&e^{i\theta_1}u_{a_{22}}(1)=e^{i\theta_2}u_{a_{23}}(1) = \dfrac{u_{f_{1}}(1)}{t_0}=:B_2
 	\end{array}\,.\right.
 \end{equation}
 
 
 Applying (ii) of the modified Neumann Vertex Condition \eqref{Robin}, we obtain the total flow equations at each vertex of the fundamental domain:
 \begin{equation}\label{Sist23}
 	\left\{
 	\begin{array}{llllll}
 		u'_{11}(0)+ u'_{12}(0)+u'_{13}(0)+t_0 u'_{f_{1}}(0) &=& \delta_{a}A_1\\
 		-u'_{11}(1)- e^{i\theta_1}u'_{12}(1)-e^{i\theta_2}u'_{13}(1) -t_0u'_{f_{2}}(1) &=& \delta_{b}B_1\\
 		
 		u'_{21}(0)+ e^{i\theta_1}u'_{22}(0)+e^{i\theta_2}u'_{23}(0) +t_0u'_{f_{1}}(0) &=& \delta_{c}A_2\\
 		-u'_{21}(1)- u'_{22}(1)-u'_{23}(1)-t_0u'_{f_{2}}(1) &=& \delta_{c}B_2
 	\end{array}
 	\right.\,.
 \end{equation}
 
 We also consider, for $\lambda \notin \sigma(H^D)$, the functions $\varphi_{0}=\varphi_{\lambda,0}$, $\varphi_{1}=\varphi_{\lambda,1}$ and we write the linear combination
 \begin{equation}
 	\left\{
 	\begin{array}{llllll}\label{LI2}
 		u_{a_{k1}}&=&A_k\varphi_{\lambda,0}+B_k\varphi_{\lambda,1}\\
 		u_{a_{k2}}&=&A_k\varphi_{\lambda,0}+e^{-i\theta_1}B_k\varphi_{\lambda,1}\\
 		u_{a_{k3}}&=&A_k\varphi_{\lambda,0}+e^{-i\theta_2}B_k\varphi_{\lambda,1}\\
 		u_{f_{1}}&=&t_0(A_1\varphi_{\lambda,0}+B_2\varphi_{\lambda,1})\\
 		u_{f_{2}}&=&t_0(A_2\varphi_{\lambda,0}+B_1\varphi_{\lambda,1})
 	\end{array}, \quad  k = 1,2
 	\right.\,\,;
 \end{equation}
 with the same notations as in previous chapters and the appropriate substitutions, we obtain the system
 $$
 M^{ab}_{c}(\theta) X=0, \quad  \mbox{com}\quad  X=[A_1\,\, B_1\,\, A_2\,\,B_2]^T
 $$ 
 with 
 $$ 
 M^{ab}_{c}(\theta)=\left(\begin{array}{cccccc}\label{matrizAab}
 	-T_1\eta-\alpha_a &	\bar F  & 0  & t_0^2 	\\
 	F &  -T_1\eta-\alpha_{b}  &  t_0^2 & 0    \\
 	0  & t_0^2 & -T_1\eta-\alpha_c & \bar F  \\
 	t_0^2 &  0   & F & -T_1\eta-\alpha_c 
 \end{array}\right)\,.
 $$
 
Let's assume that $\alpha_c=\alpha_C=0$, as admitted in \cite{OLIRO}. Having done this, the determinant $\det(M^{ab}_{c}(\theta))=:P(\eta)$ is given by 
 {\small $$
 	\begin{array}{lll}
 		\!P(\eta)\! = \!\eta^4T^4\!+\!(\!\alpha_a\!+\!\alpha_{b})T^3\eta^3\!+\!\left(\!\alpha_a\alpha_{b}\!-\!2F^2\!-\!2t_0^4\right)T^2\eta^2\!-\!(\!\alpha_a\!+\!\alpha_{b})(\!F^2\!+\!t^4)T\eta\!+\!(\!F^2\!-\!t_0^4)^2\!-\!\alpha_a\alpha_{b}F^2
 	\end{array}
 	$$}
 In the subsequent section, we will analyze the behavior of the roots of $P$. Although we do not make it explicit, we highlight that when analyzing the functions $\eta$ as a dispersion relation $\lambda$ of the operator, in fact, we are considering the observation ~\eqref{Crucial} from Chapter \ref{capPrelimina}, in which we saw that it is necessary to require that $|\eta(\lambda)|\leq 1$ in order to relate the spectra $\sigma(H)$ and $\sigma(H^{\mathrm{per}})$. That said, we will always be considering, even if implicitly, only the cases in which $r(F(\theta))\leq 1$ in our analysis of the dispersion relation.

 \subsection{Analysis of the dispersion relation.}
 
 Initially, we highlight that since $P$ is a polynomial of degree 4, there is enormous difficulty in displaying the general form of the zeros of $P$, so that the general expression of its roots, that is, with arbitrary values of $t_0$, $\alpha_a$ and $ \alpha_{b},F$ are extremely extensive and intractable. However, considering the case\footnote{Such a choice is partially justifiable, since it has some physical motivation, because thinking about the parameters~$\alpha$s as electronegativity, we would have $\alpha_{_B}<\alpha_{_C}<\alpha_{_N}$, and with the choice $\alpha_{_C}=0$, then $ \alpha_{_N}>0$ and $\alpha_{_B}=-\alpha_{_N}$ becomes reasonable.} $\alpha_{b}=-\alpha_a\neq 0$, the roots of $P(\eta)$ become simpler and are given by
 
 
 \begin{equation}\label{RAab1}
 	r^{\pm}(F)=\pm\dfrac{\sqrt{2F^2+2t_0^2+\alpha_a^2+\sqrt{16t_0^4F^2+4\alpha_a^2t_0^4+\alpha_a^4}}}{\sqrt{2}(3+t_0^2)}
 \end{equation}
 and
 \begin{equation}\label{RAab2}
 	r_{\mp}(F)=\mp\dfrac{\sqrt{2F^2+2t_0^2+\alpha_a^2-\sqrt{16t_0^4F^2+4\alpha_a^2t_0^4+\alpha_a^4}}}{\sqrt{2}(3+t_0^2)}\,.
 \end{equation}
 
 \begin{figure}[H]
 	\centering
 	\includegraphics[width=0.5\linewidth]{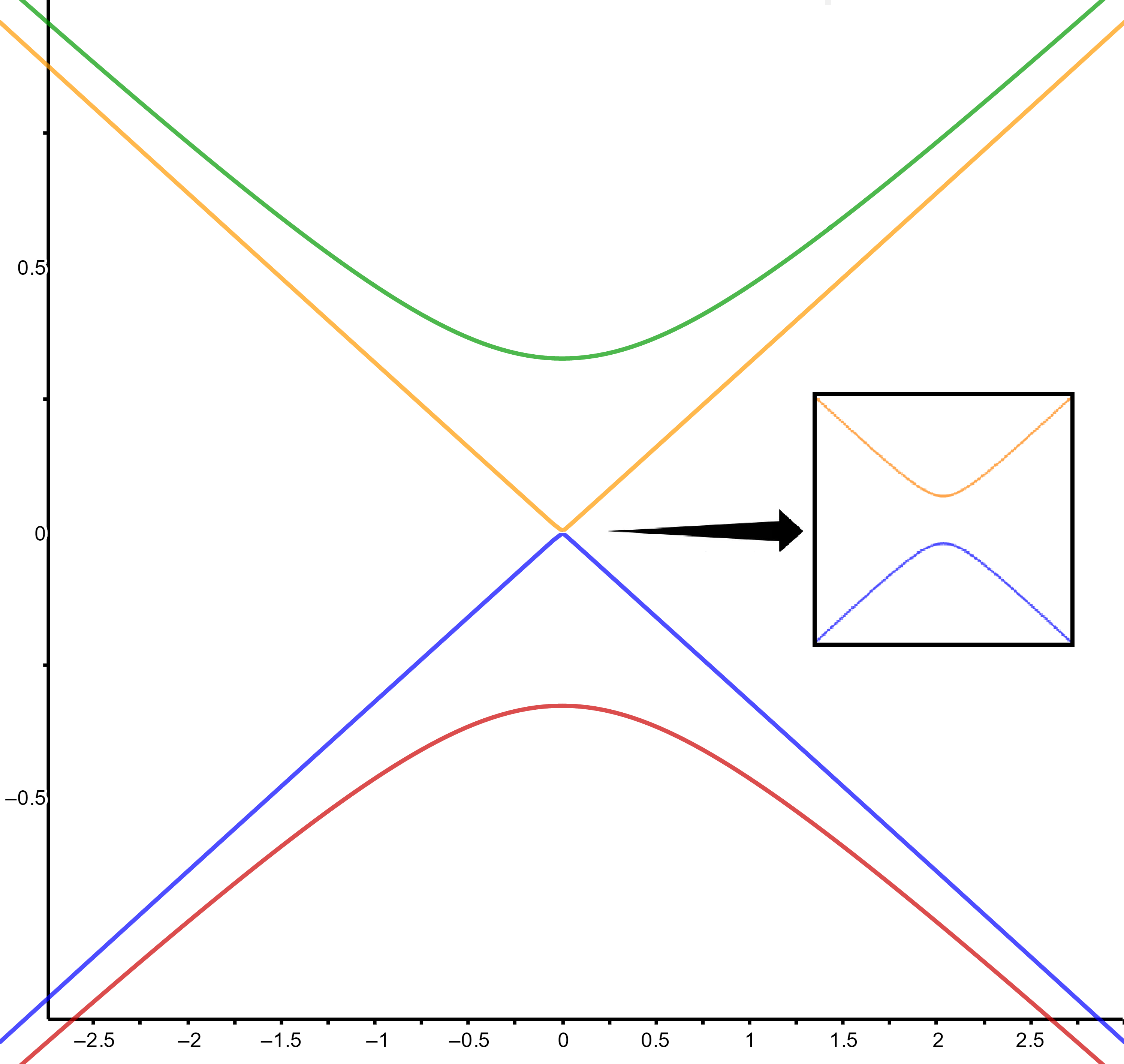}
 	\caption{Gaps in the hBN-graphene dispersion relation; $t_0=0.3\,,\alpha_a=-1$.}
 	\label{figura A}
 \end{figure}
 
 Note that, as shown in Figure \ref{figura A}, no conical ringing occurs at the zero energy level, so a spectral gap is induced by the inclusion of the boron nitride layer on top of the graphene layer. This is in complete agreement with the results found in laboratory experiments and published in the literature, namely, in \cite{FANG} and \cite{G.K.B.B.K}. In this type of stacking with bilayers of different materials, graphene carbon atoms, which until then only interacted with other equivalent carbon atoms (and always generated conical touches at the origin), began to have weak bonds with inequivalent atoms (boron and nitrogen) from the upper layer. This inequivalence of the bonds caused the rupture of the Dirac cones and the appearance of gaps at the origin. Furthermore, in \eqref{RAab2} we can determine the width of the smallest gap (that is, at the origin) formed as a function of the parameters $t_0$ and $\alpha_a$, which we will denote by $g$, and given by
 \begin{equation}\label{Gap}
 	g(\alpha_a,t_0)=\big|r_+(0)-r_-(0)\big|=\sqrt{2}\,\frac{\sqrt{2\,t_0^4+\alpha_a^2-\sqrt{4\,\alpha_a^2\,t_0^4+\alpha_a^4}}}{3+ t_0^2 }\,.
 \end{equation}
 
 In particular, for $t_0=0.3$ and $\alpha_a=-1$, we obtain, at the origin, a gap of at least $g(-1,0.3)\approx 0.0052$, illustrated in Figure \ref{figura A}. In general, for each fixed $\alpha_a$, when $t_0 \to 0^+$, the gap decreases so that $\dis\lim_{t_0\to 0^+}g(\alpha_a,t_0)=0$; and as $t_0 \to 1^-$, the gap between the valence and conduction bands increases, as shown in figure \ref{Gap2}, that is,
 $$
 g(\alpha_a,1)=\dfrac{\sqrt{2}}{4}\sqrt{2+\alpha_a^2-\sqrt{\alpha_a^4+4\alpha_a^2}}\,.
 $$ 
 
 \begin{figure}[H]
 	\centering
 	\includegraphics[width=0.5\linewidth]{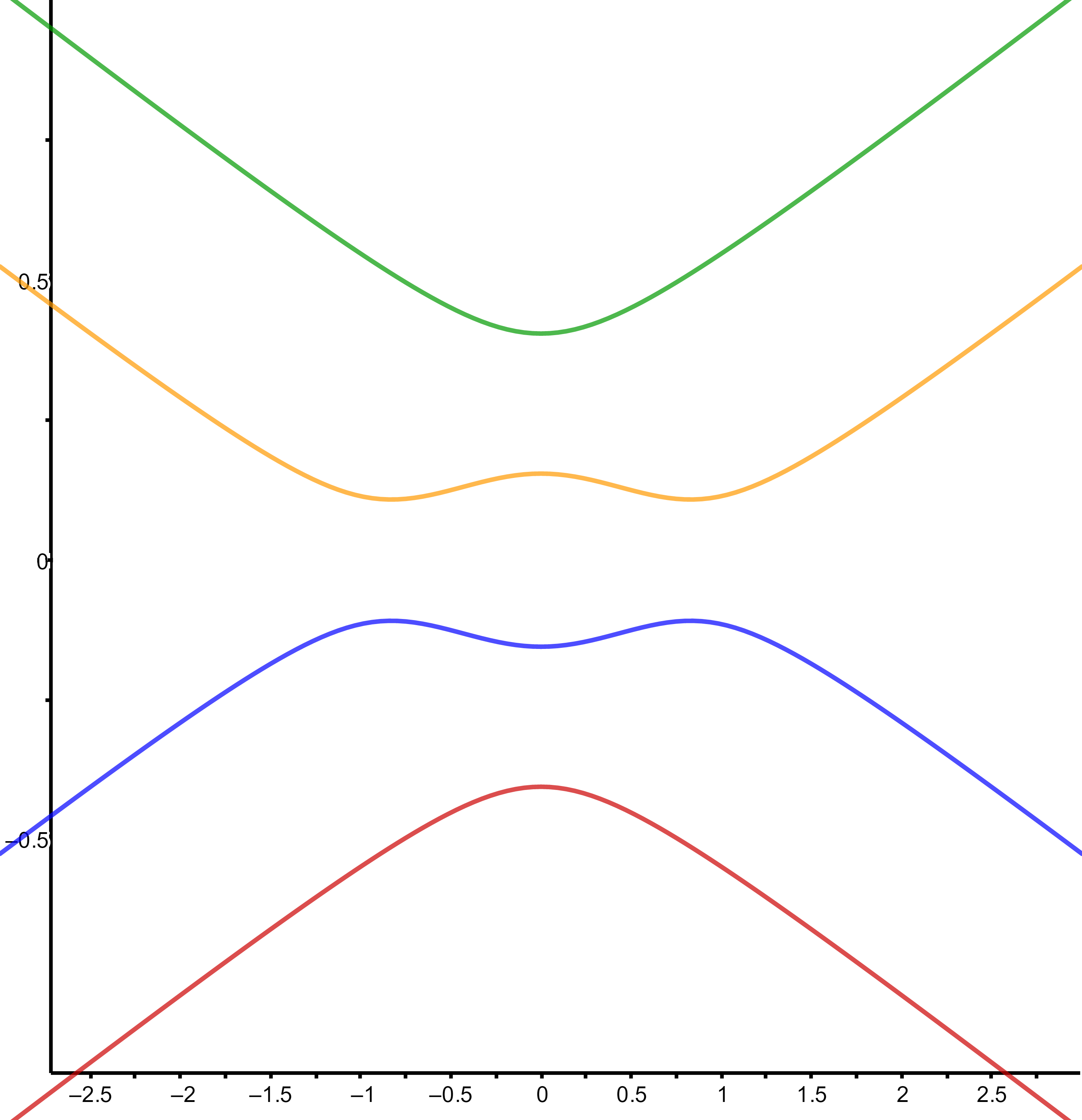}
 	\caption{Greater interaction between layers ($t_0=1)$ causes an increase in the gap; $\alpha_a=-1$.}
 	\label{Gap2}
 \end{figure}
 
 From a physical point of view, increasing $t_0$ means increasing interaction between layers. In particular, as mentioned at the beginning of the work, when $t_0=1$ the weak and strong interactions coincide.
 
 Note that, in \eqref{RAab1}, for any nonzero values of $t_0, \alpha_a, \alpha_{b}$, $r^+$ is always positive and $r^-$ is always negative, so that the two never meet. null and therefore do not intersect, as all parameters $t_0, \alpha_a, \alpha_{b}$ are not null.  
 
 \begin{theorem}[Conclusion on hBN-graphene stacking]
 	Let $\alpha_{_N}, \alpha_{_B}$ be distinct and non-null, and $0<t_0 \leq 1$ be fixed. So the dispersion relation obtained from hBN-graphene stacking always has holes at $F=0$ for $\alpha_{_B}=-\alpha_{_N}$.
 \end{theorem}
 
 In the next section, we extend the case discussed in this one and consider two new cases of physical interest: the ``sandwiches'' formed with these two hexagonal materials. In both cases, we will apply the modified Neumann conditions, the cyclic Floquet conditions and the equations of linear combinations of the solutions, which are easily adapted and applied to the case of structures with three layers (although with much longer calculations), so To avoid repetition, we will omit the calculations and display only the most essential details for the analysis.
 
 Let us initially remember that, seeking to bring our model closer to reality, it is reasonable to choose certain values for the parameters associated with the atoms in boron nitride, which we will denote by $\alpha_{_N}$ for the nitrogen atom and $\alpha_{_B}$ for boron.
 
 \section{Distinct trilayers}
 
 As briefly presented in the Introduction, the configuration formed by three distinct layers, so that the middle layer differs from the two similar outer layers, has aroused interest as it is a promising proposal that would solve the problem of graphene not being able to be used as a type of transistor
 
 We address in the two subsections \footnote{We highlight here that, although the stacking is of three layers, in our model we consider interactions only between consecutive layers.} following two types of ``sandwiches'': graphene sheet between hBN, and hBN between graphene.
 
 \subsection{The hBN-Graphene-hBN case}
 
 In this subsection we consider the ``sandwich'' formed by a monolayer of graphene between two sheets of hexagonal boron nitride, as illustrated in Figure \ref{BN-Graf-BN}.
 
 \begin{figure}[H]
 	\centering
 	\includegraphics[width=0.6\linewidth]{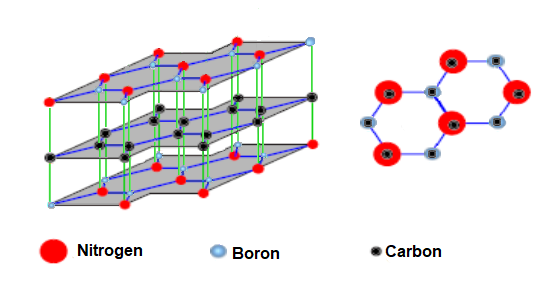}
 	\caption{The hBN-Graphene-hBN sandwich: side and top views, respectively.}
 	\label{BN-Graf-BN}
 \end{figure}
 
 There is great interest in this type of configuration, as according to the current wide discussion, the aim is to know what new properties will be obtained by including a layer of graphene (an excellent conductor) between two layers of boron nitride (an excellent insulating). In other words, mathematically we will analyze the behavior of the dispersion relation obtained from this configuration regarding the presence of Dirac cones, parabolic touches or gaps between the valence and conduction bands. Let us remember that denoting by $\alpha_{_N},\alpha_{_B}$ and $\alpha_{_C}$ the constants associated with the boundary conditions of nitrogen, boron and carbon, respectively, the matrix $M_{{BNGBN }}$ associated with this model, can be obtained in an entirely analogous way to those in the previous chapters (although in a more laborious way), and is given by 
 $$ 
 M_{{BNGBN}}\!\!=\!\!\left(\!\!\!\!\begin{array}{cccccc}
 	-T_1\eta-\alpha_{_N} &	\bar F  & 0  & t_0^2 & 0 & 0  	\\
 	F &  -T_1\eta-\alpha_{_B}  &  t_0^2 & 0 & 0 & 0  \\
 	0 & t_0^2 & -T_2\eta-\alpha_{_C} & \bar F & 0 & t_0^2  \\
 	t_0^2 &  0 & F & -T_2\eta-\alpha_{_C} & t_0^2 & 0 \\
 	0 &  0  & 0 & t_0^2 & -T_1\eta-\alpha_{_N} & \bar F \\
 	0 &  0  & t_0^2 & 0 & F & -T_1\eta-\alpha_{_B}
 \end{array}\!\!\!\!\right)
 $$
 with $T_1:= 3+t_0^2$ e $T_2:= 3+2t_0^2$. 
 
 For simplicity, taking the constant associated with the carbon atom as null, that is, $\alpha_{_C}=0$, as admitted in \cite{OLIRO}, we obtain $\det(M_{{BN.G.BN }})(\eta)=:P(\eta)=P_2(\eta)P_N(\eta)$, where
 \[
 P_2(\eta)=(3+t_0^2)\eta^2+(\alpha_{_N}+\alpha_{_B})(3+t_0^2)\eta+\alpha_{_N}\alpha_{_B}-F^2
 \] and $P_N$ a polynomial of degree 4 in $\eta$. The roots of $P_2$ are given by
 \begin{equation}\label{RAab3}
 	r_{\pm}(F)=\dfrac{-\alpha_{_B}-\alpha_{_N}\pm\sqrt{(\alpha_{_B}-\alpha_{_N})^2+4F^2}}{2(3+t_0^2)}\,,
 \end{equation}
 which clearly have gaps for any distinct values of $\alpha_{_B}$ and $\alpha_{_N}$, but have a Dirac cone \footnote{The case $\alpha_{_N}=\alpha_{_B}$ does not represents the hexagonal boron nitride material, since in hBN we assume that the atoms are distinct.} when $\alpha_{_B}=\alpha_{_N}$ (see Figures \ref{BN-G-BNa1}, \ref {BN-G-BNa01} and \ref{BN-G-BNa001}). As already highlighted in Chapter~\ref{capEstrHomAAnovo}, Observation \ref{AA-AA'}, in configurations like $AA$ or $AA'$, with two or three sheets of boron nitride, there are always gaps for $ \alpha_{_B}=-\alpha_{_N} \neq 0$, however the inclusion of a graphene sheet between two boron nitride sheets, although it did not eliminate the gap, induced a reduction in the width of the spectral gap of $\approx 0.65$ to $\approx 0.07$, that is, an order of magnitude; certainly this is the main property of this configuration.
 
 The roots of $P_N$ are given by extremely long expressions so that their general form becomes intractable. However, taking $\alpha_{_B}=-\alpha_{_N}$, we were able to carry out some analyzes so that we obtained important and interesting results, that is
 $$
 P_N(\eta)=A(t_0,F,\alpha_{_N})\eta^4+B(t_0,F,\alpha_{_N})\eta^2+C(t_0,F,\alpha_{_N})\,,
 $$ 
 being
 $$
 A(t_0,F,\alpha_{_N})=(3+t_0^2)^2(3+2t_0^2)^2\,,
 $$
 $$
 B(t_0,F,\alpha_{_N})=-\left[(5t_0^4+18t_0^2+18)F^2+8t_0^8+36t_0^6+(4\alpha_{_N}+36)t_0^4+12\alpha_{_N}^2t_0^2+9\alpha_{_N}^2\right]
 $$
 and
 $$
 C(t_0,F,\alpha_{_N})=-(F^2-2t_0^4)^2-\alpha_{_N}^2F^2 \,,
 $$
 so that, in this case, the roots of $P_N$ are given by
 \begin{equation}\label{RAab4}
 	r^{\mp}_{\pm}(F)= \mp\sqrt{ \dfrac{-B(t_0,F,\alpha_{_N}) \pm \sqrt{B^2(t_0,F,\alpha_{_N})-4A(t_0,F,\alpha_{_N})C(t_0,F,\alpha_{_N})} }{2A(t_0,F,\alpha_{_N})} }
 \end{equation}
 
 \begin{figure}[H]
 	\centering
 	\includegraphics[width=0.4\linewidth]{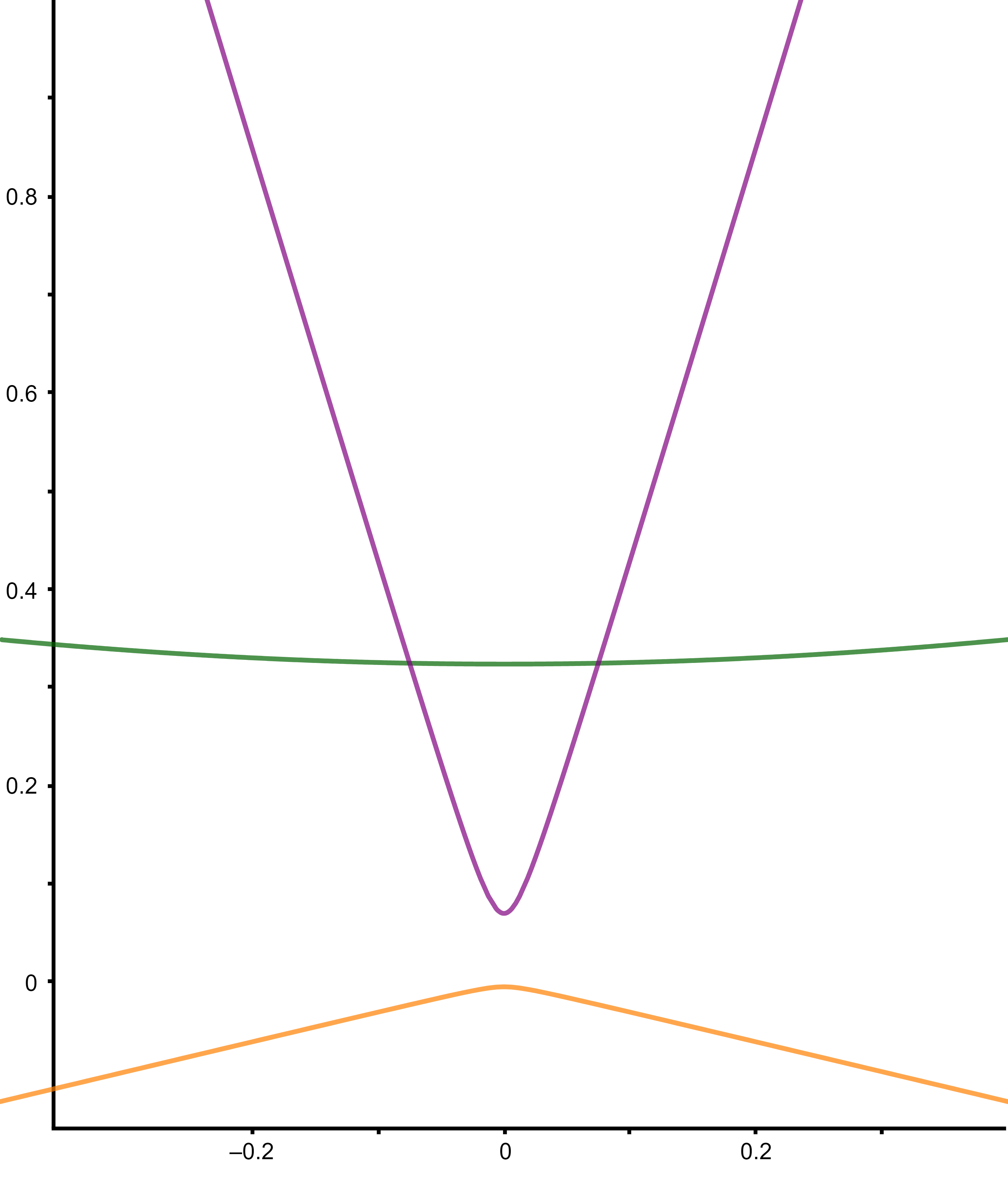}
 	\caption{Gaps at origin for $t_0=0.3$, $\alpha_{_B}=-\alpha_{_N}=1\,,\alpha_{_C}=0$. The inclusion of a graphene sheet between two hBN sheets does not eliminate the gap at $F=0$, but causes a reduction in the gap width from $\approx 0.65$ in two hBN layers to $\approx 0.074$.}
 	\label{BN-G-BNa1}
 \end{figure}
 
 \begin{figure}[H]
 	\centering
 	\includegraphics[width=0.4\linewidth]{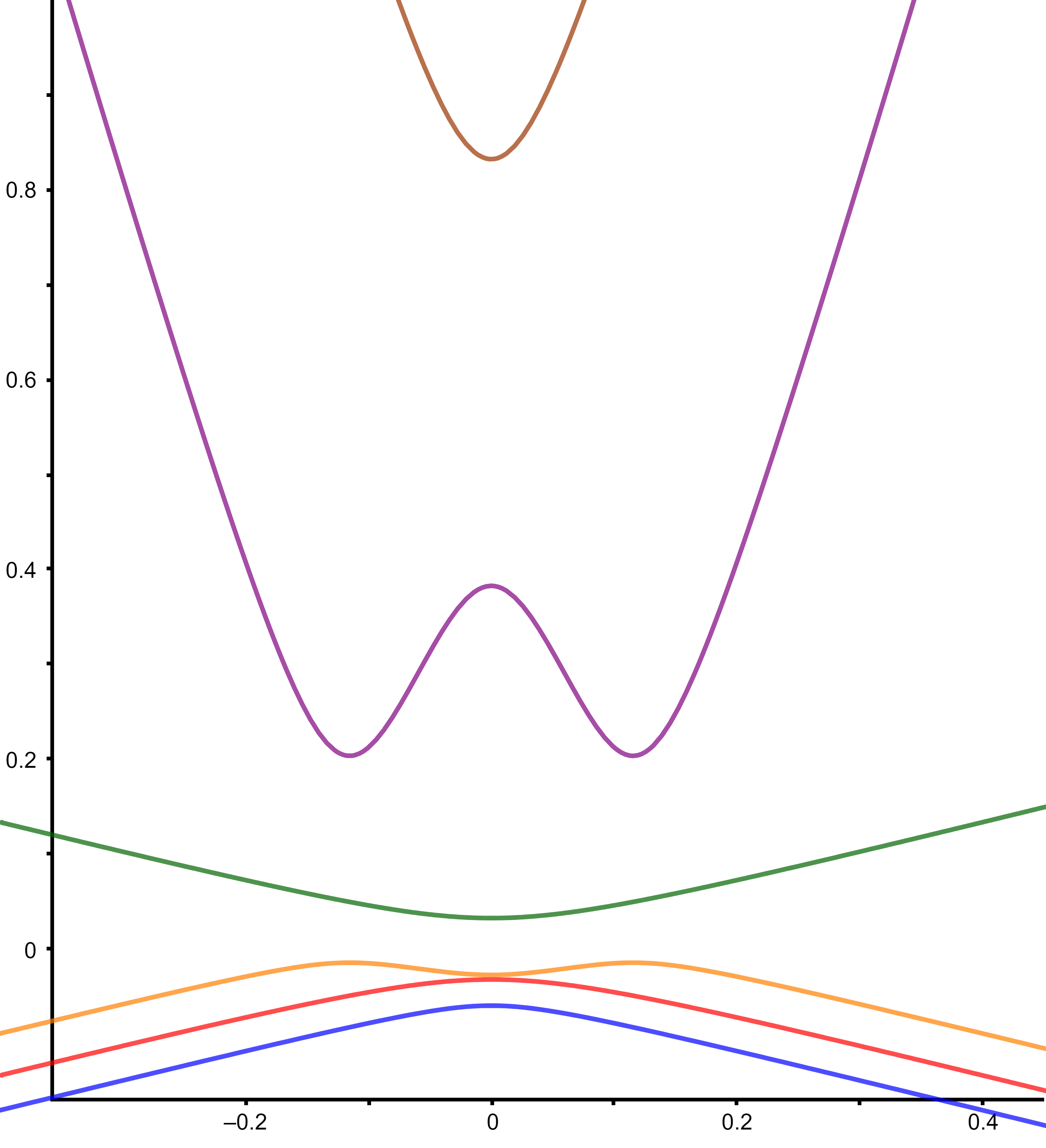}
 	\caption{Gaps in $F\!=\!0$ from $\approx 0.00483$ (the smallest) for $t_0\!=\!0.3$, $\alpha_{_B}\!=\!-\alpha_{_N}\!=\!0.1\,,\alpha_{_C}\!=\!0$.}
 	\label{BN-G-BNa01}
 \end{figure}
 
 \begin{figure}[H]
 	\centering
 	\includegraphics[width=0.4\linewidth]{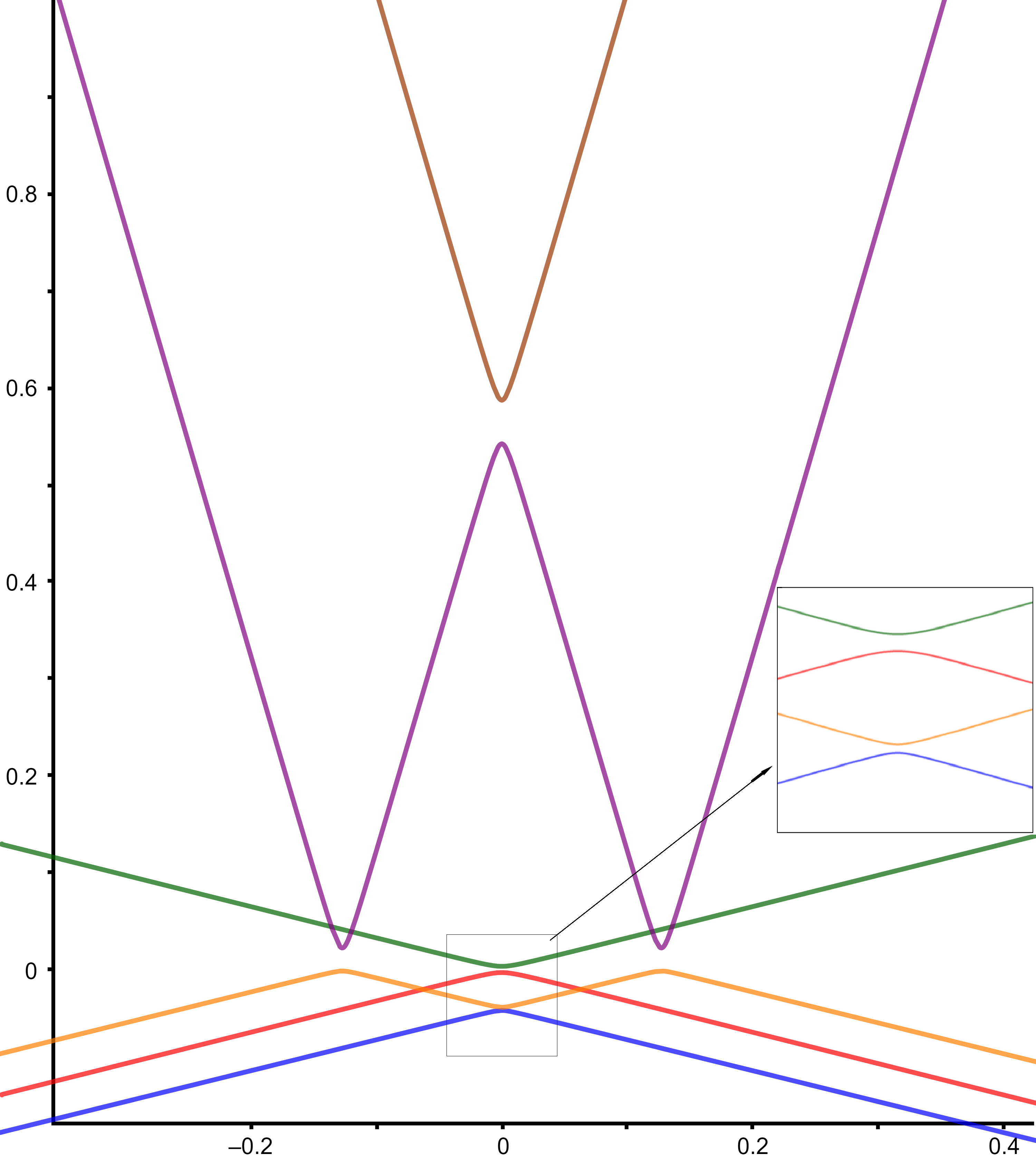}
 	\caption{Gaps in $F=0$ of $\approx \! 0.00324$, $\approx \!0.00647$ (smaller auxiliary figure on the side) and $\approx \!0.04497$ for $t_0=0.3$, $\alpha_{_B}=-\alpha_{_N}=0.01\,,\alpha_{_C}=0$.}
 	\label{BN-G-BNa001}
 \end{figure}
 
 \begin{observation}
 	Although not presented here, we analyzed stacks of the ~$AA$ or $AA'$ type, with two or three boron nitride sheets, and found that there are always gaps for $\alpha_{_B}=-\alpha_{_N} \neq 0$. Let us now note that, despite the particularities in the values $\alpha_{_B}$ and $\alpha_{_N}$, our model allows us to infer some important and interesting conclusions; namely, the inclusion of a graphene sheet between two hBN sheets does not induce conical or parabolic touches, that is, it does not eliminate the gap at $F=0$, however, it causes a reduction of an order of magnitude in the width of the opening at two layers in this graph model of hexagonal boron nitride.
 \end{observation} 
 
 \subsection{The Graphene-hBN-Graphene case}
 
 Let us now consider another type of ``sandwich'', which also arouses physical interest and is formed by a hexagonal boron nitride sheet between two layers of graphene, as shown in Figure \ref{G-BN-G}.
 
 \begin{figure}[H]
 	\centering
 	\includegraphics[width=0.6\linewidth]{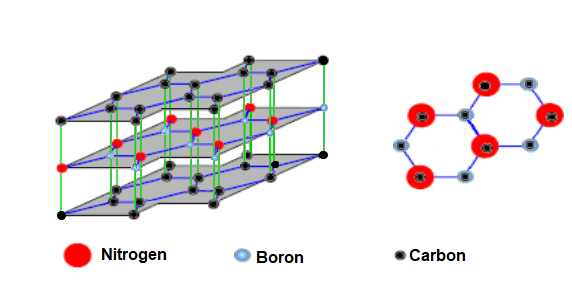}
 	\caption{Side and top views, respectively, of the Graphene-hBN-Graphene sandwich.}
 	\label{G-BN-G}
 \end{figure}
 
 In this stacking mode, the aim is to analyze the disturbance caused by the inclusion of a boron nitride monolayer, an insulating material whose dispersion relationship always has gaps, between graphene bilayers \footnote{Remember that the dispersion relationship obtained in stacking type $AA$ (or $AA'$) of three layers of graphene always has conical touches \cite{ROCHA}.} -- which we already know is an excellent conductor and whose dispersion relation has Dirac cones.
 
 The matrix $M_{{G.BN.G}}$ obtained from this stacking is given by
 $$ 
 M_{{G.BN.G}}(\eta)\!=\!\left(\begin{array}{cccccc}
 	\!\!\!-T_1\eta-\alpha_{_C} \!&\!	\bar F  \!&\! 0  & t_0^2 \!&\! 0 \!&\! 0  	\\
 	F \!&\!  -T_1\eta-\alpha_{_C}  \!& \! t_0^2 \!&\! 0 \!&\! 0 \!&\! 0  \\
 	0 \!&\! t_0^2 \!&\! -T_2\eta-\alpha_{_N} \!&\! \bar F \!&\! 0 \!&\! t_0^2  \\
 	t_0^2 \!&\!  0 \!&\! F \!&\! -T_2\eta-\alpha_{_B} \!&\! t_0^2 \!&\! 0 \\
 	0 \!&\!  0  \!&\! 0 \!&\! t_0^2 \!&\! -T_1\eta-\alpha_{_C} \!&\! \bar F \\
 	0 \!&\!  0  \!&\! t_0^2 \!&\! 0 \!&\! F \!&\! -T_1\eta-\alpha_{_C}\!\!\!
 \end{array}\right)
 $$
 with $T_1:= 3+t_0^2$ e $T_2:= 3+2t_0^2$. 
 
 Again we consider $\alpha_{_C}=0$, we obtain $\det(M_{{G.BN.G}})(\eta)=:P(\eta)=P_2(\eta)P_N(\eta) $, being 
 \[
 P_2(\eta)=(3+t_0^2)\eta^2-F^2
 \]
 and $P_N$ a polynomial of degree 4 in $\eta$. The roots of $P_2$ are given by
 \begin{equation}\label{RAab5}
 	r_{\pm}(F)=\pm\dfrac{|F|}{3+t_0^2}
 \end{equation}
 which clearly always have Dirac cones at the origin regardless of the values of $\alpha_{_B}$ and $\alpha_{_N}$. Remember that configurations of two and three graphene layers always have conical touches at the origin when stacked in the $AA$ (or $AA'$) form, but the inclusion of an hBN sheet between two graphene sheets induced gaps in four of the five Dirac cones (see below).
 
 As in the previous subsection, the roots of $P_N$ are very long and its general form becomes intractable. Again let us take $\alpha_{_B}=-\alpha_{_N}$, and we obtain 
 $$
 P_N(\eta)=A(t_0)\eta^4+B(t_0,F,\alpha_{_N})\eta^2+C(t_0,F,\alpha_{_N})\,,
 $$ 
 being
 $$
 A(t_0)=(3+t_0^2)^2(3+2t_0^2)^2\,,
 $$
 $$
 B(t_0,\alpha_{_N},F)=-\left[(5t_0^4+18t_0^2+18)F^2+8t_0^8+36t_0^6+(\alpha_{_N}+36)t_0^4+6\alpha_{_N}^2t_0^2+9\alpha_{_N}^2\right]
 $$
 and
 $$
 C(t_0,\alpha_{_N},F)=(F^2-2t_0^4)^2+\alpha_{_N}^2F^2 \,,
 $$
 so that, for $\alpha_{_B}=-\alpha_{_N}$, the roots of $P_N$ are given by 
 \begin{equation}\label{RAab6}
 	r^{\mp}_{\pm}(F)= \mp\sqrt{ \dfrac{-B(t_0,F,\alpha_{_N}) \pm \sqrt{B^2(t_0,F,\alpha_{_N})-4A(t_0,F,\alpha_{_N})C(t_0,F,\alpha_{_N})} }{2A(t_0,F,\alpha_{_N})} }\,.
 \end{equation}
 
 Independently of the values of $\alpha_{_B}$ and $\alpha_{_N}$, two roots in the graphene-hBN-graphene case always cause cones; more precisely, the graphene cone prevails in this sandwich.
 
 \begin{figure}[H]
 	\centering
 	\includegraphics[width=0.5\linewidth]{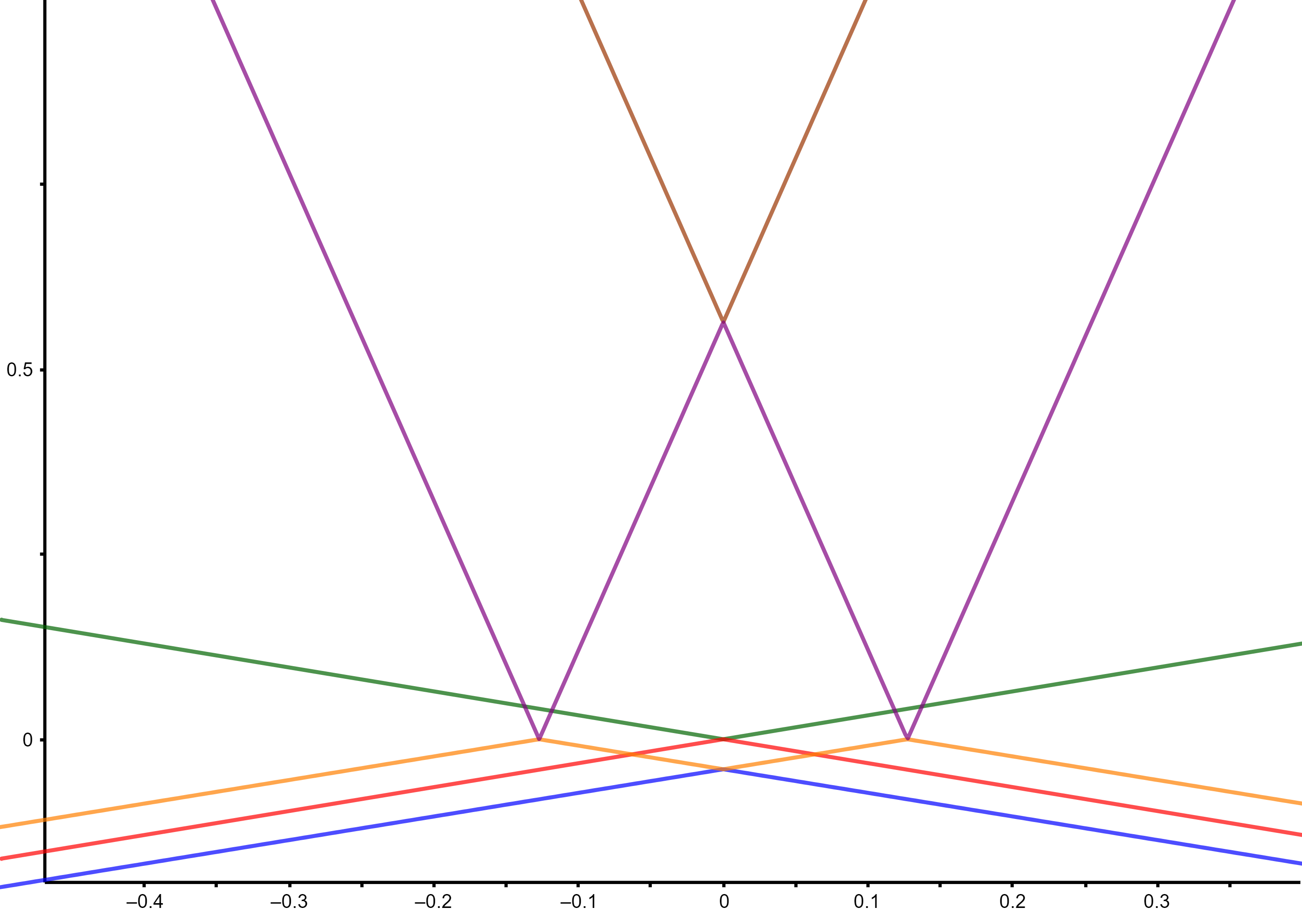}
 	\caption{Presence of five iconic touches in the interaction of equivalent atoms, that is, $\alpha_{_B}=\alpha_{_N}=0=\alpha_{_C}$\,, $t_0=0.3$. }
 	\label{G-BN-Ga0}
 \end{figure}
 
 \begin{figure}[H]
 	\centering
 	\includegraphics[width=0.5\linewidth]{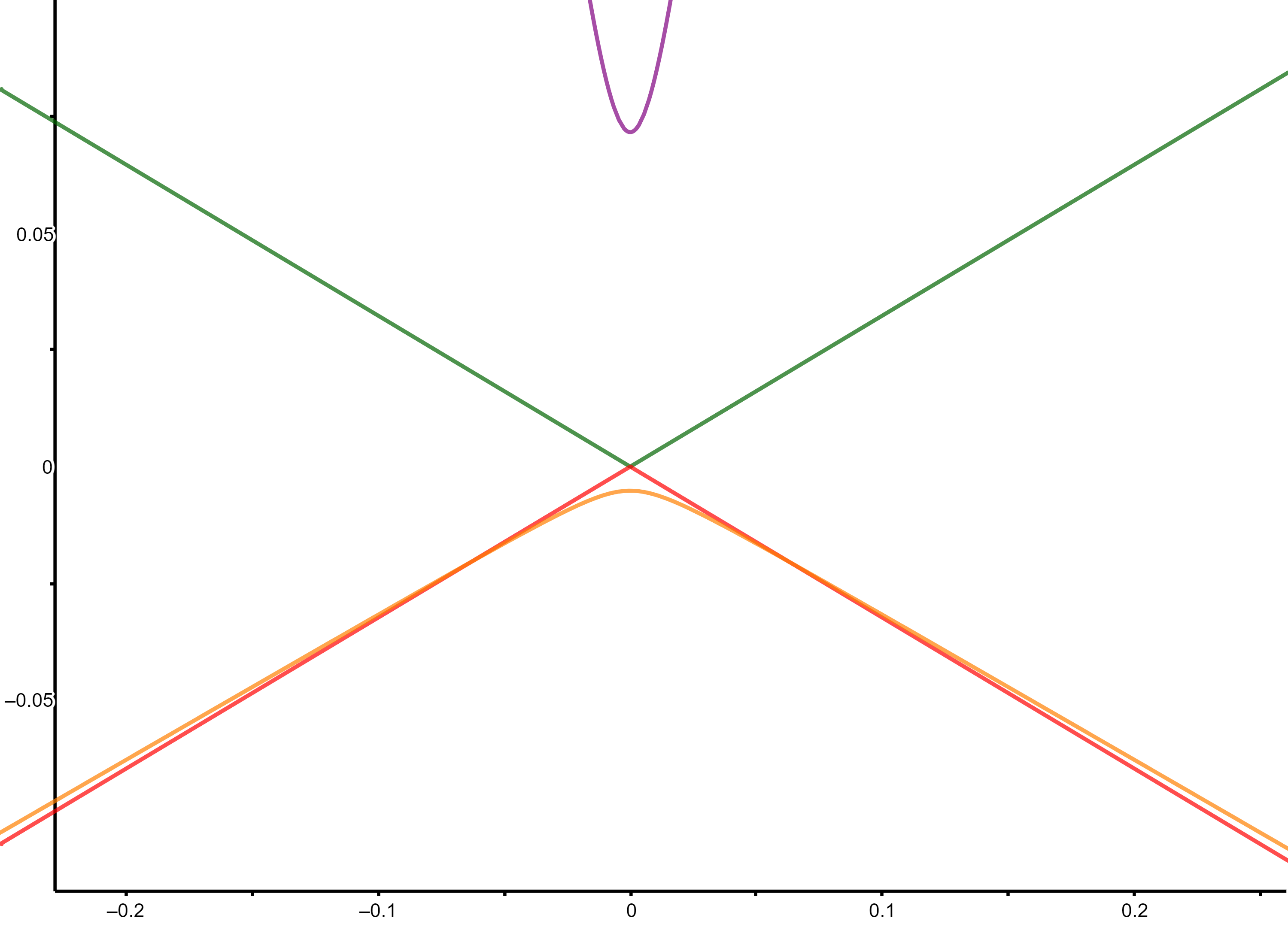}
 	\caption{$\alpha_{_B}=-\alpha_{_N}=1$ and $t_0=0.3$. Gap formation in four of the five cones due to the interaction of carbon atoms with other inequivalents of hBN, the smallest of which is $\approx 0.07684$. A graphene cone always prevails in this type of sandwich for any values of the parameters $\alpha_{_B},\alpha_{_N}$.}
 	\label{G-BN-Ga1}
 \end{figure}
 
 \begin{figure}[H]
 	\centering
 	\includegraphics[width=0.5\linewidth]{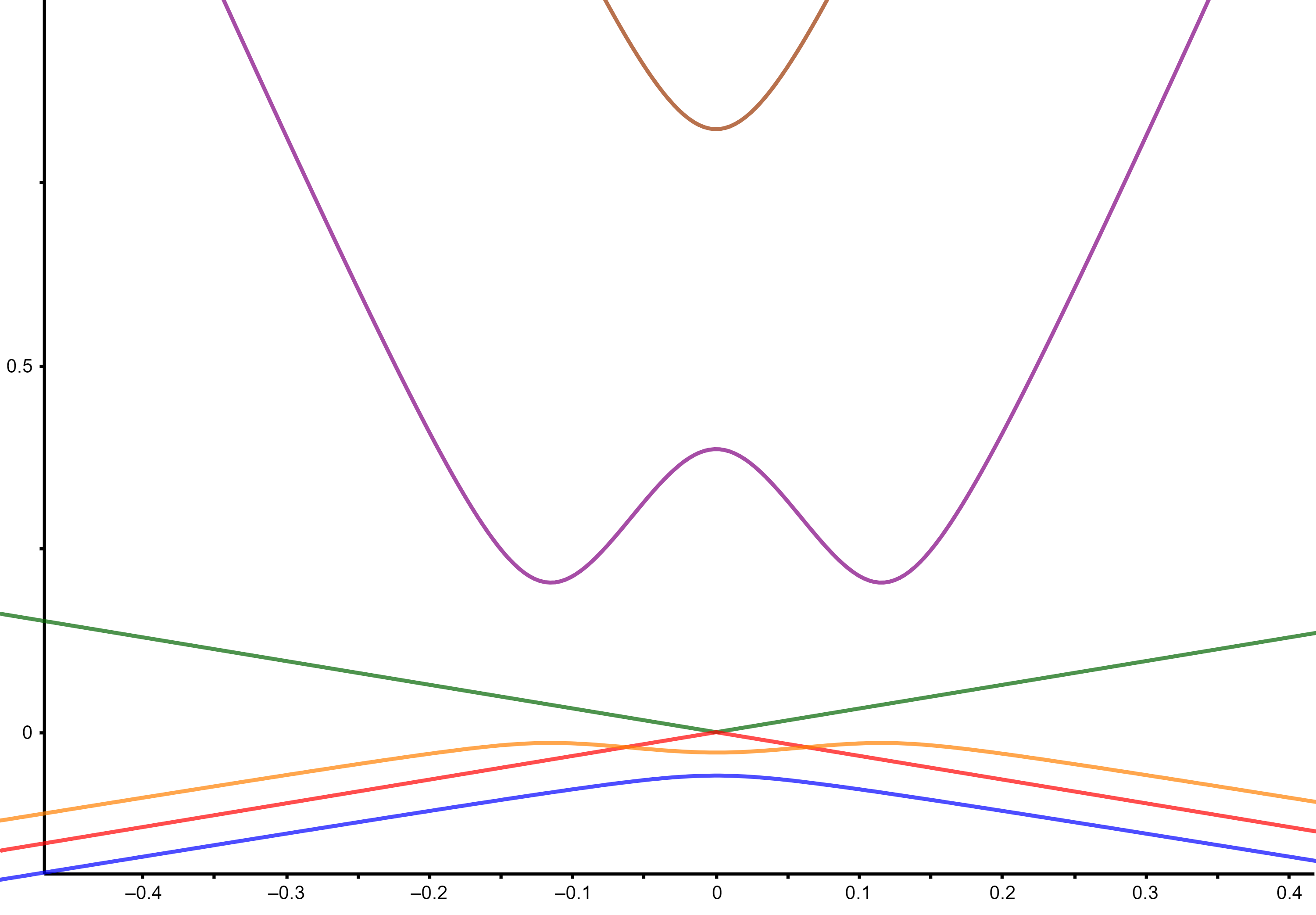}
 	\caption{$\alpha_{_B}=-\alpha_{_N}=0.1$ and $t_0=0.3$.\, Prevalence of the graphene cone: red and green curves.}
 	\label{G-BN-Ga01}
 \end{figure}
 
 \begin{figure}[H]
 	\centering
 	\includegraphics[width=0.5\linewidth]{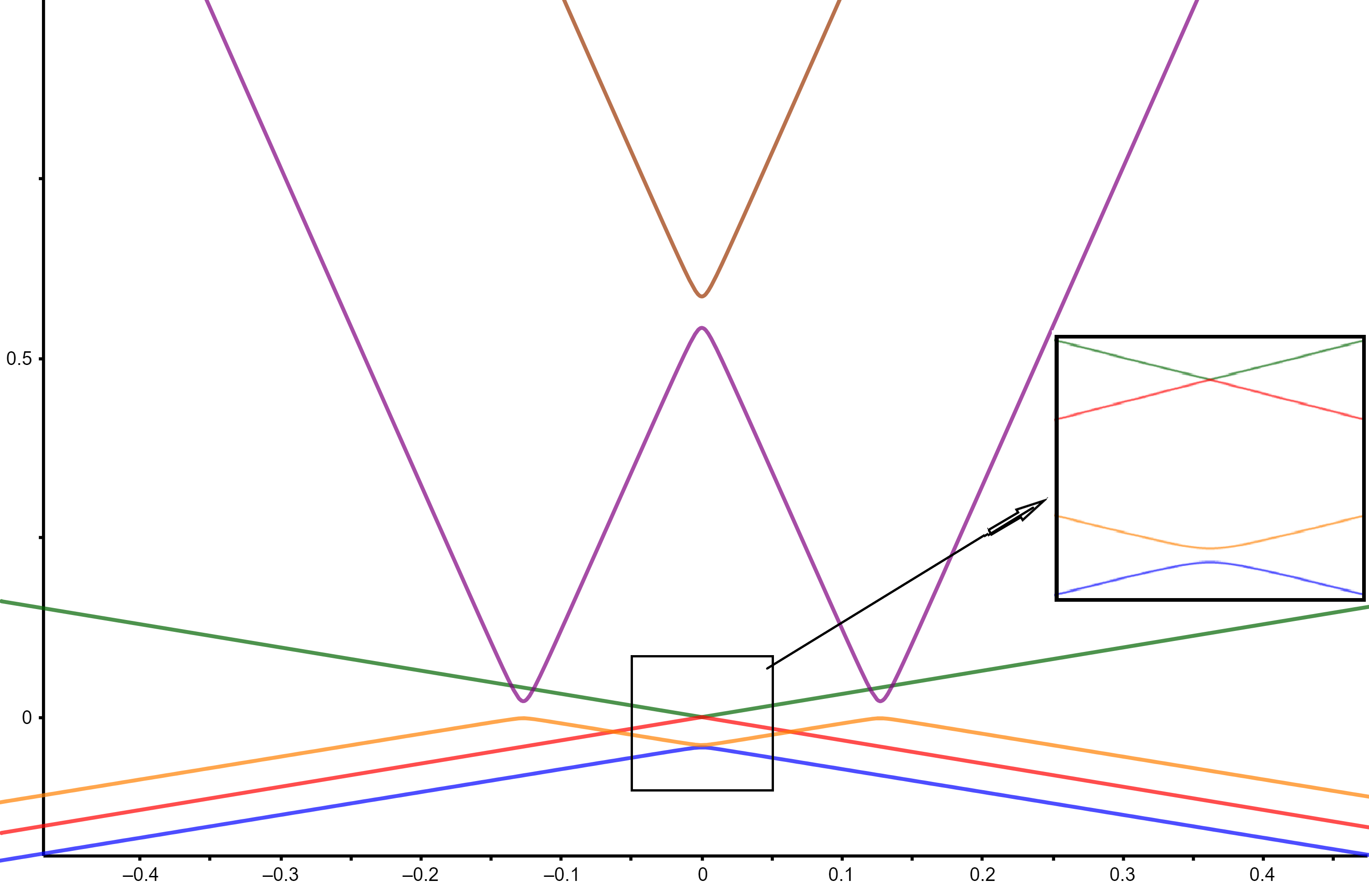}
 	\caption{$\alpha_{_B}=-\alpha_{_N}=0.01$ and $t_0=0.3$.\, The graphene cone is always maintained, while gaps are caused at other points, the smallest of which (smaller auxiliary figure on the side) is $\approx 0.00314$.}
 	\label{G-BN-Ga001}
 \end{figure}
 
 \begin{observation}
 	As in the previous subsection, we highlight an interesting fact here: the inclusion of hBN monolayer between graphene bilayers (which always have conical touches) induced gaps in some graphene Dirac points (bilayers), however a Dirac cone at the origin is always kept in this sandwich, regardless of the non-zero values of $\alpha_{_B}$ and $\alpha_{_N}$.
 \end{observation}
 
 \section{Analysis and conclusions}
 
 As mentioned in the Introduction of this work, the authors Giovannetti et al. \cite{G.K.B.B.K} carried out numerical simulations, using ``ab initio'' density functional calculations, considering bilayers in the form of van der Walls heterostructures: one made of boron nitride and the other made of graphene. The energies were found as a function of the distance between the graphene sheet and the hBN sheet, as well as gaps between the conduction and valence bands (i.e., without Dirac points); it is interpreted that the boron nitride sheets induced this gap in the graphene. The authors conclude that the appearance of these gaps offers the potential to assemble electronic systems based on these configurations.
 
 Furthermore, as demonstrated in \cite{OLIRO}, and inferred as a particular case of two or three layers of boron nitride analyzed here, two or three layers of graphene stacked in the form $AA$ do not have gaps, so that in $F=0$ \,\Big(that is, in $(\theta_1,\theta_2)=\big(\frac{2\pi}{3}, \frac{-2\pi}{3} \big)\Big)$ there are always conical touches. However, in our study, in addition to the presence of Dirac cones in the heterostructure of graphene-hBN bilayers, we also identified conical touches or gaps in trilayers arranged in a sandwich shape: graphene between boron nitride, and boron nitride between graphene. More precisely, our work allows us to conclude that the inclusion of a boron nitride layer between two graphene layers induces a gap at some Dirac points, although a conical touch is always maintained in this stacking, regardless of the parameters $\alpha_{a}$ and $\alpha_{b}$. This allows us to conclude that in any sandwich with a hexagonal material between two sheets of graphene there will always remain at least one Dirac cone. On the other hand, the inclusion of a graphene sheet between two boron nitride sheets induced a reduction in the gaps formed by hBN bilayers.
 
 
 \chapter[Hexagonal monolayers with magnetic field]{Hexagonal monolayers with magnetic field}\label{capGrafNBMagn}
 
 The electronic properties of graphene with a magnetic field have also attracted strong interest. In the mathematical articles \cite{BH,BHJ}, quantum graphs are considered as a model of graphene with a magnetic field and a complete analysis of the spectrum is provided, for any constant magnetic flux. The work \cite{BHJ} demonstrates that, considering graphene, if the magnetic flux $\phi$ is a rational multiple of $2\pi$, then the continuous spectrum is absolutely continuous and, complemented by~\cite{BH}, showing that there are Dirac cones in these cases; and if the flux $\phi$ is an irrational multiple of~$2\pi$, the spectrum is a Cantor set with zero Lebesgue measure \cite{BHJ}. At the vertices, we always start from the Kirchhoff boundary conditions (i.e., $\alpha=0$ at all vertices).
 
 We will use quantum graphs in a proposal to model hexagonal boron nitride under the effect of a magnetic field, through the Schr\"{o}dinger operator, and demonstrate that if the magnetic flux~$\phi $ is constant in the network and of the form $\phi = 2\pi\dfrac{p}{q}$ (rational multiple of $2\pi$), then there are integer values $p$ and $q$ such that, for certain boundary conditions at the vertices (starting from the Robin condition with different parameter values, our proposal to model boron nitride), the conic touches in the operator dispersion relation cease to exist so that we guarantee the existence of gaps.
 
 First, however, we present some necessary definitions and results; we will start in the following section with a very succinct review of known results, and more details can be found in \cite{BH,BHJ,PANKRAS, PANK2}.
 
 \section{Magnetic potential}
 
 Consider a graph $G$. For an edge $e \in \mathcal{E}$ we denote by $i(e) \in \mathcal{V}$ its initial vertex, and $t(e) \in \mathcal{V}$ its terminal vertex. Let us remember that each edge $e$ of $G$ can be identified with a copy of the segment $[0,1]$, such that 0 is identified with the vertex $i(e)$, 1 with the vertex $t(e )$, as constructed in Chapter 1.
 
 In this context, we will approach models of boron nitride layers (and consequently graphene) through quantum graphs considering a magnetic field in the modeling through the Schr\"{o}dinger operator; on each edge $e$ , let us take $a_e \in C^1[0,1]$ and associate, to each $e$, the operator
 $$
 L_e:=(i\partial+a_e)^2+U\,;
 $$ 
 we define the maximal operator $(g_e) \mapsto (L_eg_e) $ in the functions $ g=(g_e) \in \bigoplus_{e}{\mathcal{H}}^2[0,1]$, so that, introducing the boundary conditions at the vertices, we obtain a self-adjoint operator:
 \begin{equation}
 	\left\{
 	\begin{array}{ll}
 		g_a(v) = g_b(v) =:g(v)\,, \forall \,\, a, b \in \mathcal{E}_v\,;\\
 		\dis\sum_{e; i(e)=v}\!\!\big(g'_e(0) - ia_e(0)g_e(0)\big)-\!\! \sum_{e; t(e)=v}\!\!\big(g'_e(1) - ia_e(1)g_e(1)\big)  = \alpha(v)g(v)\,,\,\, v \in \mathcal{V}\,\, 
 	\end{array}, 
 	\right.
 \end{equation} 
 where $\alpha(v) \in \R$ is a parameter and $\mathcal{E}_v$ is the set of edges that contain $v$.
 
 More generally, now consider the 1-differential form $A(\textbf{x})=A_1(x_1,x_2)dx_1+A_2(x_1,x_2)dx_2$ and the scalar potential $A_e \in C^{\infty}(e)$ on the edge $e \in \mathcal{E}(G)$ obtained by calculating the form $A$ in the graph along the vector field generated by edge $[e] \in \mathcal{E}(G)$:
 $$
 A_e(\textbf{x})=A(\textbf{x})\cdot ([e]_1\partial_1+[e]_2\partial_2)\,.
 $$
 The integrated vector potentials are defined by
 $$
 \beta_e:=\int_e A_e(\textbf{x})d\textbf{x}\,,\,\, e \in \mathcal{E}(G)\,.
 $$
 For the hexagonal lattice, define the magnetic flux by $\phi= \int_{\hexagon}dA$ and assume that it is always constant in the hexagons $\hexagon$ of the lattice, that is,
 $$
 \beta_{f,\gamma_1\gamma_2} -\beta_{h,\gamma_1\gamma_2+1}+
 \beta_{g,\gamma_1\gamma_2+1} - \beta_{f,\gamma_1-1\gamma_2+1}+
 \beta_{h,\gamma_1-1,\gamma_2+1}-
 \beta_{g,\gamma_1\gamma_2} = \phi\,,\quad \forall\,\, \gamma \in \Z^2\,\,.
 $$
 We introduce the discrete Hilbert space
 $$
 l^2(G)=\Big\{f:\mathcal{V} \to \C; \|f \|^2=\sum_{v \in \mathcal{V}} \operatorname{deg}(v) |f(v)|^2 < \infty\Big\}\,,
 $$
 with the inner product given by
 $$ {\left\langle f, g \right\rangle}_{G} = \sum_{v \in \mathcal{V}} \operatorname{deg}(v)\bar{f}(v)g(v)\,,
 $$
 where $ \operatorname{deg}(v) =  \operatorname{outdeg}(v) +  \operatorname{indeg}(v)$ the degree of the vertex $v$, that is, the sum of the number of edges leaving $v$, $ \operatorname{outdeg}(v)$, and the number of edges arriving to $v$, $ \operatorname{indeg}(v)$.
 
 The discrete Laplacian operator $\Delta$ is defined in $l^2(G)$
 \begin{equation}\label{Laplaciano}
 	\Delta f(v)=\frac{1}{ \operatorname{deg}(v)}\!\left(\!\sum_{e\,;\,i(e)=v}\!\!\!f(t(e))+\!\!\sum_{e\,;\,t(e)=v}\!\!\!f(i(e))\right)\,.
 \end{equation}
 
 Let's now start building the continuous version of the Laplacian in $G$. The magnetic Schr\"odinger operator in such a structure can be defined as follows.
 
 Consider the Hilbert space $\mathcal{H} = \bigoplus_{e \in \mathcal{E}}{\mathcal{H}}_e$, with ${\mathcal{H}}_e = L^2[0,1]$, consisting of functions $f=(f_e),\,f_e \in {\mathcal{H}}_e$, and define in $\mathcal{H}$ the operator $H , H(f_e)=(-f''_e)$ satisfying the boundary conditions at the vertices
 \begin{equation}
 	\left\{
 	\begin{array}{ll}\label{fronteira0}
 		f_a(v) = f_b(v) \,, \forall \,\, a, b \in \mathcal{E}_v\, \\
 		\dis\sum_{e; \, i(e)=v}\!\!f'_e(0) -\!\! \sum_{e;\, t(e)=v}\!\! f'_e(1)  = 0\quad  
 	\end{array}.
 	\right.
 \end{equation}
 The operator $H$ is self-adjoint and its spectrum is closely related to the spectrum of $\Delta$: denoting by $\sigma_D=\{(\pi\,n)^2: n \in \N\}$ and, as previously, by $\sigma(\Delta)$ the spectrum of the operator $\Delta$, we have the relation
 \begin{equation}
 	\sigma(H) \backslash \sigma_D=\{ z \notin \sigma_D: \cos(\sqrt{z}) \in \sigma(\Delta)\}.
 \end{equation}
 
 Let $\Omega$ be a Borel set in $\R$ and $A$ be a self-adjoint operator; denote by $A_{\Omega}$ the part of $A$ in $\Omega$, that is, $A_{\Omega}=A1_{\Omega}(A)$ considered as an operator on the image $1_{\Omega}(A)$. Here $1_{\Omega}(A)$ \'is the spectral projection of $A$ onto $\Omega$. In \cite{PANKRAS}, Pankrashkin demonstrates the following result:
 
 \begin{proposition}\label{Prop1}(Prop.~1, page 641, \cite{PANKRAS}) For any interval $J \subset \R \backslash \sigma_D$, the operator $H_J$ \'is unitarily equivalent to the operator $\mu^{-1}(\Delta_{\mu(J)})$, with $\mu(z)=\cos(\sqrt{z})$.
 \end{proposition}
 
 Let us now take a real potential $U \in L^2[0,1]$ and fix $\alpha:\mathcal{V}\to \R$. 
 We will say that the \textit{symmetry condition} is satisfied if one of the following properties is fulfilled:\\
 $(a) \, \operatorname{indeg}(v) = \operatorname{outdeg}(v),\, \forall \,\, v \in \mathcal{V} $ \,;\quad \, $(b)$ \, $U$ is even, that is, $U(x)=U(1-x)$.\\
 
 Denote by $H$ the self-adjoint operator in $\mathcal{H} = \dis \bigoplus_{e \in \mathcal{E}}L^2[0,1]$ as
 \begin{equation}\label{Laplaciano-continuo}
 	H(f_e) =-f''_e+Uf_e\,,
 \end{equation}
 in the functions $f=\dis (f_e) \in \bigoplus_{e \in \mathcal{E}} \mathcal{H}^2[0,1]$ ($\mathcal{H}^2[0,1]$ and the usual Sobolev space in the interval) fulfilling the boundary conditions
 \begin{equation}
 	\left\{
 	\begin{array}{ll}\label{fronteira01}
 		f_a(v) = f_b(v)\,, \forall \,\, a, b \in \mathcal{E}_v\\
 		\dis\sum_{e; \, i(e)=v}\!\!f'_e(0) -\!\! \sum_{e;\, t(e)=v}\!\! f'_e(1)  = \alpha(v)f(v)\,, v \in \mathcal{V}\,. 
 	\end{array}
 	\right.
 \end{equation}
 Therefore, remembering that $\sigma_D$ denotes the spectrum of the operator $f \mapsto -f''+Uf$ in $[0,1]$ with Dirichlet boundary conditions, the Proposition \ref{Prop1} is a simple corollary of the following result.
 
 \begin{theorem}({Teor.~17, p. 652, \cite{PANKRAS}})\label{Teo17} Suppose $H$ defined as above and the potential $U$ even, and that $\alpha(v)=\alpha  \operatorname{deg}(v)$, for some $\alpha \in \R$.
 	
 	Then, for any interval $J \subset \R \backslash \sigma_D$, the operator $H_J$ is unitarily equivalent to $\mu^{-1}_{\alpha}(\Delta_{\mu_{\alpha}( J)})$, with $\Delta$ being the operator in $l^2(G)$ given by \eqref{Laplaciano} and $\mu_{\alpha}(z)=c(1;z)+\alpha \, s(1;z)$, with $c(1;z),s(1;z)$ fundamental solutions of the differential equation $-y''(t)+U(t)y(t)=zy (t)$ that meet the boundary conditions $s(0;z)=c'(0;z)=0$ and $s'(0;z)=c(0;z)=1$.
 \end{theorem}
 
 Note that, in particular, for $U=0$ and $\alpha=0$, we have $\mu_0(z)=\cos(\sqrt{z})$, which gives the Proposition \ref{Prop1}.
 
 These results can be extended to the case in which a magnetic field is considered, as per the construction in \cite{PANKRAS}, briefly presented below. Consider now the discrete magnetic Laplacian defined by
 \begin{equation}\label{LaplacianoDiscreto}
 	\Delta_{\beta}f(v)=\frac{1}{ \operatorname{deg}(v)}\!\left( \sum_{e\,;\,i(e)=v}\!\!\!e^{-i\beta_e}f(t(e))+\!\!\!\sum_{e\,;\,t(e)=v}\!\!\!e^{i\beta_e}f(i(e))\right)\,,
 \end{equation}
 where $\beta:\mathcal{E} \to \R$ is an arbitrary function, which is self-adjoint in $l^2(G)$ and limited, since $\left\| \Delta_{\beta} \right\| \leq 1 .$ Indeed,
 $$
 \begin{array}{lll}
 	\left\| \Delta_{\beta}f \right\|^2_{G} &=& 
 	\dis \sum_{v \in \mathcal{V}}\dfrac{1}{ \operatorname{deg}(v)}\left |\sum_{e:\, i(e)=v}\!\!\!e^{-i\beta_e}f(t(e)) +\!\!\sum_{e\,;\,t(e)=v}\!\!\!e^{i\beta_e}f(i(e)) \right |^2 \\
 	&\leq& \dis \sum_{v \in \mathcal{V}}\left(\sum_{e:\, i(e)=v}\!\!\!e^{-i\beta_e}|f(t(e))|^2 +\!\!\sum_{e\,;\,t(e)=v}\!\!\!e^{i\beta_e}|f(i(e))|^2 \right) \\
 	&=& \dis \sum_{w \in \mathcal{V}} \operatorname{deg}(w)|f(w)|^2 \\
 	&=& \left\| f \right \|^2_{G}\,.
 \end{array}\,.
 $$
 
 Taking a magnetic potential $a_e \in C^1[0,1]$ in each $e \in \mathcal{E}$, we associate to each $e$, the operator $\tilde{H}_e=(i\partial+a_e)^2+U$ and we define the Schr\"odinger operator $(g_e) \mapsto (\tilde{H}_eg_e) $ in functions $ g=(g_e) \in \bigoplus_{e\,\in \,\mathcal{E}}{\mathcal{H}}^2[0,1]$ which models this type of system with a magnetic field Using integration by parts, we see that $\tilde{H}_e$ does not is symmetric, and it is necessary to introduce boundary conditions at the vertices to obtain a self-adjoint operator. The standard boundary conditions for these magnetic operators, which make them self-adjoint, are:
 \begin{equation}
 	\left\{
 	\begin{array}{ll}\label{fronteira1}
 		g_a(0) = g_b(1) =:g(v)\,, \forall \,\, a, b \in \mathcal{E}_v\,;\\
 		\dis\sum_{e; i(e)=v}\!\!\big(g'_e(0) - ia_e(0)g_e(0)\big)-\!\! \sum_{e; t(e)=v}\!\!\big(g'_e(1) - ia_e(1)g_e(1)\big)  = \alpha(v)g(v)\,,\,\, v \in \mathcal{V}\,\, 
 	\end{array}, 
 	\right.
 \end{equation}
 with $\alpha(v) \in \R$ being a parameter.
 
 
 Applying the gauge transformation $\dis g_e(t)= e^{i\int^{t}_{0}a_e(s)ds}f_e(t)$ and introducing the parameter $\dis \beta(e) :=\int^{1}_{0}a_e(s)ds$, we remove the magnetic potential $a_e$ from $\tilde{H}_e$ so that $\tilde{H}_e$ is unitarily equivalent to operator $-f''_e+Uf_e$:
 \begin{equation}\label{eqGaugeTransge}
 	g_e^{-1}\left((i\partial+a_e)^2+U\right)g_e =-f''_e + Uf_e\,,
 \end{equation}
 and now the magnetic potential appears in the new boundary conditions:
 \begin{equation}
 	\left\{
 	\begin{array}{ll}\label{fronteira-magnetica}
 		f_a(0)=e^{i\beta_{e}}f_b(1) =:f(v)\,, \forall \,\, a,b \in \mathcal{E}_v\,;\\
 		f'(v)=\dis\sum_{e; i(e)=v}f'_e(0) - \sum_{e; t(e)=v}e^{\beta_e}f'_e(1)  = \alpha(v)f(v)\,,\,\, v \in \mathcal{V}\,\, 
 	\end{array}\,.
 	\right.
 \end{equation}
 
 Let $H:(f_e) \mapsto (-f''_e+Uf_e)$ be the self-adjoint operator acting on $ (f_e) \in \bigoplus_{e\in \mathcal{E}}{\mathcal{H} }^2[0,1] $ and satisfying the conditions \eqref{fronteira-magnetica}, for all $v \in \mathcal{V}$. Making some adaptations to the argument set out above, it can be shown that the Theorem \ref{Teo17} is valid in the same way if we replace the operator $\Delta$ with its magnetic version $\Delta_{\beta}$ given by \eqref{LaplacianoDiscreto}. In particular, this construction can be applied to a two-dimensional lattice with a uniform magnetic field. 
 
 Furthermore, in \cite{PANKRAS} it was observed that the $H$ operator can be studied at an abstract level using the boundary triple language and self-adjoint extensions. More precisely, using boundary triples, Pankrashkin related spectral properties of the continuous Laplacian operator $H$ in $\mathcal{H}=\bigoplus \mathcal{H}^2[0,1]$ with the discrete operator $\Delta_{ \beta}$ in $l^2(G)$. Or, the author demonstrates that
 \begin{equation}
 	\sigma(H)\backslash \sigma_D=\Big\{z \in \R \backslash \sigma_D: \mu_\alpha(z) \in \sigma(\Delta_\beta) \Big\}=\mu^{-1}_{\alpha}\big(\sigma(\Delta_{\beta})\big) \backslash \sigma_D,
 \end{equation}
 where $\mu_{\alpha}$ o discriminante de Hill  \begin{equation}\label{DiscHill}
 	\mu_{\alpha}(z) = c(1,z)+\frac{\alpha}{2}s(1,z)\,.
 \end{equation}
 In other words, for $z \in \R \backslash \sigma_D$, Pankrashkin then obtains the relationship between the spectra of $H$ and $\Delta_{\beta}$:
 \begin{equation}\label{Rel-Espec}
 	z \in \sigma(H) \Longleftrightarrow \mu_\alpha(z) \in \sigma(\Delta_{\beta}),
 \end{equation} 
 
 \begin{observation}\label{Obs-Teo-Rel-Espec}
 	As shown in \cite{PANKRAS}, the analysis of the $H$ operator above can be done using frontier triples; more precisely through the functions $c(\cdot,z),s(\cdot,z$) associated with the operator $H$ and the constant $\alpha$ of the boundary condition. The relationship determined in \eqref{Rel-Espec} is obtained under the hypothesis that the symmetry condition remains and the coupling constants comply with $\alpha(v)=\frac{\alpha}{2} \operatorname{deg}(v)$, with $\alpha \in \R$. However, if one of these conditions fails (for example, if the orientation of a single edge changes), one cannot relate the spectra of the continuous and discrete problems as seen above. 
 	
 	
 	In \cite{PANKRAS}, Pankrashkin shows that the spectrum of $H$ has a band structure if and only if the spectrum of $\Delta_{\beta}$ has a band structure, that is, the spectrum and the union of a locally finite family of segments, while in \cite{PANK2} he shows that the continuous Laplacian $H$ and the discrete Schr\"odinger operator $\Delta_{\beta}$ are unitarily equivalent, that is, they have the same spectra.
 	
 \end{observation}
 
 Based on these results, we will then apply them to the study of monolayers of hexagonal structures, such as graphene and boron nitride, with rational magnetic flux $\phi=2\pi\frac{p}{q}$ in a hexagonal lattice \ hexagon. We reinforce that the quantum graph model for graphene with parameter $\alpha(v)=0$ for every vertex~$v$, in certain magnetic fields, has already been studied in detail in \cite{BH,BHJ}, and our contribution here it is the case of graphene with $\alpha(v)$ constant, but not null, and also of hBN, always with these fields.
 
 \section{Graphene and boron nitride with magnetic field} 
 
 Let us consider the magnetic flux $\phi$ as a rational multiple of $2\pi$, that is, $\phi=2\pi\frac{p}{q}$. Therefore, the operator $\Delta_{\beta}$ is periodic \cite{BH}. Furthermore, given $\theta = (\theta_1,\theta_2) \in [-\pi,\pi)\times [-\pi,\pi) $, we consider $\Delta_{\beta}$ acting on $l^2(\hexagon)$ subject to pseudo-periodic condition
 \begin{equation}
 	u(\gamma+qE_k,v_{_{j-1}})=e^{i\theta_k}u(\gamma,v_{_{j-1}})\,,\,\, j,k=1,2\,,
 \end{equation} 
 where $E_1,E_2$ are base vectors of $\hexagon$ and $v_0,v_1$ are vertices of the fundamental domain $W_{\!\hexagon}$.
 
 We also know, from \cite{BH}, that given the discrete operator $\Delta_{\beta}$ with flow $\phi$ defined in \eqref{LaplacianoDiscreto} and acting on $l^2(\Z^2,\ C^2),$ the Floquet matrix of $\Delta_{\beta}$ is given by
 \begin{equation}\label{Matriz-p,q}
 	M_F(\theta)=\frac{1}{3}\left(\begin{array}{cc}
 		0 & I_q+e^{i\theta_1}J_{p,q}+e^{i\theta_2}K_{q} \\
 		I_q+e^{-i\theta_1}J^*_{p,q}+e^{-i\theta_2}K^*_{q}  &  0 
 	\end{array}\right)\,,
 \end{equation}
 where $J_{p,q}$ and $K_q$ are $q \times q$ matrices given by
 $$
 J_{p,q} = \operatorname{diag}\Big(\{e^{i(j-1)\phi}\}^{q}_{j=1}\Big)
 $$ and 
 \begin{equation}
 	(K_q)_{jk}=	\left\{
 	\begin{array}{ll}
 		1\,,\, k = (j+1) \operatorname{mod}(q) \\
 		0\,,\, \mbox{otherwise}
 	\end{array}
 	\right.\,.
 \end{equation}
 In particular, for $p=1,\,q=2$, we have
 \begin{equation}\label{Matriz-p=1,q=2}
 	M_F\big(\theta_1,\theta_2\big)=\frac{1}{3}\left(\begin{array}{cccc}
 		0 &  0  &  1+e^{i\theta_1} & e^{i\theta_2}\\
 		0 &  0  & e^{i\theta_2} & e^{i(\theta_1+\phi)}  \\
 		1+e^{-i\theta_1} &   e^{-i\theta_2}  &  0 & 0  \\
 		e^{-i\theta_2} &  e^{i(\theta_1+\phi)}  & 0 &  0 
 	\end{array}\right)\,.
 \end{equation}
 
 If the quasi-moment $\theta = (\theta_1,\theta_2) \in [-\pi,\pi)\times [-\pi,\pi) $ is restricted to $[0,\frac{\pi} {q})\times [-\frac{\pi}{q},\frac{\pi}{q})$, the authors demonstrate that the scattering surface of $\Delta_{\beta}$ has Dirac at energy level $0$, for any rational $p$ and $q$. To do so, they use the elegant Chambers formula (see \cite{HKL}) to help calculate the determinant of the matrix $M_F(\theta)$ associated with the operator, since this matrix is of order $q \times q$.
 
 Note that, in the boundary conditions in \eqref{fronteira-magnetica}, the same value $\alpha(v)$ was admitted at all vertices $v \in \mathcal{V}$ of the network; furthermore, in the relation \eqref{Rel-Espec} such an assumption was also made, that is, $\alpha(v)=\frac{\alpha}{2} \operatorname{deg}(v)$, with $\alpha \in \R$ constant, since, in a hexagonal network \hexagon, $\ \operatorname{deg}(v)=3$ for any vertex of the network. However, we will show that if we admit a certain variation of the flow at the vertices, that is, $\alpha(v)$ is not constant, the operator Dirac cones can disappear, as we present below.
 
 \begin{itemize} 
 	\item If $\alpha(v) = 0 $ for every vertex~$v$, we know that 
 	$$\sigma(H) \backslash \sigma_D=\{z \notin \sigma_D: \mu_0(z)=\cos(\sqrt{z} )\in \sigma(\Delta_\beta)\}\,,
 	$$
 	with $\sigma_D=\{(\pi n)^2: n \in \N\}$. 
 	With this, we see that given a cone at a point in the spectrum of $H$, this cone is preserved by applying $\mu_0(\cdot)=\cos(\cdot)$ in the spectrum of $H_\beta$ and, as the Proposition \ref{Prop1}, we conclude that, for any integer values $p, q$ such that the magnetic flux has the form $\phi=2\pi\frac{p}{q}$, the operator $\Delta_ {\beta}$ has a Dirac cone, since its spectra are unitarily equivalent.
 	
 	\item Similarly, for $\alpha(v)\neq 0$, and constant for all vertices, by Theorem \ref{Teo17}, the presence of cones still remains for arbitrary values of $p,q$ (still with $\phi=2\pi\frac{p}{q}$). In fact, initially let us note that, for $z \in \R \backslash \sigma_D$, the functions $c(\cdot,z), s(\cdot,z)$ in \eqref{DiscHill} are of class $C ^1$; therefore, the function $\mu_{\alpha}$ is also $C^1$. Thus, taking a Dirac cone of $H$, simply show that its image by the function
 	\[
 	\mu_\alpha(z)= c(\cdot,z)+\frac{\alpha}{2}\cdot s(\cdot,z)
 	\] it is still a Dirac cone of $\Delta_\beta$. In fact, more generally, suppose that $f(\theta)$ has a Dirac cone at $\theta_0$, that is, there is $\gamma>0$, namely, $f'(\theta_0)\neq 0 $, so that 
 	\begin{equation}\label{ConeMagnet0}
 		f(\theta)-f(\theta_0)=\pm\gamma\cdot|\theta-\theta_0|+\mathcal{O}\big(|\theta-\theta_0|^2\big)\,.
 	\end{equation}
 	Now, if we perform the composition with another function $g$ of class $C^1$, the cone remains if $g'(f(\theta_0))\neq 0$. In fact, according to the chain rule,
 	\begin{equation*}
 		g(y)-g(y_0)=\pm g'(y_0)|y-y_0|+\mathcal{O}\big(|y-y_0|^2\big)
 	\end{equation*}
 	and choosing $y = f(\theta)$, $y_0 = f(\theta_0)$, we get
 	\begin{equation*}
 		g(f(\theta))-g(f(\theta_0))=\pm g'\big(f(\theta_0)\big)\cdot|f(\theta)-f(\theta_0)|+\mathcal{O}\big(|f(\theta)-f(\theta_0)|^2\big)
 	\end{equation*}
 	and using \eqref{ConeMagnet0}, 
 	\begin{equation*}
 		g(f(\theta))-g(f(\theta_0)) \!=\! \pm g'(f(\theta_0))\cdot \gamma\cdot|\theta-\theta_0|+\mathcal{O}(|\theta-\theta_0|^2) \!=\! \gamma_g\cdot|\theta-\theta_0|+\mathcal{O}(|\theta-\theta_0|^2)
 	\end{equation*}
 	wint $\gamma_g := g'(f(\theta_0))\cdot \gamma \neq 0$, concluding that the cone remains after composition with $g$, as long as $g'(f(\theta_0))\neq 0$; in particular, for $g=\mu_\alpha$, we have $\gamma_{\mu_\alpha} \neq 0$ and the result follows. 
 	
 	\item However, if $\alpha(v)$ is not constant, corresponding to our modeling of boron nitride, according to Observation \ref{Obs-Teo-Rel-Espec}, Pankrashkin does not guarantee the relation \eqref{Rel-Espec}, so that it is not possible to relate the spectra of $\Delta_\beta$ in $l^2(\hexagon)$ and $H$ in $\bigoplus_{e\in \mathcal{E}}{\mathcal{H }}^2[0,1]$, as this relationship was obtained by supposing $\alpha(v)=\frac{\alpha}{2} \operatorname{deg}(v)$, with $\alpha \in \R$ taking the same value on all vertices. We will then show that for certain values of $\alpha$ the conical touches in the dispersion relation of the operator $\Delta_{\beta}$ cease to exist so that we guarantee the existence of gaps. To do so, we will use the Theorem \ref{Teo17} and the relation \eqref{Rel-Espec} above, but we will base ourselves on the technique presented in Chapter \ref{capPrelimina} (mainly due to Kuchment-Post~\cite{KP1} ), which we will adapt to the case with a magnetic field in a specific configuration; the need to work with a particular (and ``small'') period comes from the lack of an analogue to the Chambers Formula. Such adaptation and the example with rational flow in relation to $2\pi$ without Dirac cones constitute our main contributions to this model (and this chapter).
 \end{itemize} 
 
 Let us then, from now on, fix $p=1$ and consider the pseudo-periodic condition
 \begin{equation}\label{pseudo-period}
 	u(\gamma+qE_k,v_{j-1})=e^{i\theta_k}u(\gamma,v_{j-1})\,,\,\, j,k=1,2\,,
 \end{equation} 
 for, initially, the period $q=1$ and, subsequently, $q=2$. Given an edge $e$ of the network $\hexagon$, let us take $\alpha(v)$ so that
 \begin{equation}
 	\alpha(v)=	\left\{
 	\begin{array}{ll}
 		\delta_{_N}\,,\,\mbox{if}\, v=i(e)=0 \\
 		\delta_{_B}\,,\, \mbox{if}\, v=t(e)=1
 	\end{array}
 	\right.\,.
 \end{equation}
 
 In the case $q=1$, the period of the hexagonal lattice \hexagon\ coincides with that of the operator $H$ and, therefore, the modeling coincides with that already covered in Chapter \ref{capPrelimina}, for a sheet of nitride boron, so that the fundamental domain is given by $W_{\!\hexagon}=\{f,g,h, v_0,v_1 \}$, as shown in Figure \ref{Rede-Mag-q1}.
 \begin{figure}[H]
 	\centering
 	\includegraphics[width=0.5\linewidth]{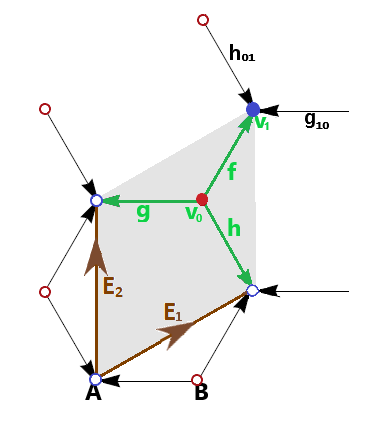}
 	\caption{The fundamental domain $W_{\hexagon}$ and the generating vectors $E_1$ and $E_2$ of the \hexagon \,network. For $q=1$, the period of the \hexagon \,network coincides with that of the $H$ operator.}
 	\label{Rede-Mag-q1}
 \end{figure}
 The continuity and flow equations at the vertices are given below.
 \begin{equation}
 	\mbox{Continuity at the vertices:}	\left\{
 	\begin{array}{cll}
 		u_{f}(0) = u_{g}(0) = u_{h}(0)&=:& A_1\\
 		e^{i\beta_f}u_{f}(1)= e^{i\beta_{g_{10}}}u_{g_{10}}(1) = e^{i\beta_{h_{01}}}u_{h_{01}}(1) &=:& B_1	
 	\end{array}
 	\right.\,,
 \end{equation}
 so that, for the edges that are outside the fundamental domain, using the cyclic Floquet conditions \footnote{As the period of the operator and the lattice coincide, the periodic Floquet relations and the pseudo-periodic ones in \eqref{pseudo-period} coincide.}, we obtain the relations $u_{g_{10}}(1)=e^{i\theta_1}u_{g}(1)$ and $u_{h_{01}}(1)=e^{i\theta_2}u_{h}(1)$;
 \begin{equation}
 	\mbox{Flux Condition:}	\left\{
 	\begin{array}{cll}\label{EqMagn0}
 		u'_{f}(0)+ u'_{g}(0)+u'_{h}(0)&=& \delta_{_N}A_1\\
 		-e^{i\beta_f}u'_{f}(1)- e^{i\beta_{g_{10}}}u'_{g_{10}}(1)-e^{i\beta_{h_{01}}}u'_{h_{01}}(1) &=& \delta_{_B}B_1	
 	\end{array}
 	\right.\,.
 \end{equation}
 
 Similar to what we saw in the initial chapters, for $\lambda \notin \sigma(H^D)$, there are functions $\varphi_0, \varphi_1$ such that 
 \begin{equation}
 	\left\{
 	\begin{array}{lll}\label{EqMagn1}
 		u_{f}&=&A_1\varphi_{0}+B_1\varphi_{1}e^{-i\beta_{f}}\\
 		u_{g}&=&A_1\varphi_{0}+B_1\varphi_{1}e^{-i(\theta_1+\beta_{g_{_{10}}})}\\
 		u_{h}&=&A_1\varphi_{0}+B_1\varphi_{1}e^{-i(\theta_2+\beta_{h_{_{01}}})}	\end{array}.\,\,
 	\right.
 \end{equation}
 
 Combining the equations \eqref{EqMagn0} and \eqref{EqMagn1}, with $\alpha_k=\dfrac{\delta_k}{\varphi'_{1}(0)}$, $k \in {B,N}$ , we obtain the matrix representation
 $$
 M_{_{\hexagon}}(\theta)=\left(\begin{array}{cc}
 	-3\mu-\alpha_{_N} & e^{-i\beta_{f}}+e^{-i(\theta_2+\beta_{h_{01}})} + e^{-i(\theta_1+\beta_{g_{10}})}\\
 	e^{i\beta_{f}}+e^{i(\theta_2+\beta_{h_{01}})} + e^{i(\theta_1+\beta_{g_{10}})}   &  -3\mu-\alpha_{_B}  
 \end{array}\right).
 $$ 
 Replacing\footnote{For simplicity, we will just use $u_e$ instead of $u_{e_{_{00}}}$, to $e = f,g,h$; ditto for $\beta_e$.} $\beta_{f_{\gamma_1,\gamma_2}}= \beta_{g_{\gamma_1,\gamma_2}} = 0$ and $\beta_{h_{\gamma_1,\gamma_2}}=-\gamma_1\phi$, with $\phi=2\pi\frac{1}{1}=2\pi$, we get the polynomial $P(\eta) := \det(M_{_{\hexagon}}(\theta))$ and the dispersion relation $r^{\pm}(\theta)$ given by 
 \begin{equation*}
 	P(\eta)=9\eta^2+3(\alpha_{_N}+\alpha_{_B})\eta+\alpha_{_B}\alpha_{_N}-3-2\cos(\theta_1)-2\cos(\theta_2)-2\cos(\theta_1-\theta_2)\,,
 \end{equation*}	
 \vspace{-0.5cm}	
 \begin{equation*}	
 	r^{\pm}(\theta)=\frac{-\alpha_{_B}-\alpha_{_N}\pm\sqrt{(\alpha_{_B}-\alpha_{_N})^2+4(3+2\cos(\theta_1)+2\cos(\theta_2)+2\cos(\theta_1-\theta_2)) }}{6}\,,
 \end{equation*}
 and we have already seen in the Chapter ~\ref{capPrelimina}, Corollary \ref{Rel-Disp-hBN}, that this dispersion relation has Dirac cones if $\alpha_{_N}=\alpha_{_B}$; see Figure \ref{Cone-Mag-q1}.
 
 \begin{figure}[H]
 	\centering
 	\includegraphics[width=0.5\linewidth]{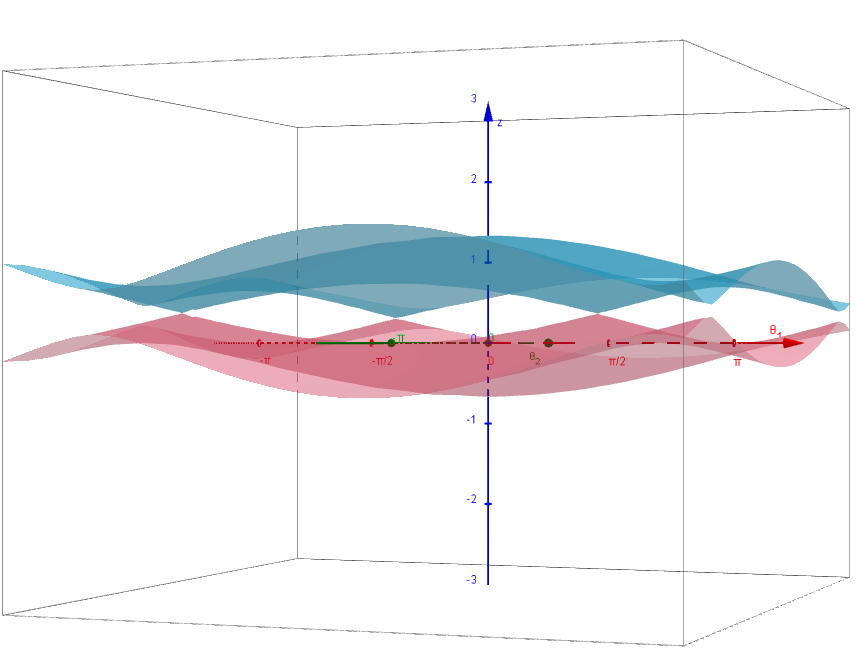}
 	\caption{For $q=1$, the cone is maintained at $(\theta_1,\theta_2) \in [0,\pi)\times[-\pi,\pi)$ for $\alpha_{_N}=\alpha_{_B}$. Here $\alpha_{_B}=\alpha_{_N}=-1.$}
 	\label{Cone-Mag-q1}
 \end{figure}
 
 In the case $q=2$, the period of the hexagonal lattice is 1, while the period of the operator is $2$ and, therefore, we will have another modeling different from that covered in Chapter \ref{capPrelimina}, for a boron nitride sheet, as shown in Figure \ref{Rede-Mag-q=2}. 
 \begin{figure}[H]
 	\centering
 	\includegraphics[width=0.9\linewidth]{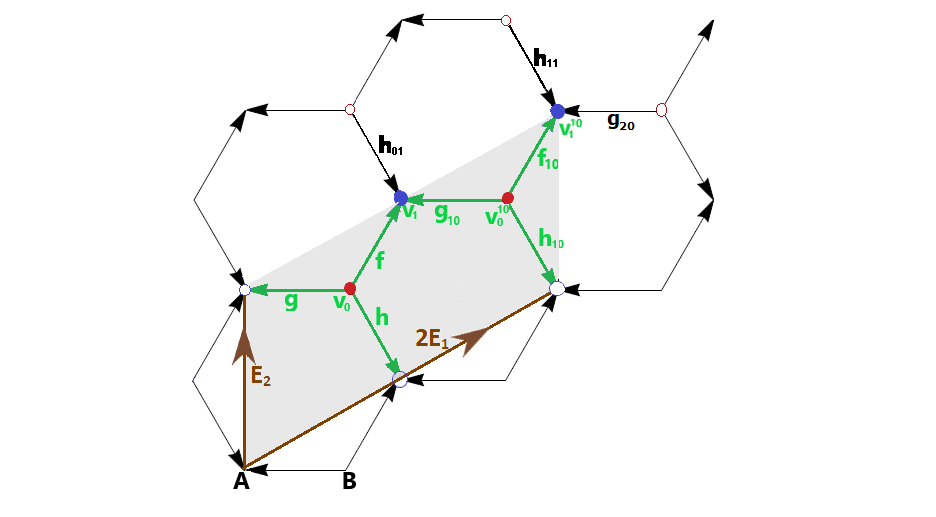}
 	\caption{Period of the operator $\Delta_\beta$ for $q=2$.}
 	\label{Rede-Mag-q=2}
 \end{figure}
 In this case, the fundamental domain is given by
 $$
 W_{\hexagon}=\big\{f,g,h,f_{10},g_{10}, h_{10}, v_0,v_1,v^{1,0}_0, v^{1,0}_1\big \}.
 $$
 The continuity and flow equations at the vertices are now given by
 \noindent\begin{equation}
 	\!\!\! \mbox{\!\!\!Continuity at the vertices:}	\left\{
 	\begin{array}{llll}
 		\!\!\!u_{f}(0) = u_{g}(0) = u_{h}(0)\!\!\!&=:& \!\!\!A_1\\
 		\!\!\!	u_{f_{10}}(0) = u_{g_{10}}(0) = u_{h_{10}}(0)\!\!\!&=:& \!\!\!A_2\\
 		\!\!\!	e^{i\beta_f}u_{f}(1)= e^{i\beta_{g_{10}}}u_{g_{10}}(1) = e^{i\beta_{h_{01}}}e^{i\theta_2}u_{h}(1) \!\!\!&=:& \!\!\!B_1\\
 		\!\!\!  e^{i\beta_{f_{10}}}u_{f_{10}}(1) = e^{i\beta_{g_{20}}}e^{i\theta_1}u_{g}(1) = e^{i\beta_{h_{11}}}e^{i\theta_2}u_{h_{10}}(1) \!\!\!&=:& \!\!\!B_2
 	\end{array}
 	\right.
 \end{equation}
 and
 \begin{equation}\label{Siste21}
 	\mbox{Flux Condition:}\left\{
 	\begin{array}{llll}
 		u'_{f_{00}}(0)+ u'_{g_{00}}(0)+u'_{h_{00}}(0)&=& \delta_{_N}A_1\\
 		-u'_{f_{00}}(1) - u'_{g_{10}}(1) - u'_{h_{01}}(1) &=& \delta_{_B}B_1\\
 		u'_{f_{10}}(0)+ u'_{g_{10}}(0) + u'_{h_{10}}(0) &=& \delta_{_N}A_2\\
 		-u'_{f_{10}}(1) - u'_{g_{20}}(1)+u'_{h_{11}}(1) &=&\delta_{_B}B_2
 	\end{array}.
 	\right.
 \end{equation}
 For $\lambda \notin \sigma(H^D)$, we can write 
 \begin{equation}
 	\left\{
 	\begin{array}{llllll}\label{CombiLinearLI2}
 		u_{f_{_{00}}}&=&A_1\varphi_{0}+B_1e^{-i\beta_{f_{_{00}}}}\varphi_{1}\\
 		u_{g_{_{00}}}&=&A_1\varphi_{0}+B_2e^{-i(\theta_1+\beta_{g_{_{20}}})}\varphi_{1}\\
 		u_{h_{_{00}}}&=&A_1\varphi_{0}+B_1e^{-i(\theta_2+\beta_{h_{_{01}}})}\varphi_{1}\\
 		u_{f_{_{10}}}&=&A_2\varphi_{0}+B_2e^{-i\beta_{f_{_{10}}}}\varphi_{1}\\
 		u_{g_{_{10}}}&=& A_2\varphi_{0}+B_1e^{-i\beta_{g_{_{10}}}}\varphi_{1}\\
 		u_{h_{_{10}}}&=& A_2\varphi_{0}+B_2e^{-i(\theta_2+\beta_{h_{_{11}}})}\varphi_{1}
 	\end{array}. \,\,
 	\right.
 \end{equation}	
 Note that the edges $g_{20}, h_{01}, h_{11}$ are outside the fundamental domain $W_{\!_{\hexagon}}$; in this case, we have the relations $u_{g_{20}}(1)=e^{i\theta_1}u_{g}(1)$, $u_{h_{01}}(1)=e^{i \theta_2}u_{h}(1)$ and $u_{h_{11}}(1)=e^{i\theta_2}u_{h_{10}}(1)$. Using this fact, combining the equations \eqref{Siste21} and \eqref{CombiLinearLI2}, with the same notations used in the previous chapters for $\eta$ and $\alpha_{_B}, \alpha_{_N}$, we obtain the matrix representation
 \begin{equation*}
 	M_{\hexagon}(\theta)\!=\!\left(\!\!\!\begin{array}{cccc}
 		-3\eta-\alpha_{_N} & e^{-i\beta_{f}}+e^{-i(\theta_2+\beta_{h_{01}})} & 0 & e^{-i(\theta_1+\beta_{g_{20}})}\\
 		e^{i\beta_{f}}+e^{i(\theta_2+\beta_{h_{01}})} &  -3\eta-\alpha_{_B}   & e^{i\beta_{g_{10}}} & 0  \\
 		0 &  e^{-i\beta_{g_{10}}}  &  -3\eta-\alpha_{_N}  & e^{-i\beta_{f_{10}}}+e^{-i(\theta_2+\beta_{h_{11}})}  \\
 		e^{i(\theta_1+\beta_{g_{20})}} &  0  & e^{i\beta_{f_{10}}}+e^{i(\theta_2+\beta_{h_{11}})} &  -3\eta-\alpha_{_B}   
 	\end{array}\!\!\!\!\right)
 \end{equation*}
 and considering $\beta_{f_{\gamma_1,\gamma_2}}= \beta_{g_{\gamma_1,\gamma_2}} = 0$ and $\beta_{h_{\gamma_1,\gamma_2}}=-\gamma_1\phi$, with $\phi=\pi$, we rewrite the matrix
 $$
 M_{\hexagon}\big(\theta_1,\theta_2\big)=\left(\begin{array}{cccc}
 	-3\eta-\alpha_{_N} & 1+e^{-i\theta_2} & 0 & e^{-i\theta_1}\\
 	1+e^{i\theta_2} &   -3\eta-\alpha_{_B}   & 1 & 0  \\
 	0 &  1  &  -3\eta-\alpha_{_N}  & 1-e^{-i\theta_2}  \\
 	e^{i\theta_1} &  0  & 1-e^{i\theta_2} &  -3\eta-\alpha_{_B}  
 \end{array}\right).
 $$
 Putting	
 $P(\eta):=\det\Big(M_{\hexagon}\big(\theta_1,\theta_2\big)\Big)$, we get 
 \begin{eqnarray*}
 	P(\eta)\!\!&=&\!\!81\eta^4+54(\alpha_{_N}+\alpha_{_B})\eta^3+9\big[\alpha_{_N}^2+\alpha_{_B}^2+4\alpha_{_N}\alpha_{_B}-6\big]\eta^2-6(\alpha_{_N}+\alpha_{_B})(3-\alpha_{_N}\alpha_{_B})\eta\\
 	&+&\alpha_{_B}^2\alpha_{_N}^2-6\alpha_{_N}\alpha_{_B}+3+2\cos(\theta_1-2\theta_2)-2\cos(\theta_1)-2\cos(2\theta_2)\,.
 \end{eqnarray*}
 \noindent \textbf{Case $\alpha_{_N}=\alpha_{_B}$:} Here, we get
 \begin{eqnarray*}
 	P_{B}(\eta)\!\!\!&\!\!=\!\!&\!\!\!81\eta^4\!+\!108\alpha_{_B}\eta^3\!+\!54(\alpha_{_B}^2\!-\!1)\eta^2\!+\!12(\alpha_{_B}^2\!-\!3)\alpha_{_B}\eta\!+\!\alpha_{_B}^4\!-\!6\alpha_{_B}^2\\
 	\! &\!+\!&\!3\!+\!2\cos(\theta_1\!-\!2\theta_2)\!-\!2\cos(\theta_1)\!-\!2\cos(2\theta_2),
 \end{eqnarray*}	
 whose roots are
 $$
 r^{\pm}_{\mp}(\theta_1,\theta_2)=\frac{\alpha_{_B}}{3}\pm\frac{\sqrt{3\mp\sqrt{2}\sqrt{3+\cos(\theta_1)+\cos(2\theta_2)-\cos(\theta_1-2\theta_2)}\,\,} \,}{3}\,,
 $$
 which, according to its graphical representation in Figure \ref{Cone-Mag-q=2} (with $\alpha_{_B}=\alpha_{_N}=-1$), we see that the conic touch is maintained for $(\theta_1,\theta_2) \in [0,\frac{\pi}{2})\times[-\frac{\pi}{2},\frac{\pi}{2})$ for $\alpha_{_B}$ and $\alpha_{_N}$ equal; a fact that was already expected, since $\alpha(v)$ would be a constant in this case.
 
 \begin{figure}[H]
 	\centering
 	\includegraphics[width=0.7\linewidth]{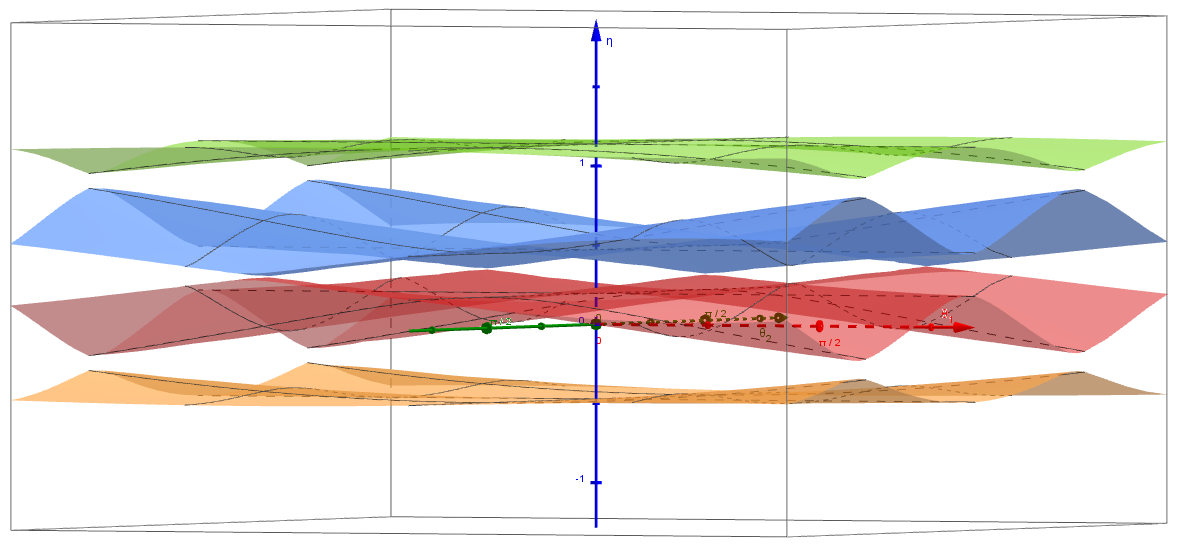}
 	\caption{For $q=2$, the cone is maintained at $(\theta_1,\theta_2) \in [0,\frac{\pi}{2})\times[-\frac{\pi}{2},\frac{\pi}{2})$ for $\alpha_{_N}=\alpha_{_B}$, as expected, since in this case, $\alpha(v)$ is a constant.}
 	\label{Cone-Mag-q=2}
 \end{figure}
 
 \noindent \textbf{Case $\alpha_{_N}=-\alpha_{_B}$.} In this case, we obtain a biquadratic polynomial in $\eta$  	\begin{eqnarray*}
 	P_B(\eta)&\!\!=\!\!& 81\eta^4-6(3\alpha_{_B}^2+1)\eta^2
 	+\alpha_{_B}^4+6\alpha_{_B}^2+3+2\cos(\theta_1-2\theta_2)-2\cos(\theta_1)-2\cos(2\theta_2)
 \end{eqnarray*}
 whose roots are
 $$
 \eta^{\pm}_{\mp}(\theta_1,\theta_2)=\pm\frac{\sqrt{\alpha_{_B}^2+3\mp\sqrt{2}\sqrt{3+\cos(\theta_1)+\cos(2\theta_2)-\cos(\theta_1-2\theta_2)}\,\,} \,}{3}\,,
 $$
 and from Figure \ref{Gap-Mag-q=2} (with $\alpha_{_B}=1$ and $\alpha_{_N}=-1$), we see that gaps (open) are formed at the points where conical touches occurred, or that is, for non-constant $\alpha(v)$, the operator dispersion relation does not have Dirac cones (even with rational magnetic flux with respect to $2\pi$).
 
 \begin{figure}[H]
 	\centering
 	\includegraphics[width=0.7\linewidth]{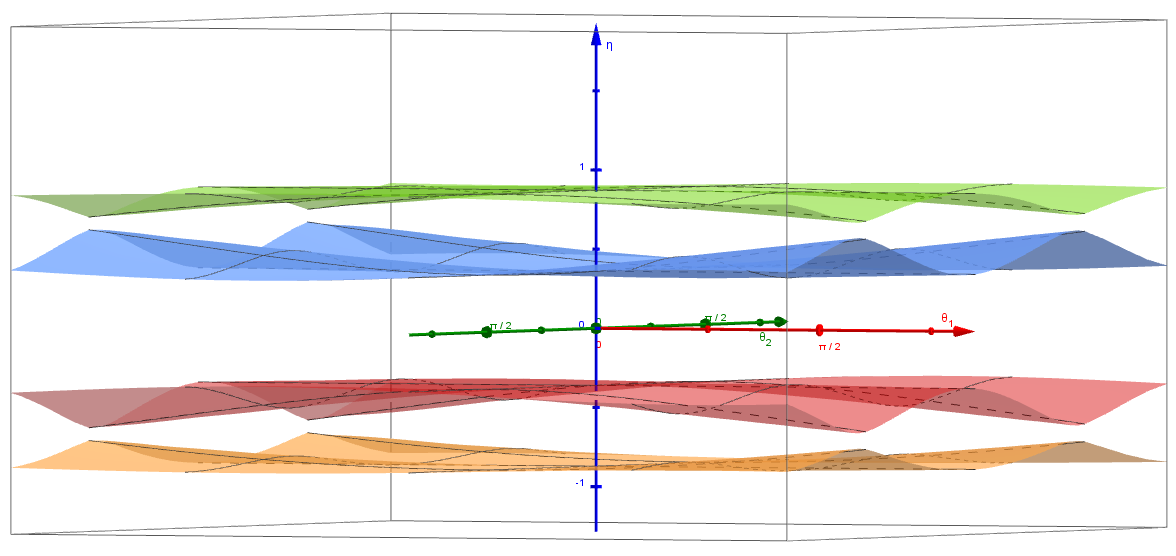}
 	\caption{Graphical representation of the roots $\eta^{\pm}_{\mp}$ with $\alpha_{_B}=1, \alpha_{_N}=-1$. For $q=2$, in $(\theta_1,\theta_2) \in [0,\frac{\pi}{2})\times[-\frac{\pi}{2},\frac{\pi }{2})$ the cone is not maintained for $\alpha_{_N}=-\alpha_{_B}$, that is, for non-constant $\alpha(v)$, the operator dispersion relation has an (open) gap.}
 	\label{Gap-Mag-q=2}
 \end{figure}
 
 In general, for arbitrary values of real $\alpha_{_B}$ and $\alpha_{_N}$, on each edge $e \in \hexagon$, we summarize these facts in the following theorem.
 
 \begin{theorem}[Magnetic Graphene and hBN] Let $\alpha_{_B}, \alpha_{_N}$ be any real numbers, $(\theta_1,\theta_2) \in [0,\frac{\pi}{2})\times[-\frac{\pi}{2},\frac{\pi}{2})$, $a_e \in C^1[0,1]$ is a magnetic potential and $U \in L^2[0,1]$ is a symmetric potential. Consider the operator
 	\[
 	H^{\phi}_{\hexagon}:(f_e) \mapsto (-f''_e+Uf_e), \quad (f_e) \in  \bigoplus_{e\in \mathcal{E}}{\mathcal{H}}^2[0,1],
 	\] with constant magnetic flux $\phi$ in the hexagons of the hexagonal lattice, and boundary conditions
 	\begin{equation*}
 		\left\{
 		\begin{array}{ll}
 			f_a(0)=e^{i\beta_{e}}f_b(1) =:f(v)\,, \forall \,\, a,b \in \mathcal{E}_v\,;\\
 			\dis\sum_{e; i(e)=v}f'_e(0) - \sum_{e; t(e)=v}e^{\beta_e}f'_e(1)  = \alpha(v)f(v)\,,\,\, v \in \mathcal{V}\,, 
 		\end{array}\,
 		\right.\,
 	\end{equation*}
 	where $\beta_e:= \int^1_0\!\!a_e(s)ds$.
 	Suppose $\phi=\pi$; then the dispersion relation of the operator $H^{\phi}_{\hexagon}$ is given by
 	\begin{equation*}
 		\eta^{\pm}_{\mp}(\theta_1,\theta_2)=-\frac{\alpha_{_N}+\alpha_{_B}}{6}
 		\pm\frac{\sqrt{(\alpha_{_N}-\alpha_{_B})^2+12\mp4\sqrt{2}\sqrt{3+\cos(\theta_1)+\cos(2\theta_2)-\cos(\theta_1-2\theta_2)\,\,} \,}}{6}\,.
 	\end{equation*}
 	From this, it can be concluded that 
 	\begin{itemize}
 		\item[(a)] If $\alpha(v)$ is constant at the vertices $v \in \hexagon$, $H^{\phi}_{\hexagon}$ has Dirac cones; 
 		\item[(b)] If $\alpha(v)$ is not constant at the vertices $v \in \hexagon$, then $H^{\phi}_{\hexagon}$ has gaps.
 	\end{itemize}
 	
 	In particular, for $\phi=\pi$ or $\phi=2\pi$, magnetic graphene has a Dirac cone, while magnetic hexagonal boron nitride has a gap between the bands.
 \end{theorem}
 
 Note that taking $(\theta_1,\theta_2) \in [0,\frac{\pi}{q})\times[-\frac{\pi}{q},\frac{\pi}{q}) $, we saw that for $q=1$, according to the Chapter \ref{capPrelimina} and also in this section, the Dirac Cones, if they exist, occur when $\theta_1=-\theta_2$, while for $q=2 $ conic touches, when they exist, occur for $\theta_1=-2\theta_2$. In fact, putting
 \[
 F(\theta_1,\theta_2) := 3+\cos(\theta_1) +\cos(2\theta_2)-\cos(\theta_1-2\theta_2),
 \] we see that $F(\theta_1,\theta_2)=0$ if, and only if, $\theta_1= \pi+ 2m\pi$ and $\theta_2= -\frac{\pi}{2}+n\pi$ , \, $m,n \in \Z$, that is, $F$ cancels only under the condition $\theta_1=-2\theta_2$.
 
 \section{Analysis and conclusions}
 
 According to the development explained above, of these models with magnetic field, if $\alpha(v)$ is not constant, it is not possible to guarantee the relation \eqref{Rel-Espec}, that is, it is not guaranteed that it is possible to relate the spectra from $\Delta_\beta$ into $l^2(\hexagon)$ and $H$ into $ \bigoplus_{e\in \mathcal{E}}{\mathcal{H}}^2[0,1]$ , since such a relation was obtained by supposing $\alpha(v)=\frac{\alpha}{2} \operatorname{deg}(v)$, with $\alpha \in \R$ taking the same value in all the vertices (according to our proposal, they would all be models with a single type of atom at all vertices, like graphene). 
 
 We then show that there are integer values $p$ and $q$, and for certain values of $\alpha$ taking two values on the hexagonal lattice $\hexagon$ (thus a model for boron nitride), in which the conical touches on the dispersion relation of the operator $H^{\phi}_{\hexagon}$ cease to exist and we also guarantee the existence of gaps. To do so, we use the Theorem \ref{Teo17} and the relation \eqref{Rel-Espec}, however we are mainly based on the technique presented in the Chapter \ref{capPrelimina}, which we adapt to the case with a magnetic field in a specific configuration in which we used a particular period of the operator, which differed from the period of the hexagonal lattice. It was convenient to use this technique due to the fact that we do not have an analogue to the Chambers Formula used in \cite{BH}, which also restricts us to small periods.
 
 We then conclude that the hexagonal quantum graph model, with the continuous Laplacian operator on the edges and with a magnetic field, which models the magnetic hexagonal boron nitride, does not always have a Dirac cone if the modeling considers the flow $\phi=2\pi\frac{p}{q}$ (rational multiple of $2\pi$) constant in this network. In the case of boron nitride, we take for the parameters in the boundary conditions the values $\alpha(v)=\delta_{_N}$ if $v$ is the initial vertex of the edges, and $\alpha(v)=\delta_{_B}$ if $ v$ for final vertex. On the other hand, this freedom of choosing different values for $\alpha(v)$ at the vertices does not occur in the case of graphene, since this material has the same constant $\alpha(v)=\delta_{_C}$ at all vertices of the hexagonal lattice and, consequently, the operator proposed to model graphene will always present Dirac cones, for any values of $p$ and $q$ such that the magnetic flux $\phi$ in each hexagon is a rational multiple of $2\pi$. These adaptations to $\delta_{_C}\ne0$ and the example with rational flow in relation to $2\pi$ without Dirac cones constitute our main contributions to this model and also to this last chapter of the thesis.
 
 \chapter{Final considerations}
 
 In Chapter \ref{capPrelimina}, we presented the modeling of a sheet of hexagonal material with two types of alternating atoms at the vertices and concluded that its dispersion relationship will have a Dirac cone if the atoms are equal and will have gaps, otherwise. In particular, graphene monolayer always has a conical touch, while hexagonal boron nitride always has gaps. 
 
 Then, in Chapter \ref{capEstrHomAA}, we extend the study in Chapter \ref{capPrelimina} to stacked bilayers of the $AA$ type; we conclude, according to our modeling, that in materials of this nature there will be a Dirac cone only if the atoms are equal, and there will be parabolic touches if the parameters comply with the relation $\alpha_{b}=\pm\,2t_0^2\,+\,\alpha_a $; in particular, two sheets of hBN, stacked like $AA$, do not have Dirac cones, but may have parabolic touches. 
 
 On the other hand, if in the configuration of Chapter \ref{capEstrHomAA} the stacking in type $AA'$ is considered, we conclude in Chapter \ref{capEstrHomAAnovo} that this material will not have conical or parabolic touches if the atoms are distinct, but they have parabolic touch if the atoms are equal. In particular, in the $AA'$ stacking, hBN bilayers always have holes, while graphene bilayers always have Dirac cones.
 
 Heterostructures were considered in Chapter \ref{capHetNBG}; we confirmed the presence of Dirac cones in the system composed of a layer of graphene and another of hBN, a fact already covered in the physics literature. We also identified conical touches or gaps in the graphene between boron nitride and boron nitride between graphene sandwiches, respectively. In the former, graphene caused a reduction in the openings of existing gaps in the hBN bilayers, while in this case, the hBN induced a gap in some Dirac points of the graphene and a conical touch is always maintained in this configuration.
 
 Finally, in Chapter \ref{capGrafNBMagn}, we use quantum graphs in a proposal for modeling hexagonal boron nitride under the effect of a magnetic field, through the Schr\"{o}dinger operator, and demonstrate that if the magnetic flux $\phi $ is constant in the lattice hexagons and of the form $\phi = 2\pi\dfrac{p}{q}$, there are integer values $p$ and $q$ such that, for certain boundary conditions $\alpha(v)$ at the vertices $v$ of the network, gaps are formed at the conic points of the dispersion relation of the operator defined in this network.
 
 Although we present a modest and simplified mathematical modeling proposal for hBN and graphene, we consider the results obtained analytically to be consistent and satisfactory: we confirm some facts already known in the physics literature obtained, in general, numerically or through laboratory experiments, and in some In some cases (such as $AA$ stacking) we have behaviors obtained from this type of model that would be interesting to confirm.



\end{document}